\documentclass[preprint,aps,superscriptaddress,preprintnumbers,amsmath,
amssymb,floatfix,12pt,nofootinbib]{revtex4}

\usepackage{color} 
\usepackage{subfig}
\usepackage{listings}

\usepackage{graphicx}

\lstdefinestyle{mystyle}{
	backgroundcolor=\color{backcolour},   commentstyle=\color{codegreen},
	keywordstyle=\color{magenta},
	numberstyle=\tiny\color{codegray},
	stringstyle=\color{codepurple},
	basicstyle=\footnotesize,
	breakatwhitespace=false,    
	basicstyle=\footnotesize\ttfamily,     
	breaklines=true, 
	lineskip={-1.5pt}, 	                
	captionpos=b,                    
	keepspaces=true,                                
	numbersep=5pt,                  
	showspaces=false,                
	showstringspaces=false,
	showtabs=false,                  
	tabsize=4,
	xleftmargin=30pt, 
	xrightmargin=30pt
}

\newcommand{\bea}{\begin{eqnarray}}
\newcommand{\eea}{\end{eqnarray}}

\definecolor{codegreen}{rgb}{0,0.6,0}
\definecolor{codegray}{rgb}{0.5,0.5,0.5}
\definecolor{codepurple}{rgb}{0.58,0,0.82}
\definecolor{backcolour}{rgb}{0.95,0.95,0.92}

                                {\end{figure}}

                                {\end{table}}

 \def\spose#1{\hbox to 0pt{#1\hss}}
 \def\gtapprox{\mathrel{\spose{\lower 3pt\hbox{$\mathchar"218$}}
 \raise 2.0pt\hbox{$\mathchar"13E$}}}

\newcommand*{\CPT}{\raise0.4ex\hbox{$\chi$}PT}
\newcommand*{\chpt}{\raise0.4ex\hbox{$\chi$}PT}
\newcommand*{\schpt}{S\raise0.4ex\hbox{$\chi$}PT}

\def\eqref#1{{(\ref{#1})}}

\def\hat{\widehat}

\def\bea{\begin{eqnarray}}
\def\eea{\end{eqnarray}}

\begin{document}
\bibliographystyle{apsrev}

\pacs{ }
\title{Lattice quantum gravity and asymptotic safety}

\author{J.~Laiho} 
\email[]{jwlaiho@syr.edu}
\affiliation{Department of Physics, Syracuse University,
Syracuse, New York, USA}

\author{S.~Bassler}
\affiliation{Department of Physics, Syracuse University,
Syracuse, NY, USA}

\author{D.~Coumbe}
\affiliation{The Niels Bohr Institute, Copenhagen University, Blegdamsvej 17, DK-2100 Copenhagen, Denmark}

\author{D.~Du}
\affiliation{Department of Physics, Syracuse University,
Syracuse, NY, USA}

\author{J.~T.~Neelakanta}
\affiliation{Department of Physics, Syracuse University,
Syracuse, NY, USA}

\begin{abstract}
We study the nonperturbative formulation of quantum gravity defined via Euclidean dynamical triangulations (EDT) in an attempt to make contact with Weinberg's asymptotic safety scenario.  We find that a fine-tuning is necessary in order to recover semiclassical behavior.  Such a fine-tuning is generally associated with the breaking of a target symmetry by the lattice regulator; in this case we argue that the target symmetry is the general coordinate invariance of the theory.  After introducing and fine-tuning a nontrivial local measure term, we find no barrier to taking a continuum limit, and we find evidence that four-dimensional, semiclassical geometries are recovered at long distance scales in the continuum limit.  We also find that the spectral dimension at short distance scales is consistent with 3/2, a value that could resolve the tension between asymptotic safety and the holographic entropy scaling of black holes. We argue that the number of relevant couplings in the continuum theory is one, once symmetry breaking by the lattice regulator is accounted for.  Such a theory is maximally predictive, with no adjustable parameters.  The cosmological constant in Planck units is the only relevant parameter, which serves to set the lattice scale.  The cosmological constant in Planck units is of order 1 in the ultraviolet and undergoes renormalization group running to small values in the infrared.  If these findings hold up under further scrutiny, the lattice may provide a nonperturbative definition of a renormalizable quantum field theory of general relativity with no adjustable parameters and a cosmological constant that is naturally small in the infrared.
\end{abstract}

\date{\today}

\maketitle

\section{Introduction}\label{sect:intro}

The most straightforward implementation of general relativity as a quantum field theory encounters serious difficulties.  There are two main problems.  The first is that a coupling expansion of general relativity is nonrenormalizable, which can be seen already at the level of power counting.  The gravitational coupling $G$ has dimension $-2$, so that higher order corrections in a perturbative expansion come with ever-increasing powers of $\mu/m_{\rm Planck}$, with $\mu$ the cutoff scale.  The divergences have been verified by explicit calculation; they first appear at two loops \cite{Goroff:1985th} in the case of pure gravity and at one loop \cite{'tHooft:1974bx} in the presence of matter.  Although perturbative quantum gravity still makes sense as a low energy effective theory \cite{Donoghue:1997hx}, the infinite number of low energy constants required leads to a loss of predictive power.  

The second problem is that the small value of the vacuum energy is not explained.  There is an energy density, which gravity should see, associated with the quantum fluctuations of every field in the system; for example, effective field theory arguments suggest that the fluctuations in the gravitational field contribute to the vacuum energy at graviton energies nearly up to the Planck scale.  This would lead to a cosmological constant that is around 120 orders of magnitude larger than what has been observed \cite{Weinberg:2000yb}. 

To address the first problem, Weinberg suggested decades ago that quantum gravity might be asymptotically safe \cite{Weinberg:1980gg}, i.e. that gravity might be effectively renormalizable when formulated nonperturbatively.  In this scenario, the renormalization group flow of gravitational couplings has a nontrivial fixed point, with a finite dimensional ultraviolet critical surface of trajectories attracted to the fixed point at short distances.  Asymptotically free theories, such as quantum chromodynamics (QCD), are a special case of asymptotically safe theories, where the fixed point is the Gaussian fixed point.  Such theories become noninteracting in the high energy limit.  For gravity, because the theory is perturbatively nonrenormalizable, the theory is not asymptotically free, but may still possess a nontrivial fixed point, such that it is asymptotically safe.  The couplings at the fixed point are not necessarily small, so that an investigation of this scenario for gravity requires nonperturbative methods.

The lattice is a powerful tool for performing nonperturbative calculations in quantum field theory and is the standard for precision calculations of QCD in the nonperturbative regime \cite{Agashe:2014kda, Aoki:2013ldr}.  In a lattice formulation of an asymptotically safe field theory, the fixed point would appear as a second-order critical point, the approach to which would define a continuum limit.  The divergent correlation length characteristic of a continuous phase transition would allow one to take the lattice spacing to zero while keeping observable quantities fixed in physical units.  

Euclidean dynamical triangulations (EDT) is a particular implementation of a lattice regularization of quantum gravity \cite{Ambjorn:1991pq, Agishtein:1991cv}.  
This approach has been successful in two dimensions, where gravity coupled to conformal matter corresponds to bosonic string theory; the results from the lattice agree with the continuum calculations in noncritical string theory wherever they can be compared \cite{Ambjorn:2002uk}.  The perfect agreement between lattice and continuum methods is a powerful test of the dynamical triangulations approach, at least in two dimensions.  Early studies of four-dimensional EDT, however, were not so successful.   Numerical calculations showed that the model has two phases, neither of which looks much like semiclassical general relativity in four dimensions  \cite{Ambjorn:1991pq,  deBakker:1994zf, Ambjorn:1995dj, Catterall:1994pg, Egawa:1996fu}.  Furthermore, the critical point separating these two phases, though initially thought to be second order, turned out to be first order, thus ruling out the usual approach of taking a continuum limit \cite{Bialas:1996wu, deBakker:1996zx}\footnote{For more recent studies confirming that the phase transition is first order, see Refs. \cite{Ambjorn:2013eha, Coumbe:2014nea, Rindlisbacher:2015ewa}.}.

Early work that included a nontrivial local measure term with a new arbitrary coupling identified a region of the extended phase diagram with properties that appeared to be different from the previously identified phases \cite{Bruegmann:1992jk, Bilke:1998vj}; this was dubbed the crinkled region.  
However, follow-up studies showed that the crinkled region did not have desirable features, like a dimension close to four, suggesting that the crinkled region was most likely a part of the collapsed phase with particularly large finite-size effects \cite{Ambjorn:2013eha, Coumbe:2014nea}.  The presence of a crossover rather than a genuine phase transition between the collapsed phase and the crinkled region further supported this conclusion.

There has been progress on the lattice using a variation on the dynamical triangulations approach known as causal dynamical triangulations (CDT).  The difficulties encountered in the original EDT approach led Ambj{\o}rn and Loll to introduce a causality constraint on the set of triangulations over which the sum in the path integral is performed \cite{Ambjorn:1998xu}, an approach they called CDT.  This approach distinguishes between spacelike and timelike links on the lattice so that an explicit foliation of the lattice into spacelike hypersurfaces of fixed topology can be introduced.  A number of interesting results have been obtained from numerical simulations using this approach, including the emergence of a four-dimensional semiclassical phase resembling Euclidean de Sitter space \cite{Ambjorn:2004qm, Ambjorn:2005qt}, and a fractal (spectral) dimension that runs as a function of the distance scale probed \cite{Ambjorn:2005db}.  However, the foliation requires the breaking of space-time symmetry, casting doubt on whether CDT maintains general covariance and has general relativity as its classical limit.  Recent work (in three dimensions) suggests that it is possible to relax the restriction of a fixed foliation in CDT \cite{Jordan:2013awa}, but a different treatment of spacelike and timelike links is still necessary.  It is thus important to revisit the explicitly space-time symmetric EDT formulation to try to address this issue.

In this work we revisit the original EDT approach with a novel interpretation and new calculations, building on and significantly extending previous work \cite{Laiho:2011ya, Coumbe:2014nea}.  We provide evidence that there exists a set of parameter tunings in the original EDT model that allows one to obtain semiclassical geometries in the infrared, where the theory must have a four-dimensional, semiclassical limit if it is to recover classical general relativity.  Our previous work did not implement this fine-tuning \cite{Coumbe:2014nea}, leading to the negative results quoted there.  

The key idea introduced in this paper is that a fine-tuning of the bare parameters is necessary in order to recover the correct continuum limit, in analogy to simulations of lattice QCD with two degenerate-mass flavors of Wilson fermions.  Wilson's solution to the well-known fermion doubling problem is to introduce a term that raises the masses of all the species of fermions except for one to the cutoff scale \cite{Wilson:1975id}.  This term also introduces a hard breaking of chiral symmetry, so that a fine-tuning of the fermion mass is necessary in order to recover QCD with light or massless quarks.  This symmetry is not broken anomalously (for two flavors), but remains a symmetry of the theory even in the presence of quantum effects.  We argue that a similar fine-tuning is necessary for gravity.  The need to fine-tune a bare lattice parameter only occurs when there is a target symmetry that is broken by the regulator, which leads us to question what symmetry it is that the regulator is breaking.  A good candidate for this symmetry is general coordinate invariance.  There are various checks that we have correctly identified the target symmetry, and we present them in this work. 

We argue that a parameter associated with the nonuniform local measure must be tuned to restore coordinate invariance.  Once this fine-tuning is performed, we find evidence that the correct infrared physics is recovered at sufficiently fine lattice spacings.  There also appears to be no obstacle to taking the continuum limit, another indication that the asymptotic safety scenario for gravity is realized in our model.  We also argue that after accounting for the fine-tuning that is necessary due to the breaking of the symmetry by the lattice regulator, there is only one relevant coupling in the theory of pure gravity without matter.  A theory with one relevant coupling, i.e., a theory with a one-dimensional UV critical surface, has no free parameters aside from one dimensionful constant that defines the units of mass or length.  Such a theory has no adjustable parameters and is thus maximally predictive.  Our work suggests that the single relevant parameter in pure gravity is the cosmological constant times Newton's constant $G\Lambda$ and that it decreases to small values under renormalization group running to the infrared.  

There is an argument against the viability of the asymptotic safety scenario due to Banks \cite{Banks:2010tj}, which compares the entropy expected of a renormalizable quantum field theory with that of a theory dominated by black hole degrees of freedom at high energy.  We find that this tension between asymptotic safety and black hole entropy scaling could be resolved by the fractal dimension observed in our model at short distances.  

This paper is organized as follows.  In Sec.~\ref{sec:symmetry} we review the symmetries of general relativity and discuss the implications for dynamical triangulations, in particular the need to perform a fine-tuning.  We also discuss the key questions that must be addressed to have a solution to the main problems of quantum gravity using lattice methods.  In Sec.~\ref{sec:edt} we review the EDT formulation, discuss its numerical implementation, and review the results for the phase diagram obtained so far.  We also discuss what would be needed to address the key questions using this formulation.  In Sec.~\ref{sec:classical} we show that once a fine-tuning is introduced in EDT, we recover geometries with a global scaling dimension close to four, which resemble Euclidean de Sitter space.  The agreement with de Sitter space improves as the continuum limit is approached.  In Sec.~\ref{sec:continuum} we discuss how one can determine a relative lattice spacing in our approach.  We also show that the existence of a continuum limit is plausible, and we present results for the spectral dimension at both long and short distance scales.  We discuss the implications of our results for the short-distance spectral dimension in the asymptotic safety scenario.  In Sec.~\ref{sec:running} we present our argument that there is only one relevant parameter in the continuum quantum gravity theory.  Section~\ref{sect:others} compares our results with that of other methods.  We conclude in Sec.~\ref{sec:concl}.  Finally, an Appendix gives details of the parallel algorithm that we introduce, which has enabled us to produce several of the ensembles generated in this work.

\section{General coordinate invariance and lattice quantum gravity}\label{sec:symmetry}

\subsection{General coordinate invariance}

The fundamental symmetry of general relativity is general coordinate invariance (or more precisely, diffeomorphism invariance), which implies that space-time coordinates have no independent physical meaning.  Under an infinitesimal diffeomorphism, $x^\mu\to x^\mu-\xi^\mu$, the metric $g_{\mu\nu}$ transforms as
\bea  \delta g_{\mu\nu}= \nabla_\mu\xi_\nu + \nabla_\nu \xi_\mu.
\eea

Because continuum diffeomorphism invariance is a continuous space-time symmetry, it is difficult to preserve in a lattice formulation.  Dynamical triangulations is an attempt to avoid this problem by working directly on the space of invariant geometries, without explicitly introducing a metric.  Each physical geometry is mapped to a single discretized geometry which best describes it, so that states are not overcounted or undercounted in a path-integral approach.  This correct counting is argued to hold if the edge lengths of the lattice building blocks are held fixed \cite{Romer:1985rc}, as in the dynamical triangulations approach.  Dynamical triangulations should then behave like the standard, gauge-invariant, Wilson lattice formulation of a gauge theory, where gauge fixing and the introduction of a Fadeev-Popov determinant are not needed.  This is born out in two-dimensional dynamical triangulations, where nontrivial dynamics appears in the conformal sector of the quantum theory.  The 2D dynamical triangulations approach agrees perfectly with continuum analytical calculations, even though the continuum calculations involve gauge fixing and the introduction of a Fadeev-Popov determinant \cite{Ambjorn:2002uk}.  Thus, the diffeomorphism invariant theory is recovered in the continuum limit of dynamical triangulations in two dimensions.

In four dimensions there are genuine field degrees of freedom and nontrivial dynamics in the classical theory, unlike in two dimensions, and it may be that the full continuum diffeomorphism invariance is more difficult to recover.  Although the action is expressed in terms of coordinate invariant quantities, and the only remnant of a coordinate system is the vertex labeling, the discretization implies that there is no notion of continuum diffeomorphism invariance.  The question is whether working directly with geometric invariant quantities is sufficient to recover the continuum limit of a diffeomorphism invariant continuum theory, or if the discretization is more problematic in four dimensions.  We conjecture that the approach of dynamical triangulations recovers diffeomorphism invariance even at finite lattice spacing if the lattice action is classically perfect.  Classically perfect actions have been studied in lattice QCD, where such a condition on the action is sufficient to guarantee that chiral symmetry is maintained even at finite lattice spacing \cite{Hasenfratz:1993sp}.  This approach to maintaining chiral  symmetry in lattice QCD comes at the expense of including an infinite number of higher dimensional terms in the action, though in practice the action must be truncated, and this approach to chiral fermions suffers some residual breaking of chiral symmetry, depending on how many terms are kept in the truncation.  In two-dimensional gravity, the action in dynamical triangulations with only a cosmological constant term is already classically perfect because the Einstein-Hilbert term is a topological invariant, with all metrics trivially satisfying the classical equations of motion.  Since this is no longer true in the four-dimensional case, additional complications may arise in an attempt to recover the 4D continuum theory.

In this work we will settle for the restoration of the full, continuum, four-dimensional diffeomorphism invariance only in the continuum limit, in analogy to the restoration of chiral symmetry in two-flavor QCD with Wilson fermions.  We present evidence that this is possible with the introduction of a nontrivial (local) path integral measure.  Our motivation for introducing the measure term is as follows.  It has been shown by Fujikawa that the local measure for the gravitational path integral is fixed if one imposes the Becchi-Rouet-Stora-Tyutin (BRST) symmetry associated with general coordinate transformations \cite{Fujikawa:1983im}.  The local measure includes a Jacobian factor of the form $\prod_x \sqrt{g}^\beta$, with $\beta$ fixed by the BRST symmetry.  This local measure term ensures that there are no anomalies associated with general coordinate transformations.  Since coordinate invariance in gravity plays the analogous role of gauge invariance in a gauge theory, a consistent quantum theory requires that this symmetry must be maintained even at the quantum level.  However, if the lattice regulator breaks the continuum coordinate invariance, it is possible that the parameter $\beta$, which would otherwise be fixed, must become a function of lattice spacing and be fine-tuned in order to restore the continuum symmetry.  We present evidence that such a tuning is necessary.

\subsection{Key questions for the lattice theory}

We would like to establish whether the EDT formulation realizes the asymptotic safety scenario for gravity, and in this subsection we outline what would be needed to decide the issue.  There are important problems that need to be addressed.  One of these is the existence of a nontrivial fixed point.  In the standard scenario, this requires the existence of a continuous phase transition, where correlation lengths diverge so that the lattice spacing can be sent to zero while physical length scales remain finite.  Another crucial requirement is the recovery of four-dimensional, semiclassical physics at long distances, so that the continuum theory reproduces classical general relativity in the appropriate limit.  General relativity is well tested, and any quantum theory of gravity must reproduce the successes of the classical theory, or else it is ruled out by observations.  Another requirement, if gravity is to be realized as a consistent quantum field theory, is that the theory must be unitary. 

Yet another important problem that the asymptotic safety scenario for gravity must contend with is the question of how it can be compatible with holography.  There is an argument due to Banks \cite{Banks:2010tj, Shomer:2007vq} against the possibility that gravity could be a renormalizable quantum field theory because of an incompatibility between the holographic (Bekenstein-Hawking) entropy scaling of black holes and that expected of a renormalizable theory at high energies.  This argument compares the density of states at high energy expected for a theory of gravity versus that expected of a conformal field theory.  Since a renormalizable quantum field theory is a perturbation of a conformal field theory by relevant operators, the asymptotic density of states of a renormalizable field theory must match that of a conformal field theory.  The local nature of the conformal theory implies that the scaling of energy and entropy must be extensive in the spatial volume, and given this and the fact that a finite-temperature conformal field theory has no dimensionful scales other than the temperature, dimensional analysis implies that the entropy $S$ and energy $E$ scale as
\bea  S\sim (RT)^{d-1}, \ \ \ \ \ E\sim R^{d-1}T^d,
\eea
where $R$ is the radius of the spatial volume, $T$ is temperature, and $d$ is the space-time dimension.  It then follows that the entropy of a renormalizable theory must scale with energy as
\bea\label{eq:CFT}    S\sim E^{\frac{d-1}{d}}.
\eea
For gravity, the argument goes, the high energy spectrum is dominated by black holes.  The $d$-dimensional Schwarzschild solution in asymptotically flat space-time has a black hole with event horizon of radius $r^{d-3}\sim G M$, where $G$ is Newton's constant and $M$ is the mass of the black hole.  The Bekenstein-Hawking entropy formula tells us that $S\sim r^{d-2}$, leading to
\bea\label{eq:GR}   S\sim E^{\frac{d-2}{d-3}}.
\eea
This scaling disagrees with that of Eq.~(\ref{eq:CFT}) for $d=4$.  Various attempts to address this concern have appeared in the literature, with some questioning the assumptions leading to Eq.~(\ref{eq:GR}) \cite{Percacci:2010af, Falls:2012nd}.

Even if gravity is renormalizable nonperturbatively, such a theory does not necessarily address the cosmological constant fine-tuning problem.  Attempts to solve this fine-tuning problem in the literature are numerous \cite{Weinberg:1988cp, Weinberg:2000yb}, but it might be fair to say that none of them are fully satisfactory.  One way in which this problem may be addressed in the asymptotic safety scenario is if the gravity sector of the theory contains only one parameter that is relevant in the renormalization group sense.  In the case of pure gravity, for example, this happens if the ultraviolet critical surface is one dimensional, and it would mean that the theory has no adjustable parameters, requiring only a single dimensionful input, say Newton's constant, to set the scale.  Once this scale is set, all of the physics at low energies would be a prediction, including, in principle, the cosmological constant.  In the case where matter is added to gravity, assuming that the theory is again asymptotically safe, and assuming that the number of relevant couplings in the matter sector is $n$, if gravity adds only one more relevant coupling, then the theory would be fully determined by a measurement of Newton's constant and the $n$ parameters of the matter sector, and the cosmological constant would again be a prediction.  Whether this scenario is realized for a suitable matter sector, and if so, how it relates to low energy effective field theory attempts to understand the cosmological constant problem, remain open questions.

\section{Lattice formulation}\label{sec:edt}

\subsection{The model}

In Euclidean quantum gravity the partition function is formally given by a path integral sum over all geometries
\bea\label{eq:part}  Z_E = \int {\cal D}[g] e^{-S_{EH}[g]},
\eea
where the Euclidean Einstein-Hilbert action is
\bea  \label{eq:ERcont} S_{EH} =  -\frac{1}{16 \pi G}\int d^4x \sqrt{g} (R - 2\Lambda),
\eea
with $R$ the Ricci curvature scalar, $g$ the determinant of the metric tensor, $G$ Newton's constant, and $\Lambda$ the cosmological constant.
    
In dynamical triangulations, the path integral is formulated directly as a sum over geometries, without the need for gauge fixing or the introduction of a metric.  The dynamical triangulations approach is based on the conjecture that the path integral for Euclidean gravity is given by the partition function \cite{Ambjorn:1991pq, Bilke:1998vj}
\bea\label{eq:Z} Z_E = \sum_T \frac{1}{C_T}\left[\prod_{j=1}^{N_2}{\cal O}(t_j)^\beta\right]e^{-S_{ER}}
\eea
where $C_T$ is a symmetry factor that divides out the number of equivalent ways of labeling the vertices in the triangulation $T$.  The term in brackets in Eq.~(\ref{eq:Z}) is a nonuniform measure term, where the product is over all two-simplices (triangles), and ${\cal O}(t_j)$ is the order of triangle $j$, i.e. the number of four-simplices to which the triangle belongs.  This corresponds in the continuum to a nonuniform weighting of the measure in Eq.~(\ref{eq:part}) by $\prod_x \sqrt{g}^{\beta}$, and in our simulations $\beta$ is treated as a free parameter.  It has been suggested that the nonuniform measure term might be important in four dimensions \cite{Bruegmann:1992jk}; in the next subsection we discuss the role that it plays in our calculations.

  In four dimensions the discretized version of the Einstein-Hilbert action is the Einstein-Regge action \cite{Regge:1961px}
  \begin{equation} \label{eq:GeneralEinstein-ReggeAction}
S_{E}=-\kappa\sum_{j=1}^{N_2} V_{2}\delta_j+\lambda\sum_{j=1}^{N_4} V_{4},
\end{equation}
  \noindent where $\delta_j=2\pi-{\cal O}(t_j)\arccos(1/4)$ is the deficit angle around a triangular hinge $t_j$, with ${\cal O}(t_j)$ the number of four-simplices meeting at the hinge, $\kappa=\left(8\pi G \right)^{-1}$, $\lambda=\kappa\Lambda$, and the volume of a $d$-simplex is 
\begin{equation} \label{eq:SimplexVolume}
V_{d}=\frac{\sqrt{d+1}}{d!\sqrt{2^{d}}}a^d,
\end{equation}
\noindent where the equilateral $d$-simplex has a side of length $a$.  After performing the sums in Eq.~(\ref{eq:GeneralEinstein-ReggeAction}) one finds
\begin{equation}\label{eq:DiscAction}
S_{E}=-\frac{\sqrt{3}}{2}\pi\kappa N_{2}+N_{4}\left(\kappa\frac{5\sqrt{3}}{2}\mbox{arccos}\frac{1}{4}+\frac{\sqrt{5}}{96}\lambda\right),
\end{equation}
where $N_i$ is the number of simplices of dimension $i$.  Introducing a new parametrization, we rewrite the Einstein-Regge action in the simple form 
\bea\label{eq:ER}  S_{ER}=-\kappa_2 N_2+\kappa_4N_4,
\eea
where we have made the identifications $\kappa_2=\frac{\sqrt{3}a^2}{16G}$ and $\kappa_4=\frac{5\sqrt{3}a^2}{16\pi G}\arccos(\frac{1}{4})+\frac{\sqrt{5}}{96}\frac{\Lambda a^4}{8\pi G}$.
These relations allow us to go back and forth between $\kappa_4$ and $\kappa_2$, which are convenient for numerical simulations, and $G$ and $\Lambda$, the couplings familiar from general relativity.

Geometries are constructed by gluing together four-simplices along their ($4-1$)-dimensional faces.  The four-simplices are equilateral, with constant edge length $a$.  The set of all four-geometries is approximated by gluing together four-simplices, and the dynamics is encoded in the connectivity of the simplices.  Most early simulations of EDT used a set of triangulations that satisfies the combinatorial manifold constraints, so that each distinct $4$-simplex has a unique set of 4+1 vertex labels.  The combinatorial manifold constraints can be relaxed to include a larger set of degenerate triangulations in which distinct four-simplices may share the same 4+1 (distinct) vertex labels \cite{Bilke:1998bn}.  The reason we use degenerate triangulations is that it leads to a factor of $\sim$10 reduction in finite-size effects compared to combinatorial triangulations \cite{Bilke:1998bn}.  There appears to be no essential difference in the phase diagram between degenerate and combinatorial triangulations in four dimensions \cite{Coumbe:2014nea}, suggesting that if a second-order transition could be identified, the two sets of triangulations would be in the same universality class.  In two dimensions, where analytical results are available, the two are in the same universality class, again with degenerate triangulations having finite-size effects that are around an order of magnitude smaller.

We sum over geometries with fixed global topology $S^4$ using Monte Carlo methods.  Since the local update moves that we use are topology preserving, in order to restrict the geometries to $S^4$ it is sufficient to start from the minimal four-sphere at the beginning of the Monte Carlo evolution.  As is standard in such simulations, we tune $\kappa_4$, corresponding to the bare cosmological constant, to its critical value in order to take the infinite volume limit.  Although this tuning allows us to take the infinite lattice-volume limit, it is not sufficient to take the infinite physical-volume limit.  In four dimensions the update moves are only ergodic if the lattice four-volume is allowed to vary.  However, it is convenient to keep $N_4$ approximately fixed.  We work at (nearly) fixed four-volume by including a term in the action $\delta\lambda |N_4^f-N_4|$ to keep the four-volume close to a fiducial value $N_4^f$.  This does not alter the action at values of $N_4=N_4^f$, but serves to keep the volume fluctuations about $N_4^f$ from growing too large to be practical to simulate.  Since the algorithm we use is only truly ergodic if the volume fluctuations are unrestricted, this introduces a systematic error for $\delta \lambda \neq 0$.  However, for sufficiently small $\delta \lambda$ the effect is expected to be negligible.  We take $\delta \lambda=0.04$ in our runs, and we find no significant difference in a smaller set of test runs when we reduce $\delta \lambda$ to 0.02. 

\subsection{The path integral measure}

The measure that we have chosen in Eq.~(\ref{eq:Z}) is the discrete analogue of the local measure term suggested by continuum methods \cite{Fujikawa:1983im, Fradkin:1974df, Unz:1985wq}.  It was shown by Fujikawa that the local measure is fixed if one imposes the BRST symmetry associated with general covariance, and includes a term of the form $\prod_x \sqrt{g}^\beta$.  The parameter $\beta$ can be fixed by stipulating that no gauge anomalies associated with general coordinate transformations can arise \cite{Fujikawa:1983im}.  Since the purpose of coordinate invariance is to ensure that physical states are correctly counted in a path-integral formulation, it is plausible that this symmetry should also completely fix the measure.  However, it is not clear to us how to translate the continuum calculations of $\beta$ \cite{Fujikawa:1983im, Fradkin:1974df, Unz:1985wq} into the lattice formulation, i.e. what power of $\beta$ is implicit in the entropic factors that emerge in the sum over geometries and what power must be explicitly included as part of the action.  However, since we argue that our lattice formulation breaks the relevant continuum symmetry, it may then be necessary to fine-tune $\beta$ as a function of lattice spacing, rather than to keep it fixed.  We present evidence in the remainder of this paper that $\beta$ is the only additional coupling that needs to be added to the EDT formulation and that this additional tuning is sufficient to take the continuum limit.

It is possible that the measure term that we are including in our calculations has an alternative interpretation.  The quantity in brackets in Eq.~(\ref{eq:Z}) can be exponentiated and promoted to a term in the action.  Such a term could be identified with an infinite sum of higher derivative curvature invariants, although there is no sense in which the lowest order $R^2$ term dominates, and the direct relation to continuum higher curvature terms is complicated.  It may be that such a term is needed because the ultraviolet critical surface of the continuum theory is greater than one dimension, as suggested by other approaches \cite{Codello:2007bd,Codello:2008vh,Benedetti:2009rx,Falls:2014tra}.  In this work we advocate that the interpretation of this expression as a measure term is the more appropriate one, in which case the fine-tuning involved is understood as arising from the breaking of a continuum symmetry by the lattice regulator.  

Independent of the picture advocated in this work, there is an argument in favor of the interpretation of the term $\prod_{j=1}^{N_2}{\cal O}(t_j)^\beta$ as a local measure term.  The addition of gauge fields to EDT has been considered \cite{Bilke:1998vj}, and it was found that there was a significant difference in the results of the simulations depending on whether one added the gauge fields on the links or on the direct lattice.  By universality, one would not expect these two approaches to lead to different theories, at least in the presence of a continuous phase transition where one can define a continuum limit.  It turns out that the two ways of adding gauge fields to EDT are equivalent if a nonuniform weighting of the triangulations is applied \cite{Ambjorn:1999ix}.  This weighting is a product of local volume factors, and corresponds precisely to the term introduced in this work and identified as a measure factor.  In this context, the term acts like a Jacobian arising because of the transformation of the gauge field from the dual to the direct lattice.  This further strengthens our confidence that associating this contribution with a local measure term is the most natural interpretation. 

Continuum arguments suggest that when the local measure is chosen correctly, it cancels all divergences of the form $\delta^{(4)}(0)$ from the Lagrangian effective action \cite{Fradkin:1974df}.  Although these divergences are set equal to zero when using a scale-invariant continuum regulator such as dimensional regularization \cite{'tHooft:1978id}, this is not possible in general when the regulator does not maintain scale invariance, as is the case with a lattice regulator.  These divergences lead to an additive renormalization of the cosmological constant.  This suggests that when varying $\beta$ as a function of lattice spacing in the simulations, the bare cosmological constant receives an unphysical power divergent additive renormalization that needs to be subtracted in order to recover the correct running.  We find evidence that this is the case.

\subsection{Details of the numerical implementation}

The numerical methods used to evaluate the partition function in Eq.~(\ref{eq:Z}) are by now well established \cite{Ambjorn:1997di}.  We have developed a parallel variant of the standard algorithm, which we call parallel rejection, in order to speed up the simulations.  Parallel rejection involves splitting the run into parallel streams each running the standard scalar algorithm.  The scalar algorithm to perform the Monte Carlo integration of the dynamical triangulations partition function consists of an ergodic set of local moves, known as the Pachner moves, which are used to update the geometries \cite{Agishtein:1991cv, Gross:1991je}, and a Metropolis step, which is used to accept or reject the proposed move.  The parallel rejection algorithm is implemented in a way that satisfies detailed balance, and this has been tested by demonstrating that it gives identical results, configuration by configuration, to the scalar algorithm.  Parallel rejection takes advantage of, and partially compensates for, the low acceptance of the Metropolis step in the range of parameters in which we are interested and is described in more detail in the Appendix.  Although one could split the runs up into many separate, trivially parallel streams, this straightforward approach is impractical when there are long thermalization and autocorrelation times.  This parallelization buys us almost an order of magnitude speed-up over the scalar algorithm on the latest multicore machines.

\subsection{Phase diagram}

\begin{figure}
\begin{center}
\includegraphics[scale=.55]{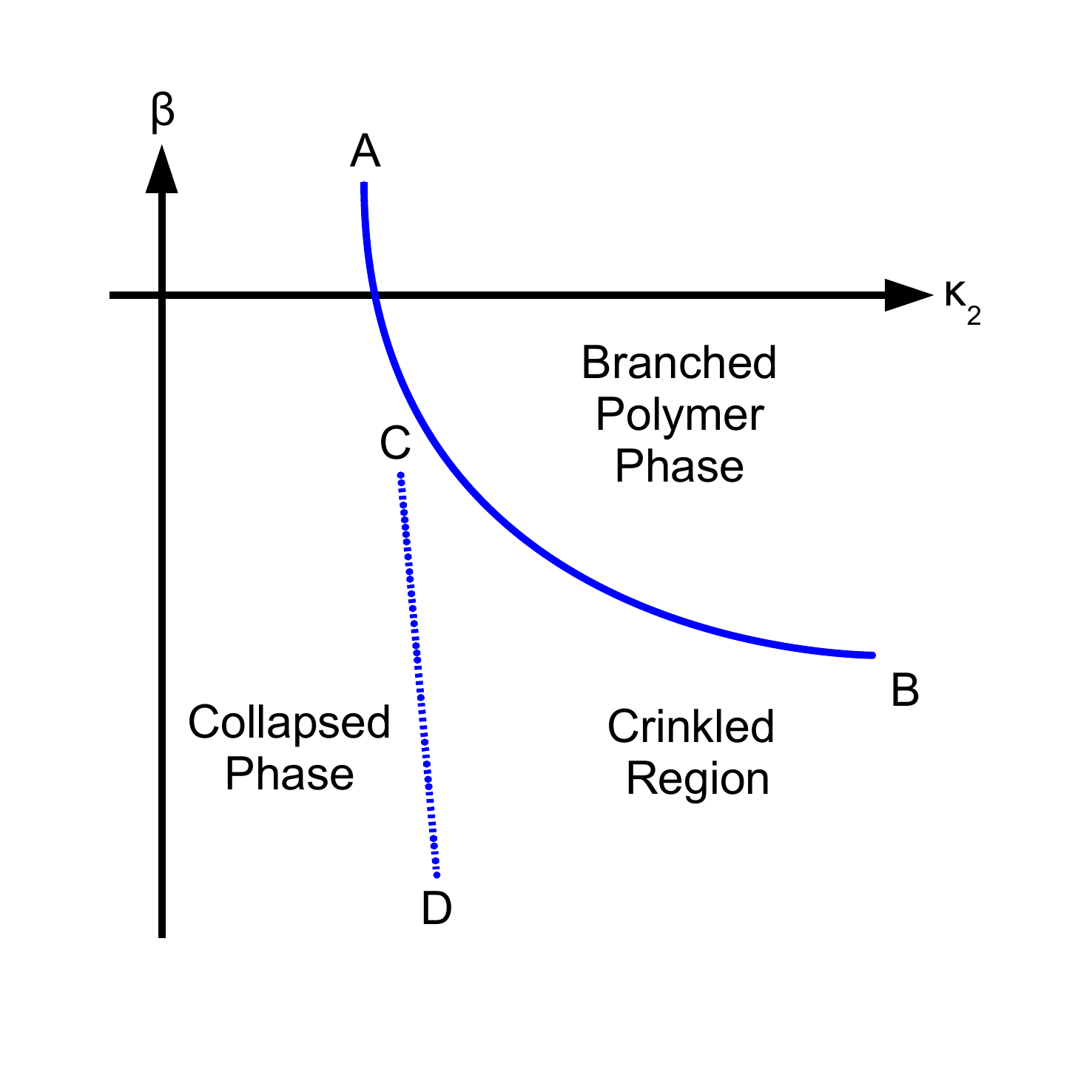}
\vspace{-3mm}
\caption{Schematic of the phase diagram as a function of $\kappa_2$ and $\beta$. \label{fig:phase1}}
\end{center}
\end{figure}

\begin{figure}
\begin{center}
\includegraphics[scale=.9]{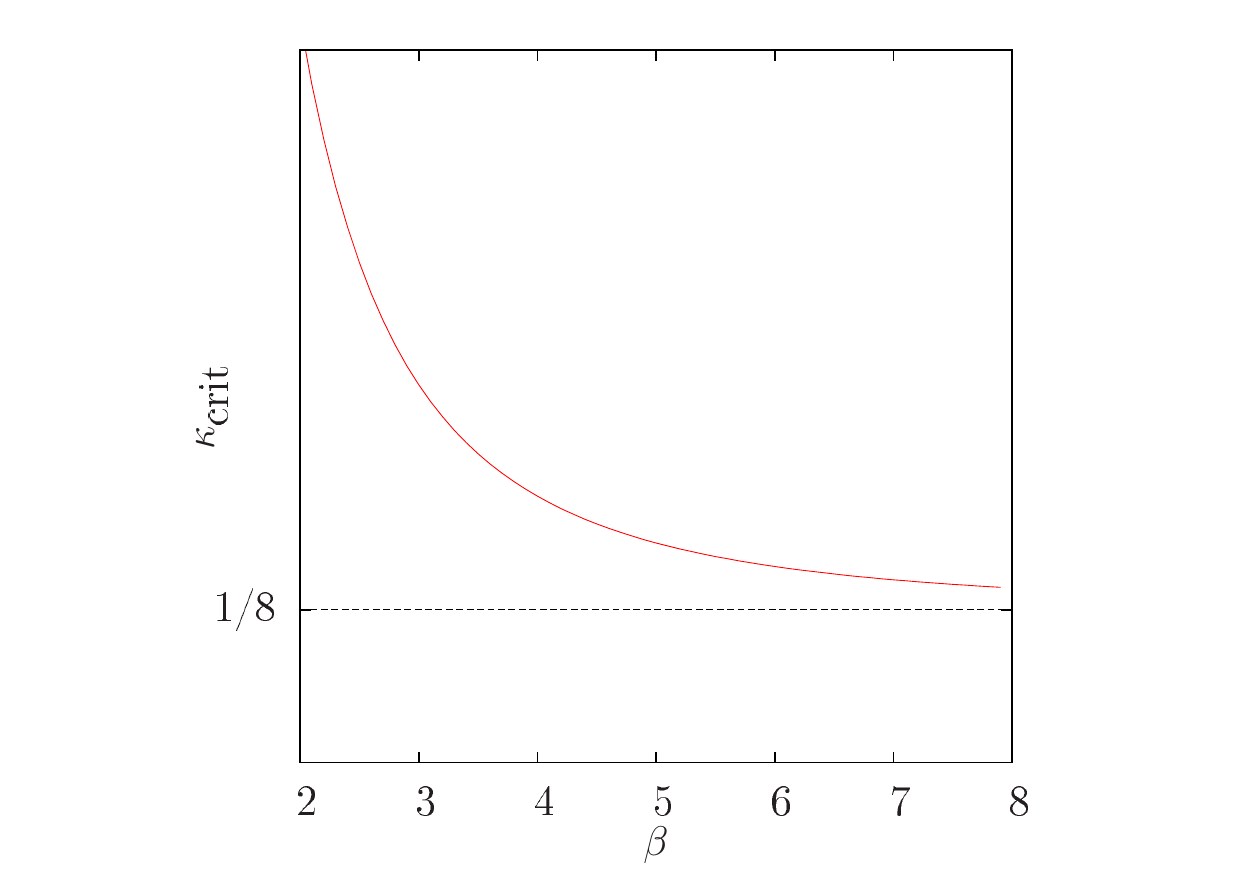}
\vspace{-3mm}
\caption{The tuned $\kappa_{\rm crit}$ as a function of $\beta=6/g^2_0$ illustrating the phase diagram of QCD with improved Wilson fermions. \label{fig:phaseQCD}}
\end{center}
\end{figure}

The parameter $\kappa_4$ must be adjusted to take the infinite volume limit, leaving a two-dimensional parameter space in $\kappa_2$ and $\beta$.  Two recent studies of this phase diagram, including the nontrivial measure term, were carried out in Ref.~\cite{Ambjorn:2013eha} (for combinatorial triangulations) and in Ref.~\cite{Coumbe:2014nea} (for degenerate triangulations); the two works are in good agreement on the qualitative features of the EDT phase diagram.  We begin by reviewing these features here.  Figure~\ref{fig:phase1} shows the phase diagram in the $\kappa_2$, $\beta$ plane.  There is a first-order transition line $\overline{AB}$ separating two phases, the branched polymer phase and the collapsed phase.  There is a region, the crinkled region, that shares features of both phases, but looks like the collapsed phase for sufficiently large volumes.  There does not appear to be a distinct phase transition between the crinkled region and the rest of the collapsed phase, but rather a crossover, as indicated by the dashed line $CD$ in Fig~\ref{fig:phase1}.

The phases can be characterized by their fractal dimensions.  Two useful definitions of fractal dimension are the Hausdorff dimension and the spectral dimension.  The Hausdorff dimension $D_H$ is determined from the scaling of the radius of a sphere with volume $V$ in the limit that $r\to 0$, and is given by
\bea  D_H = \lim_{r\to 0} \frac{\log{(V(r))}}{\log(r)}.
\eea
The spectral dimension $D_S$ is defined by a diffusion process, and is related to the return probability $P_r(\sigma)$ for a random walk on a geometry to return to its starting point after $\sigma$ diffusion steps.  The spectral dimension can be obtained from
\bea\label{eq:spec}  D_S(\sigma) = -2 \frac{d \log{\langle P_r(\sigma) \rangle}}{d\log{\sigma}}.
\eea

The model is in the collapsed phase for sufficiently small values of $\kappa_2$.  The collapsed phase is characterized by a large, and possibly infinite, Hausdorff dimension.  The spectral dimension in this phase is also large, and possibly infinite in the infinite volume limit.  In this phase there are a small number of highly connected vertices, so that a large number of simplices share a few common vertices.  Thus, in this phase there are vertices with large coordination number, such that the entire lattice geometry can be traversed in just a few lattice spacings.  The effective curvature in this phase is large and negative \cite{Smit:2013wua}.  The geometric properties of the collapsed phase suggest that it is a poor candidate for a theory of gravity.

For sufficiently large values of $\kappa_2$ and $\beta$ the model is in the branched polymer phase.  In this phase the geometry has many minimal necks with polymerlike baby universe branchings.  This phase has a highly irregular geometry, with a fractal treelike structure even on large scales.  The Hausdorff dimension of a branched polymer is expected to be 2, a result that is confirmed by our previous calculations \cite{Coumbe:2014nea}.  The spectral dimension of the branched polymer phase is expected to be $D_S=4/3$, a result that has also been numerically confirmed \cite{Ambjorn:2013eha,Coumbe:2014nea}.  Once again, the properties of this phase make it a poor candidate for recovering four-dimensional space-time.

Given the unphysical behavior over most of the phase diagram and our earlier symmetry arguments, a practical question remains.  How do we identify the correct fine-tuning for $\beta$?  The analogy between gravity and lattice QCD with Wilson fermions suggests a way forward.  Figure~\ref{fig:phaseQCD} shows a schematic of the phase diagram of QCD with improved Wilson fermions.  The coupling $\beta$, in standard lattice QCD notation, is inversely proportional to the bare strong coupling $g_0^2$, and the coupling $\kappa$ is inversely proportional to the bare fermion mass.  For generic values of $\kappa$, the theory does not resemble QCD with massless or nearly massless quarks.  Only at the (nonzero) critical $\kappa$ does the theory approach the massless quark limit, and only in the continuum limit with $\kappa$ tuned close to its critical value do the effects of chiral symmetry breaking discretization effects vanish.  At $\kappa_{\rm crit}$, there is a first-order phase transition; the second-order critical point is approached as $g_0\to 0$, ($\beta\to \infty$)\footnote{In 2-flavor lattice QCD with Wilson fermions, there are actually two possible phase structures at coarse coupling, depending on the precise form of the lattice realization of the gauge and fermion fields.  One of these scenarios has a second-order transition line at $\kappa_{\rm crit}$ separating the physical phase from an unphysical phase, called the Aoki phase, in which there are massless pions, but also undesirable effects due to the spontaneous breaking of parity and flavor symmetries \cite{Aoki:1983qi}.  The other scenario is called the first-order scenario, or Sharpe-Singleton \cite{Sharpe:1998xm} scenario.  This scenario has a first order line at $\kappa_{\rm crit}$ separating physical and unphysical phases.  Even at the critical value for $\kappa$, the pions are not massless, but acquire masses that are proportional to the lattice spacing.  For the purposes of comparison to this work, it is the latter scenario that provides the better analogy to the EDT phase diagram.}.  In dynamical triangulations, there is also a first-order line, which softens as $\kappa_2$ is taken larger.  Given this, the similarity of the phase diagram for QCD using improved Wilson fermions to that of dynamical triangulations is evident.  This similarity suggests that we ought to follow the same prescription in our gravity calculations: tune the bare parameters to values close to the first-order transition line, and then follow it towards a second-order critical point, if one exists.  Interestingly, early works in EDT have pointed out that the Hausdorff dimension appears to be close to four near the transition \cite{deBakker:1994zf, Egawa:1996fu}.  The following sections report our study of the EDT model close to the first-order transition line, where we present evidence that this analogy is a good one.

\subsection{Numerical tests}

Here we discuss the numerical tests that the EDT approach must pass if it is to realize the asymptotic safety scenario.  We first turn to a study of whether the lattice theory can realize the classical theory in the continuum, long distance limit.  A basic test that must be passed for this to be the case is that the dimension of the theory measured at long distance scales must be four.  Even if the geometries have a fractal structure at short distance scales, all of the fractal dimensions must agree with four at sufficiently large scales in order to reproduce our world.  Thus, both the Hausdorff dimension and the spectral dimension must be compatible with four at large scales.  Beyond this, the lattice theory must also reproduce the solutions to the classical Einstein equations in the classical limit, both the cosmological solutions and Newton's law in the appropriate limit.  In this work we focus on the cosmological solution, leaving the interaction of matter needed to test Newton's law to future work.

To ensure that the asymptotic safety scenario holds, we must identify a continuous phase transition, the approach to which defines a continuum limit.  We need to determine the relative lattice spacing to decide how to take the continuum limit, and to test whether this corresponds to our expectations based on the properties of the phase diagram, i.e. whether the latent heat decreases along the first-order line as the continuum limit is approached.  It is also interesting to study the running of the spectral dimension with distance scale and compare our results for the spectral dimension at short distances with the results from other approaches.

Finally, we can study the dimension of the ultraviolet critical surface, which tells us the number of free parameters that must be taken from experiment in order to make predictions using the theory.  If, as discussed in Sec.~\ref{sec:symmetry}, the ultraviolet critical surface of the theory is one dimensional, that would ensure that the continuum theory has no adjustable parameters, with only a single measurement of a dimensionful quantity needed to set the lattice scale.  In particular, the cosmological constant in Planck units would then be a prediction of the theory, at least in principle.  We must be careful here, since the breaking of a continuum symmetry by the regulator generically leads to the appearance of extra relevant parameters in the lattice theory as compared to the continuum theory, thus leading to an overestimate of the dimension of the ultraviolet critical surface.  


\section{Evidence for classical behavior}\label{sec:classical}

\begin{figure}[]
  \begin{center}
  \subfloat{\includegraphics[width=0.65\textwidth]{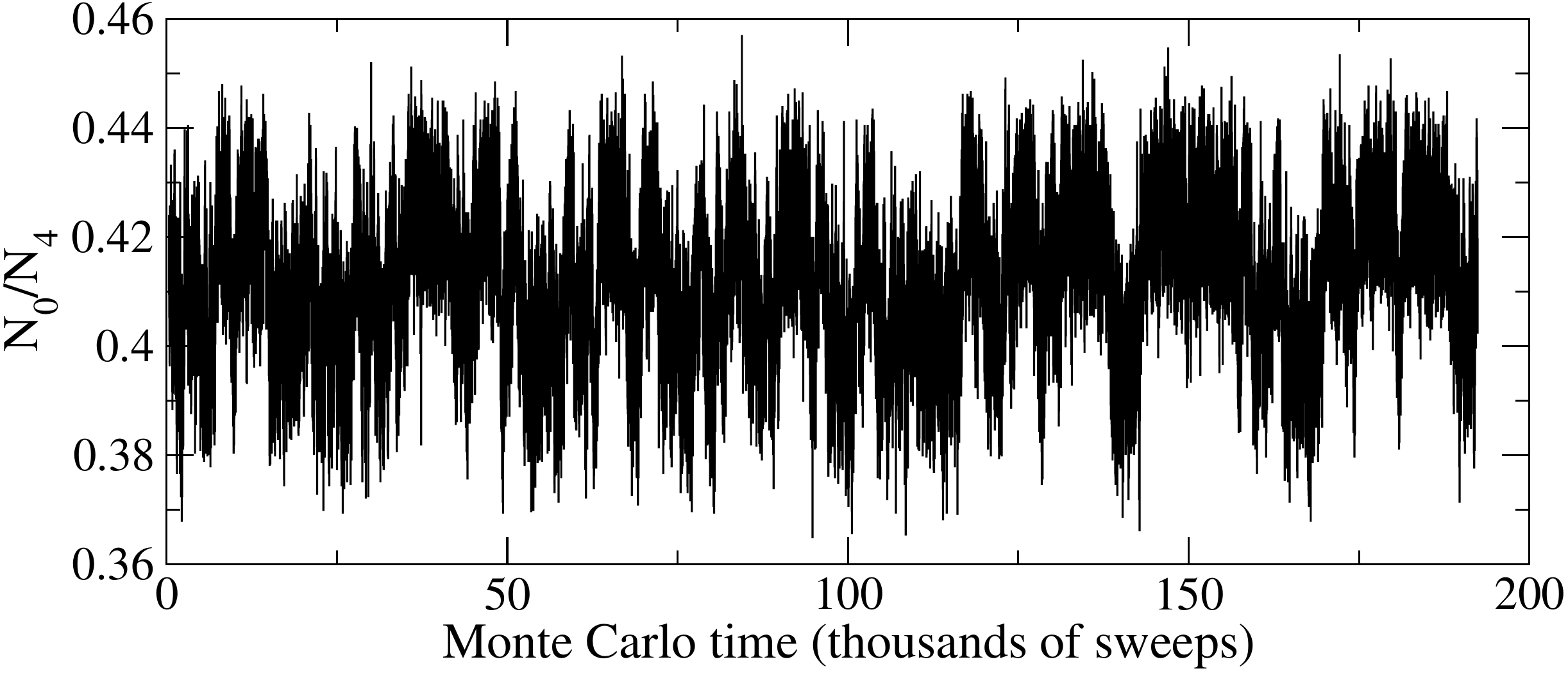}}\\
  \subfloat{\includegraphics[width=0.65\textwidth]{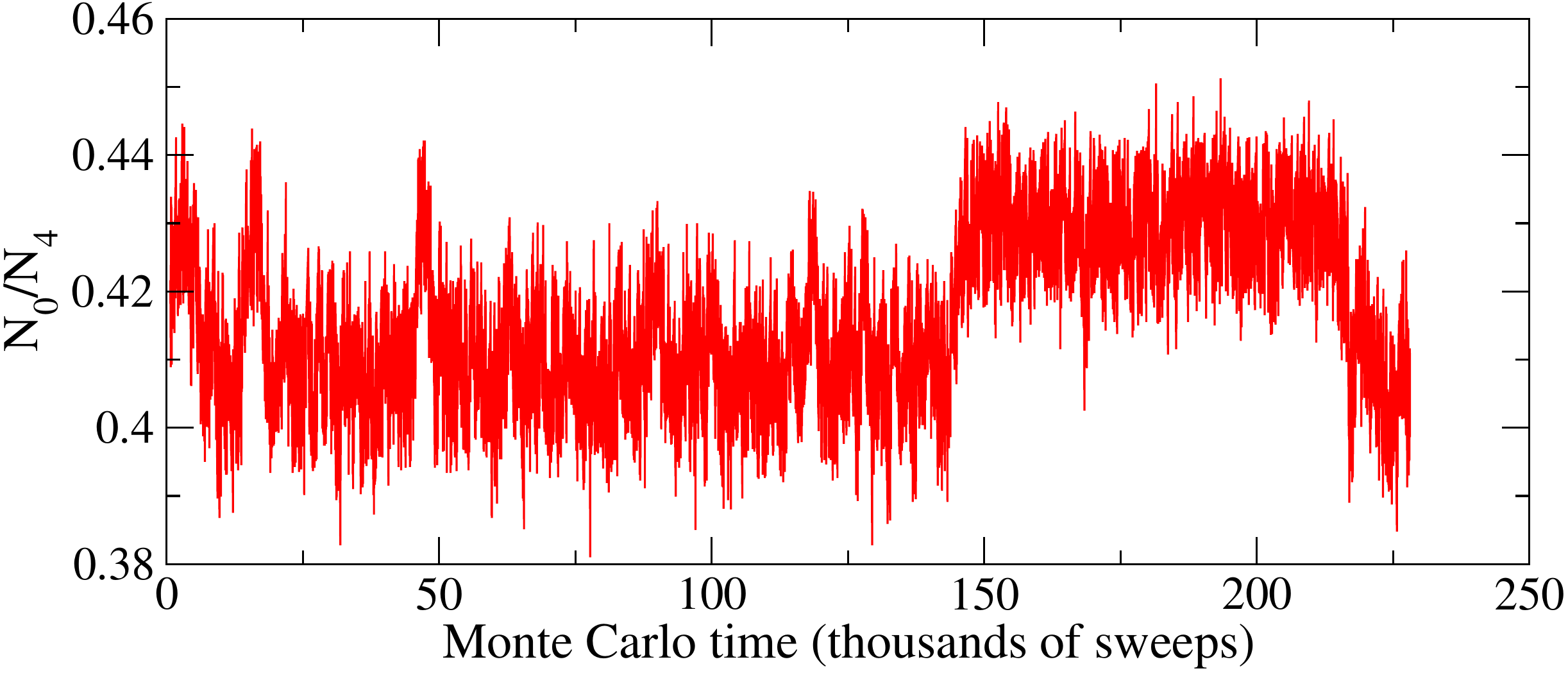}}\\
\subfloat{\includegraphics[width=0.65\textwidth]{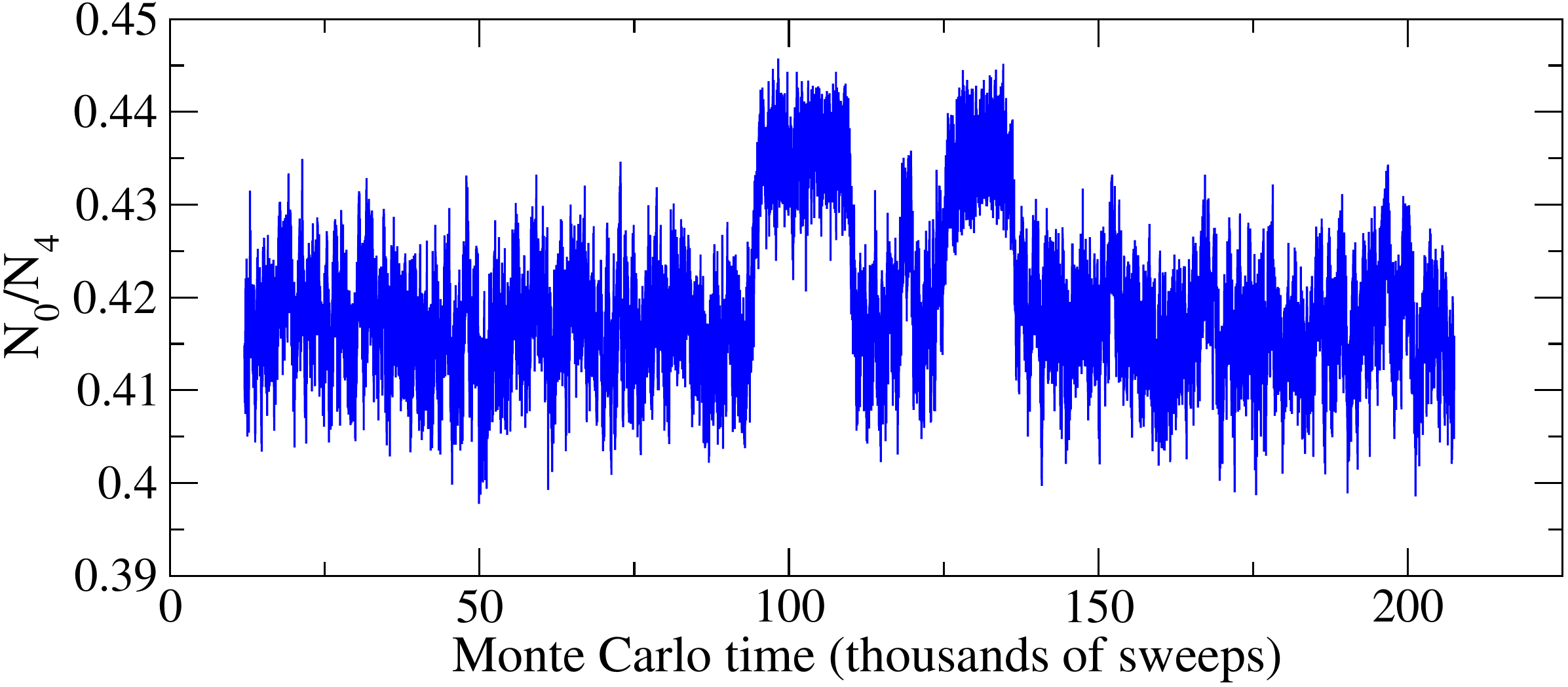}}\\
\end{center}
  \caption{The Monte Carlo time history of $N_0/N_4$ for three different volumes at $\beta=0$.  From top to bottom the volumes are 4k, 8k, and 16k simplices.  For the 4k and 8k volume runs, a sweep was $10^8$ attempted moves, while for the 16k run, a sweep was $4\times 10^8$ attempted moves.  The tunneling rate is much greater for the 4k ensemble and decreases dramatically with increasing volume.  \small }
\label{fig:N0history}
\end{figure}

\begin{figure}[]
  \centering
  \includegraphics[width=0.59\textwidth]{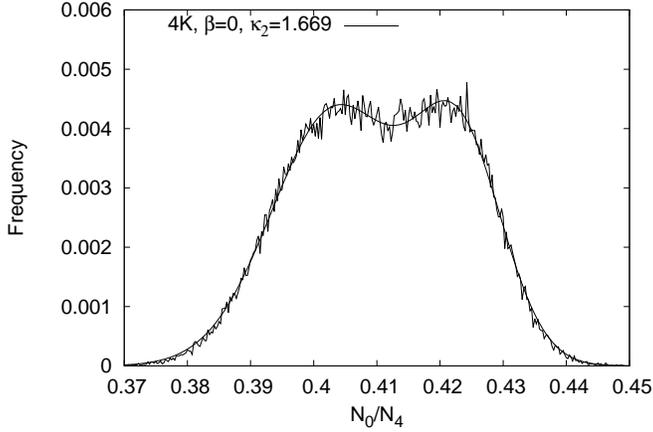}
  \vspace{-0.7cm}
  \caption{Histogram of $N_0/N_4$ values for the tuning run of the 4k ensemble at $\beta=0$.  Also shown is a fit to a double Gaussian. \small }
\label{fig:hist}
\end{figure}

\begin{figure}[]
  \centering
  \includegraphics[width=0.56\textwidth]{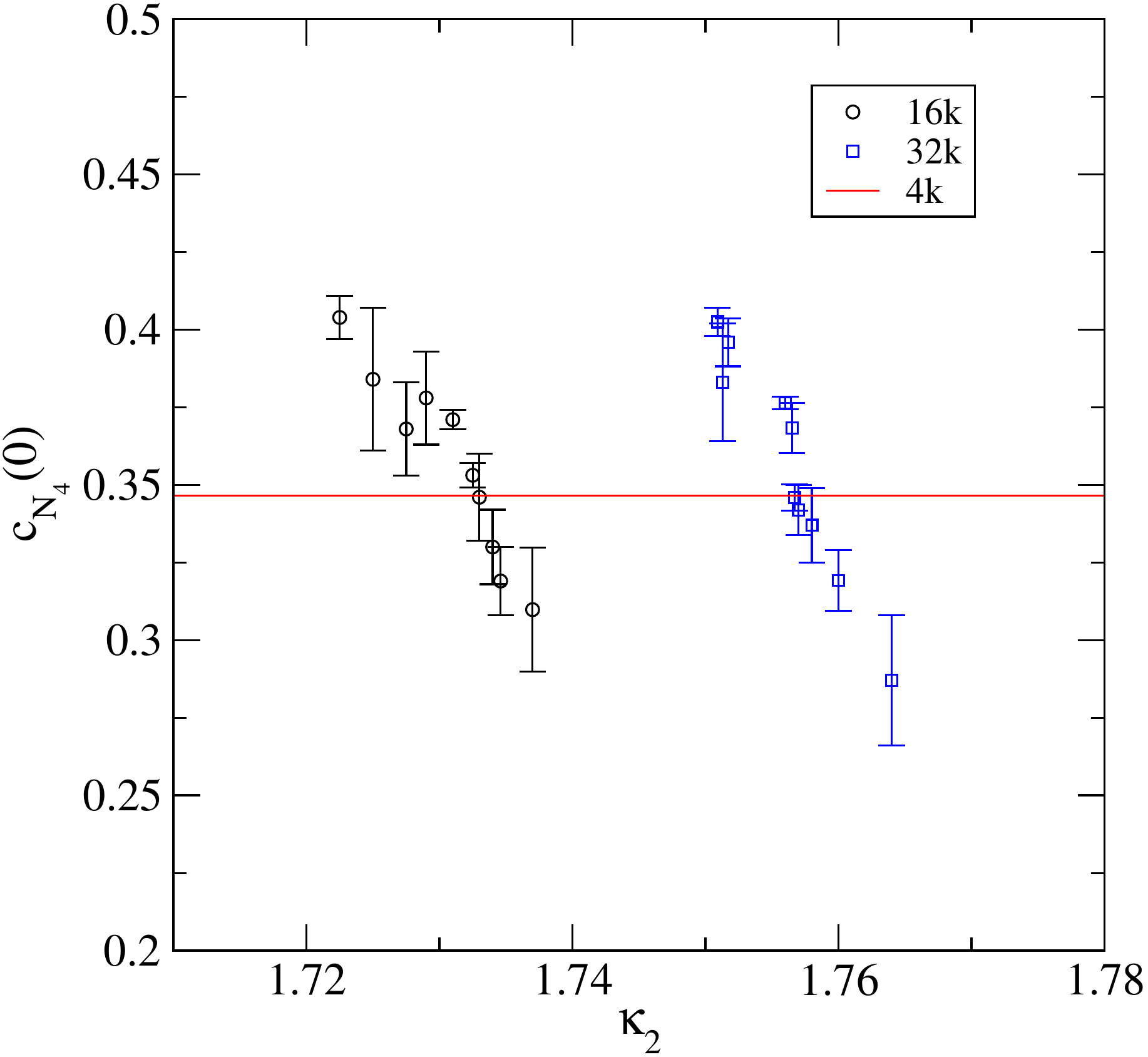}
  \vspace{-0.3cm}
  \caption{Values of $c_{N_4}(0)$ as a function of $\kappa_2$ close to the critical $\kappa_2$ value at $\beta=0$ for two different volumes, 16k and 32k.  The transition is located at the point where the slope of $c_{N_4}(0)$ versus $\kappa_2$ is a maximum.  The red line is the value of $c_{N_4}(0)$ determined on the 4k ensemble.  The Hausdorff dimension assumed in the rescaling of $c_{N_4}(0)$ is four, so that the location of the critical $\kappa_2$ value (the maximum slope of the curve) should coincide with the line if $D_H=4$.  \small }
\label{fig:tuning}
\end{figure}

In order to investigate the hypothesized phase structure of EDT, we generate a number of runs close to the transition.  In lattice QCD with Wilson fermions, only one of the phases on one side of the first-order phase transition is physical.  Since a first-order transition is characterized by tunneling between two metastable states, if we simulate too close to the phase transition then for some of the Monte Carlo time the run will be in the wrong, unphysical metastable phase.  We find in our gravity runs that semiclassical behavior does emerge when the first-order transition line is approached from the left.  It turns out that close to the transition the behavior of the collapsed phase is qualitatively different from its behavior deep in the collapsed phase.  In order to minimize contamination from the wrong phase, we restrict our Monte Carlo averages to the times when the run is in the correct phase.  Figure~\ref{fig:N0history} shows the histories of the number of vertices divided by the number of four-simplices $N_0/N_4$ for three of our ensembles as a function of Monte Carlo time.  The quantity $N_0$ serves as a good order parameter \cite{Catterall:1994pg, deBakker:1996zx} to identify which phase the run is in as a function of Monte Carlo time.  To obtain expectation values in the desired phase, one must average only the parts of the run when the value of $N_0/N_4$ shows that the run is on the side of the phase transition associated with the collapsed phase.  The smaller value of $N_0/N_4$ corresponds to the collapsed phase, since in this phase there is a higher connectivity between the simplices, with a small number of vertices shared among a large number of simplices.  The larger the volume, the sharper the transition, and the easier it is to distinguish between the two phases, as can be seen in Fig.~\ref{fig:N0history}.

We use the histogram of the time history of the order parameter $N_0$ to estimate the location of the transition.  The location of the (pseudo)critical line varies as a function of volume.  Thus, we need to tune the bare parameters at many volumes in order to carry out a finite-size scaling analysis as close as practical to the first-order phase transition line.  We tune the bare parameter $\kappa_2$ until the peaks in the $N_0$ histogram are roughly the same size.  Figure~\ref{fig:hist} shows an example of this histogram for a tuning run on our smallest volume.  For our largest volume ensembles this is too difficult so we use a different procedure for locating the transition that makes use of a global observable, rather than a local observable like $N_0$.  This is because there are systematic effects associated with determining the location of the transition from the histogram of $N_0$ that require a large number of tunnelings between metastable states to reduce, and the incidence of tunnelings is exponentially suppressed as the volume increases.  Our global observable is the peak in the volume-volume correlator introduced in the next subsection.  Figure~\ref{fig:tuning} shows this quantity for tuning runs on our largest volume lattices at $\beta=0$.  The point with the steepest slope is the location of the phase transition.  For comparison, the solid line shows where the transition is expected to be if the global Hausdorff dimension is four, as determined on our 4k ensemble at the same lattice spacing.  We find good agreement between the location of the transition by this method and the desired four-dimensional scaling.

\begin{table}
\begin{center}
\caption{The parameters of the ensembles generated for this work.  The first column shows the relative lattice spacing as determined in Sec.~\ref{sec:continuum}, with the ensembles at $\beta=0$ serving as the fiducial lattice spacing.  The quoted error is a systematic error associated with finite-volume effects.  The second column is the value of $\beta$, the third is the value of $\kappa_2$, the fourth is the number of four-simplices in the simulation, and the fifth is the number of configurations sampled.}
\label{tab:ensembles}
\begin{tabular}{ccccc}
\hline \hline
\ \ $a_{\rm rel}$ \ \ & \ \ $\beta$ \ \ \ & \ $\kappa_2$ \ & \ \ \ \ \ $N_4$ \ \ & \ \ \ Number of configurations \\
\hline
1.47(10) & 1.5 & 0.5886 & \ \ 4000 & 1815  \\
1.23(9) & 0.8 & 1.032 & \ \ 4000 & 523  \\
1 & 0.0 & 1.669 &         \ \ 4000 & 920  \\
1 & 0.0 & 1.7024 &     \ \  8000 & 2500  \\
1 & 0.0 & 1.7325 &     \ \ 16000 & 1387  \\
0.81(6) & $-0.6$ & 2.45 & \ \ 4000 & 1560  \\
0.72(5) & $-0.8$ & 3.0 &  \ \ 8000 & 1975  \\
\hline
\end{tabular}
\end{center}
\end{table}

Table~\ref{tab:ensembles} shows the parameters for the EDT ensembles that we have generated.  We consider multiple volumes at $\beta=0$ in order to study the finite-size scaling in this region, and we consider many different points along the first-order transition line in order to study the lattice spacing dependence of our geometries.

\begin{subsection}{Global Hausdorff Dimension}

\begin{figure}[]
  \centering
  \includegraphics[width=0.56\textwidth]{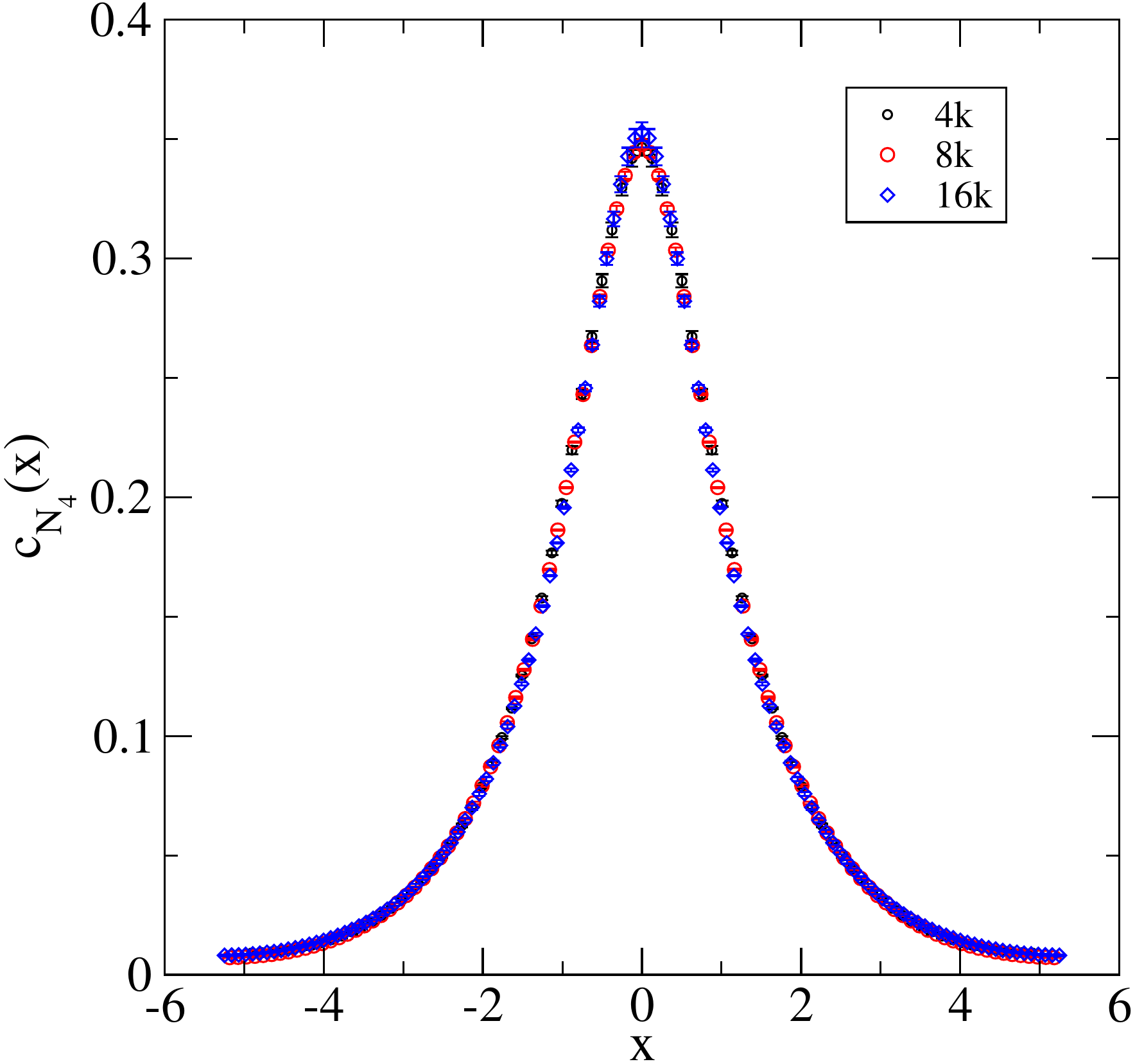}
  \caption{\small Scaling of the volume-volume distribution as a function of the rescaled variable $x=\delta/N_{4}^{1/D_{H}}$ using lattice volumes of 4K, 8K and 16K four-simplices at $\beta=0$.}
\label{fig:dh4}
\end{figure}

\begin{figure}[]
  \centering
{\includegraphics[width=0.56\textwidth]{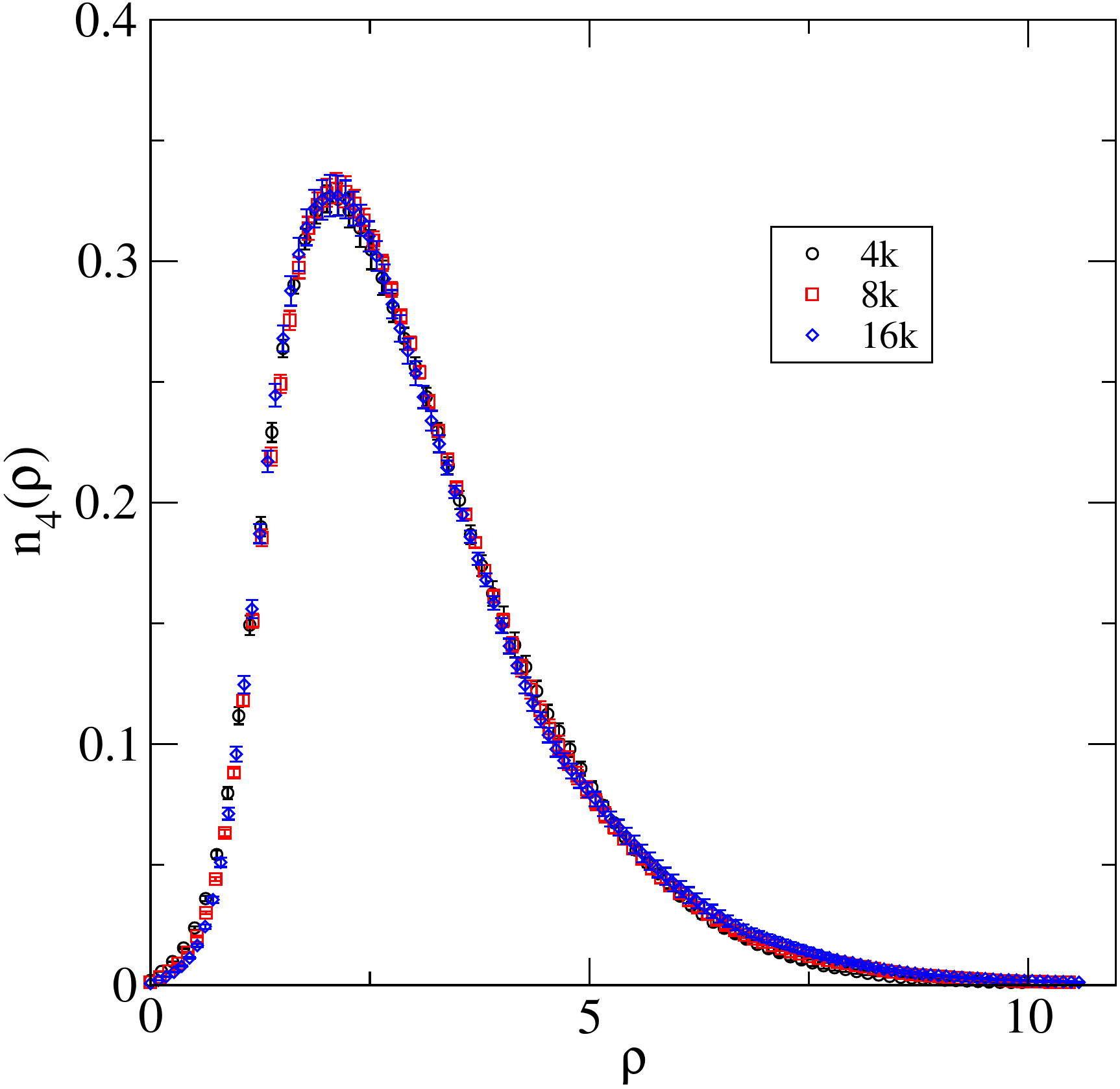}}
\caption{\small Scaling of the function $n_4(\rho)$ as a function of rescaled variable $\rho$ for three lattice volumes at $\beta=0$.}
\label{fig:n4}
\end{figure}

We study the global Hausdorff dimension using the standard technique of finite-size scaling.  In particular, we look at the scaling of two different correlation functions.  These correlation functions give us a measure of the shape of the lattice geometries, and can also be compared to classical expectations.  We first consider the three-volume correlator introduced in Ref.~\cite{Coumbe:2014nea}, which is similar to the one introduced in Ref. \cite{Ambjorn:2005qt} to study CDT,

\begin{equation}
C_{N_{4}}\left(\delta\right)=\sum_{\tau=1}^t\frac{\left\langle N_{4}^{\rm shell}(\tau)N_{4}^{\rm shell}(\tau+\delta)\right\rangle }{N_{4}^{2}}.
\end{equation}

\noindent \begin{math}N_{4}^{\rm shell}(\tau)\end{math} is the total number of four-simplices in a spherical shell one four-simplex thick, a geodesic distance \begin{math}\tau\end{math} from a randomly chosen simplex, and $t$ is the maximum number of shells. \begin{math}N_{4}\end{math} is the total number of four-simplices and the normalization of the correlator is chosen such that 
\bea \sum_{\delta=-t}^{t}C_{N_{4}}\left(\delta\right)=1.
\eea
If we rescale \begin{math}\delta\end{math} and \begin{math}C_{N_{4}}\left(\delta\right)\end{math}, defining \begin{math}x=\delta/N_{4}^{1/D_{H}}\end{math}, then the universal distribution
\bea c_{N_{4}}\left(x\right)=N_{4}^{1/D_{H}}C_{N_{4}}\left(N_{4}^{1/D_{H}}x\right),
\eea
should be independent of the lattice volume.  One can then determine the fractal Hausdorff dimension, \begin{math}D_{H}\end{math}, as the value that leaves \begin{math}c_{N_{4}}\left(x\right)\end{math} invariant under a change in four-volume \begin{math}N_{4}\end{math}.  We also consider the quantity $N_{4}^{\rm shell}(\tau)$ directly.  This quantity can be interpreted as the three-volume of the lattice as a function of Euclidean time $\tau$.  Rescaling $\tau$ and $N_{4}^{\rm shell}(\tau)$, and defining $\rho=\tau/N_4^{1/{D_H}}$ and 
\bea n_4(\rho)=1/N_4^{1-1/{D_H}}N_4^{\rm shell}(N_4^{1/D_H}\rho),
\eea
 we expect $n_4(\rho)$ to also be independent of four-volume.  

Figure~\ref{fig:dh4} shows the rescaled correlation function $c_{N_4}(x)$ for three different volumes at the first-order transition for $\beta=0$.  The data sets are binned to reduce errors from autocorrelations, after which a single elimination jackknife procedure is used to estimate the statistical errors.  The peak in $c_{N_4}(x)$ is used to test an ensemble for thermalization, because it probes the long-distance fluctuations of the lattice geometries.  Our studies of this quantity show that the runs have equilibrated and that we have several autocorrelation times worth of measurements for this quantity on each ensemble.  We follow this same blocked jackknife procedure for estimating statistical errors throughout this work.  The value of $D_H$ is taken to be four in our rescalings to produce $c_{N_4}(x)$.  As one can see from Fig.~\ref{fig:dh4}, the overlap of the curves is very good, indicating that the Hausdorff dimension is indeed close to four.  A fit designed to maximize the overlap between the curves with $D_H$ a free parameter leads to a value $D_H=4.1\pm 0.3$, where the error is statistical only.  Figure~\ref{fig:n4} shows the rescaled correlator $n_4(\rho)$, again with the Hausdorff dimension set to four in the rescaling; this correlator also shows excellent agreement between the rescaled curves, providing strong evidence that the Hausdorff dimension close to the transition is near four.

\end{subsection}

\subsection{The de Sitter solution}

\begin{figure}
\begin{center}
\includegraphics[scale=.7]{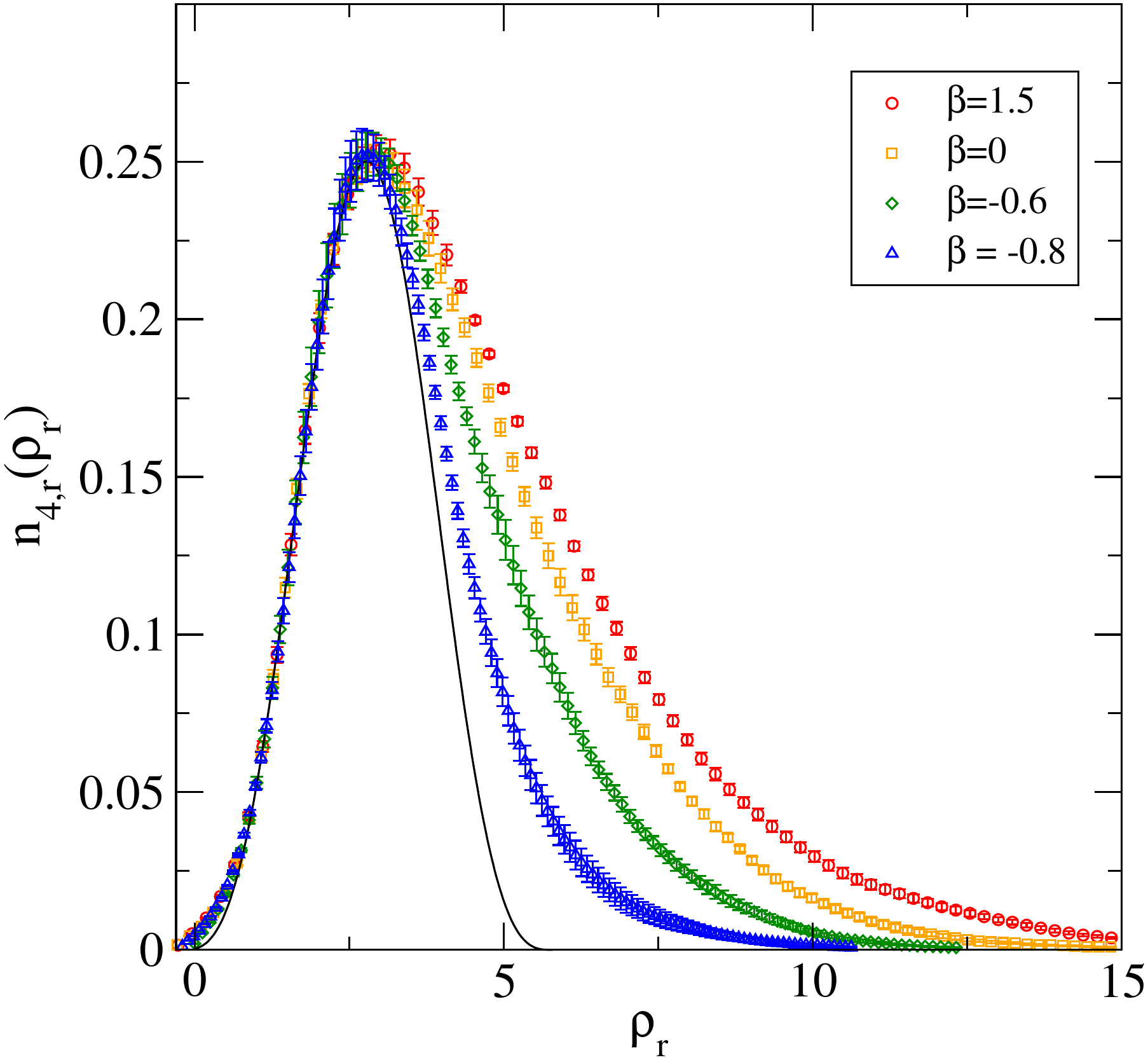}
\vspace{-3mm}
\caption{$n_{4,r}(\rho_r)$ versus rescaled Euclidean time $\rho_r$ for four different lattice spacings and best fit curve to data at $\beta=-0.6$.  The solid curve is the fit to the de Sitter solution, where the fit range is between $\rho_r$=1.1 and 2.8.    \label{fig:desitter}}
\end{center}
\end{figure}

The correlator $n_4(\rho)$ can also be interpreted as the spatial three-volume of our lattice geometries as a function of Euclidean time.  In this case, it is straightforward to compare our results to the classical expectation.  Figure~\ref{fig:desitter} shows the result for $n_4(\rho)$ from various ensembles along with the classical curve for Euclidean de Sitter space.  Euclidean de Sitter space in four dimensions is a four-sphere, in which case we would find
\bea\label{eq:desitter}  N_4^{\rm shell} =  \frac{3}{4}N_4\frac{1}{s_0 N_4^{1/4}}\sin^3\left(\frac{i}{s_0 N_4^{1/4}}\right),
\eea

\noindent with $s_0$ a free parameter, and $i$ the Euclidean time in lattice units.  Figure~\ref{fig:desitter} shows a comparison between the numerical lattice data for $N_4^{\rm shell}$ at several different lattice spacings and the theoretical expectation, Eq.~(\ref{eq:desitter}).  The qualitative agreement is quite good before the classical turning point, but afterwards the lattice data has a long tail that is absent in the classical solution.  Note that the lattice data for $n_4(\rho)$ and $\rho$ have been further rescaled so that the data points before the classical turning point are in agreement.   The long tail in the lattice data appears to be caused by baby universes that are connected to the mother universe by cutoff size necks; we describe our reasoning for this later in the subsection.  Our method for determining the relative lattice spacing at different couplings is discussed in detail in the next section, but anticipating those results, we are able to order the ensembles from coarsest to finest.  Intriguingly, the tail decreases in size as the continuum limit is taken, suggesting that the baby universes are a cutoff effect.  In contrast, the long tail is hardly affected by changing the volume at fixed lattice spacing, as can be seen in Fig.~\ref{fig:n4}.  This is important, since a prerequisite for a quantum theory of gravity is that it reproduce the classical theory at long distances.  Although one might expect that the long-distance behavior in a theory should not be modified by cutoff effects, this can happen when the lattice regulator breaks a symmetry that is important at long distances, such as chiral symmetry in the case of Wilson fermions.  The pion sector of QCD is not correctly described at coarse lattice spacings for Wilson fermions, even though this is the lightest particle in the physical spectrum with the longest Compton wavelength.  Only in the continuum limit do these unphysical effects go away \cite{Bernard:2004ab}.  

It is noteworthy that as we take the continuum limit of the EDT theory this brings our results for the shape of the lattice into better agreement with the results of the classical theory, and also with CDT.  In the CDT approach it has been shown that four-dimensional Euclidean de Sitter space emerges as a solution \cite{Ambjorn:2008wc}.  For CDT, precise agreement with Eq.~(\ref{eq:desitter}) is found in what is called the ``extended" phase of that theory.  For both EDT and CDT to be in the same universality class, it is necessary that both theories must reproduce the same results in the continuum limit, although they may disagree at finite lattice spacing.  The decrease of the large tail in Fig.~\ref{fig:desitter} as the continuum limit of the EDT theory is approached is evidence that CDT and EDT could in fact agree in the continuum limit.  Further nontrivial evidence that the two approaches may be in the same universality class is presented in the following section.

Our fits to the de Sitter space solution are conveniently parametrized by the modified formula
\bea\label{eq:desitter2}  N_4^{\rm shell} = \frac{3}{4} \eta N_4\frac{1}{s_0 N_4^{1/4}}\sin^3\left(\frac{i}{s_0 N_4^{1/4}}+b\right)
\eea
where in addition to $s_0$, we introduce the free parameters $\eta$ and $b$.   Equation~(\ref{eq:desitter2}) describes Euclidean de Sitter space in the limit that $b=0$ and $\eta=1$.  We include these parameters because we find that in order to get good fits a small offset $b$ is needed, and $\eta$ accounts for the fraction of the lattice volume that is in the mother universe.  The classical theory curve in Fig.~\ref{fig:desitter} shows the result of a fit to Eq.~(\ref{eq:desitter2}) to the $\beta=-0.6$ ensemble.  The fit is to a range of $i$ values from 9 to 22.  Smaller $i$ values show a small but statistically significant deviation from the classical solution, and data points past the classical turning point show large deviations because of the long tail.  The correlated $\chi^2$ per degree of freedom for this fit is 0.66; this corresponds to a $p$-value of 0.77.  Thus, within a window, the data is well described by the classical expectation.  

\begin{figure}
\begin{tabular}{c c}

\includegraphics[scale=.20]{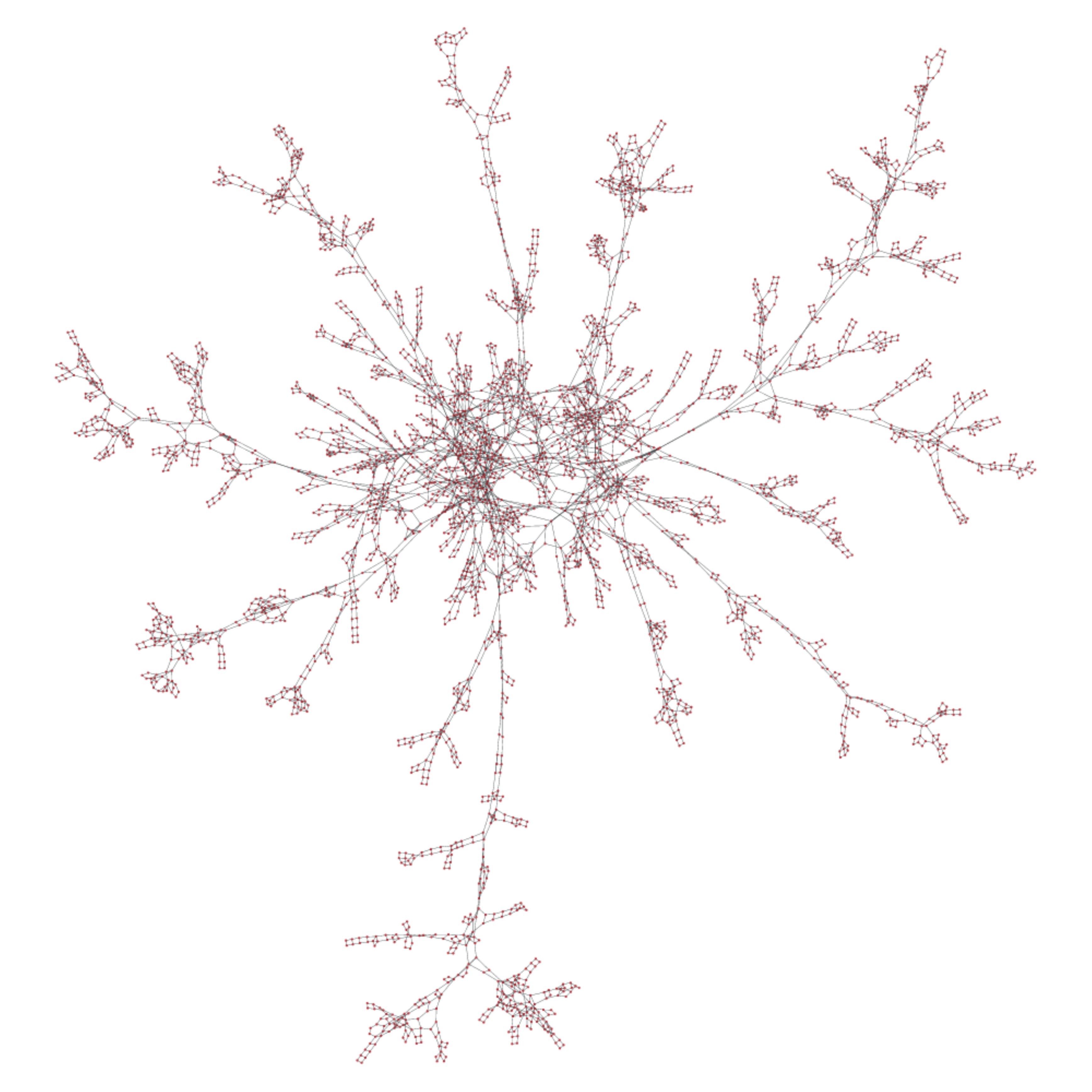}  & \ \ \ \ 
\includegraphics[scale=.20]{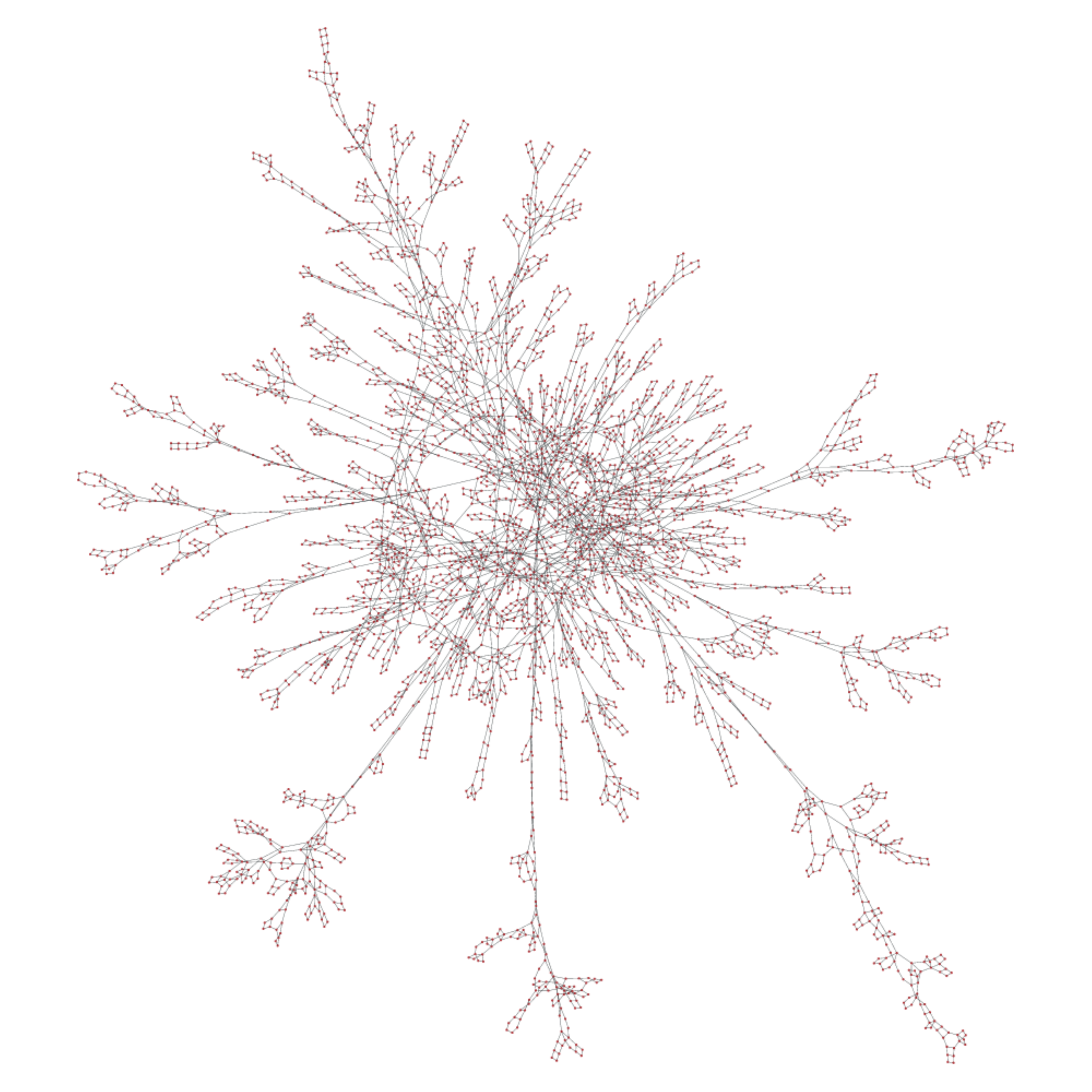}  \\ \ \ \ \ \ \ \ 
\includegraphics[scale=.20]{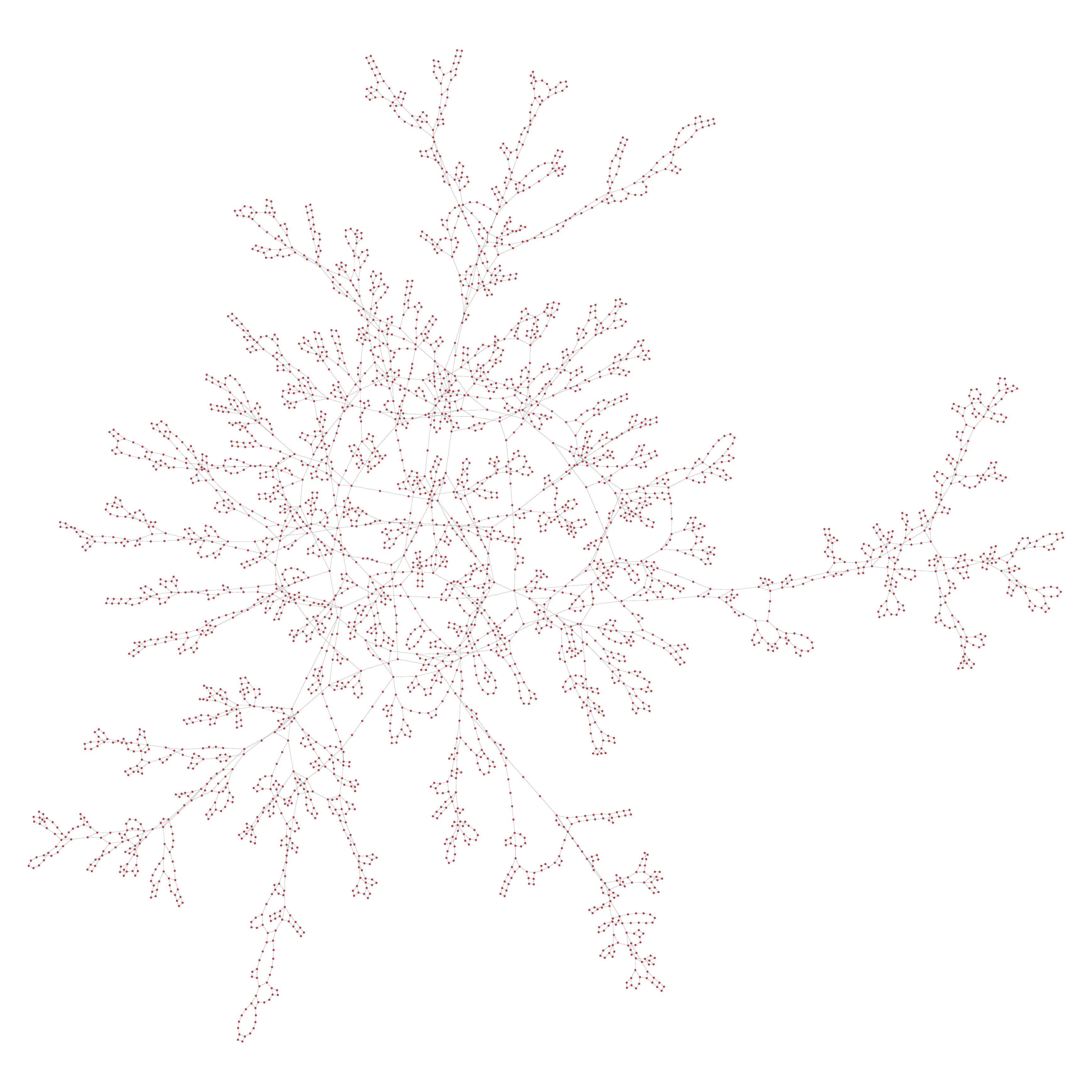} \ \ \ \ \ \ &
\includegraphics[scale=.22]{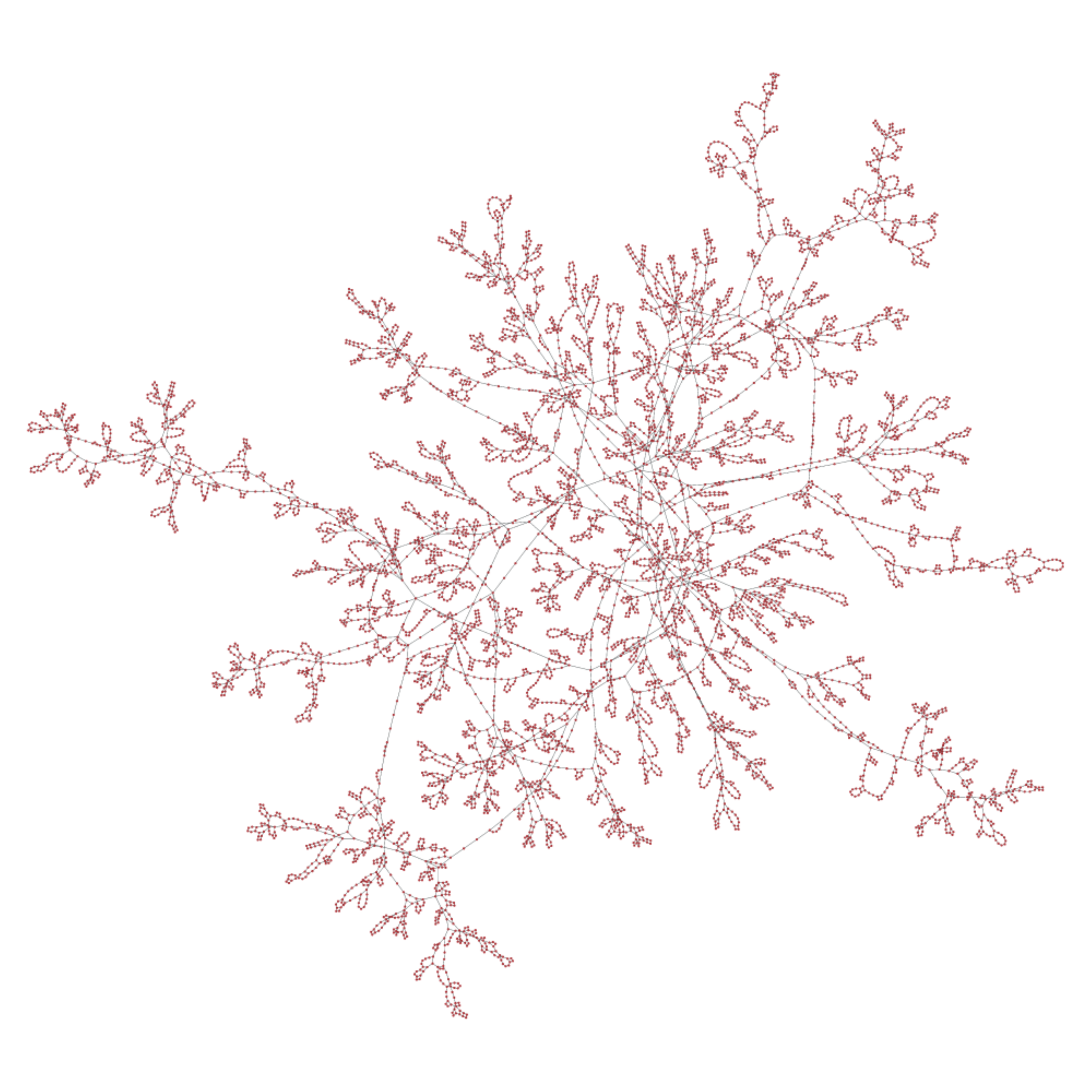} \ \ \ \ \ 
\end{tabular}
\caption{ Visualizations of the geometries at four different lattice spacings.  The image to the top left is a very coarse lattice with $\beta=1.5$, the image to the top right is a coarse lattice with $\beta=0$, the image to the bottom left is a fine lattice with $\beta=-0.6$, and the image to the bottom right is our finest lattice with $\beta=-0.8$.  The three coarser lattices have 4k four-simplices, while the finest has 8k.  The dots correspond to the four-simplices, and the lines show the connections between nearest neighbors.   \label{fig:network}}

\end{figure}

We have been attributing the large deviations from the classical expectation at very long distances to the presence of baby-universe-like structures.  We can gain further insight into the behavior of the geometries by using a network visualization tool \cite{peixoto_graph-tool_2014}.  Figure~\ref{fig:network} shows four different configurations, one at a very coarse lattice spacing (top left), one at a coarse lattice spacing (top right), another at a finer lattice spacing (bottom left), and another at a still finer lattice spacing (bottom right).  They have the same lattice volume of 4k four-simplices, except for the fourth, which has 8k four-simplices.  The image to the upper left is of a configuration from an ensemble at a very coarse lattice spacing, and one can see that there is a concentration of simplices in the central mother universe region, with long branched baby-universe-like structures coming out of that region with little connectivity to the central region.  In contrast, as the lattice spacing becomes finer the region one might ascribe to the mother universe is larger in comparison, with a more uniformly distributed set of simplices, and with baby universes of smaller extent branching off of the central region.  Although this picture is mostly qualitative, it is still useful.  In the classical theory with a discretization of Euclidean de Sitter space, the visualization tool would produce a set of nodes that fill a circle uniformly.  If one imagines a boundary circle centered on the mother universe containing roughly 90$\%$ of the nodes for each of the configurations in Fig.~\ref{fig:network}, it is clear that the finest lattice would fill the circle more uniformly than the coarsest.  There is a limitation of this visualization in that it provides a snap-shot of only a single configuration in an ensemble; it is important to keep in mind that physical observables require an ensemble average.  The configurations shown in Fig.~\ref{fig:network} are, however, typical of the ensembles to which they belong.  This visualization supports the picture that we have of the long tails in Fig.~\ref{fig:desitter} being due to baby-universe-like structures branching off of a central mother universe, and that these structures become a smaller fraction of the total universe as the continuum limit is approached, suggesting that the asymmetric tail in Fig.~\ref{fig:desitter} is a lattice artifact.


\section{Towards the continuum limit}\label{sec:continuum}

As discussed in the previous sections, we find that the physical phase is that to the left of the first-order critical line (line $AB$ in Fig.~\ref{fig:phase1}), as long as the simulations are done sufficiently close to the critical line.  We find that as $\beta$ is decreased and we follow the critical line to larger values of $\kappa_2$, the latent heat of the phase transition, as measured by the separation of the peaks in the histogram of our order parameter $N_0$, decreases.  Thus, the transition becomes softer for larger $\kappa_2$ and raises the hope that the first-order line ends at a higher-order critical point.  It is difficult to test this because the cost of simulating along the critical line for large $\kappa_2$ is high due to long autocorrelation times and large finite-size effects.  This is to be expected for simulations at very fine lattice spacings.  Because of these difficulties, it is hard to determine whether there is a second-order end point for some large (perhaps infinite) value of $\kappa_2$, or whether there is no such fixed point.  Whether the location of the putative fixed point is at finite or infinite $\kappa_2$ has important implications for our scenario, and it is therefore important to look at what studies have been done that could decide the issue.  The reason this distinction is important is as follows.  If the theory has a fixed point at finite $\beta$ and $\kappa_2$, or at $\kappa_2\to\infty$, then either way, the theory would be asymptotically safe.  However, if the fixed point is at finite $\beta$ and $\kappa_2$, this would have implications for the idea that a fine-tuning is required because the lattice regulator breaks a continuum symmetry.  If there was a second-order critical end point at finite $\kappa_2$, then there would be many trajectories that would approach the fixed point from different directions, not just the fine-tuned one that follows the first-order line \cite{Ambjorn:pc}.  Thus, a fixed point at finite $\kappa_2$ would argue against our interpretation of the fine-tuning associated with the model, while a fixed point that is approached as $\kappa_2\to\infty$ would argue in its favor.  Although asymptotic safety would hold either way, the two scenarios have different implications for the number of relevant couplings in the theory, and thus our ability to make predictions.  We cannot rule out either scenario, but we believe that the evidence favors the idea that a fine-tuning is needed to restore a continuum symmetry, which would imply that the continuum limit is approached as $\kappa_2\to\infty$.

Studies of the EDT phase diagram are inconclusive on this issue.  In three dimensions a similar looking phase diagram appears when a nontrivial measure term of the type considered in this work is included \cite{Renken:1997na}.  A strong coupling expansion, which is valid as the three-dimensional analogue of our $\kappa_2$ is taken to infinity, allows one to characterize the geometries in that limit.  Although the calculation does not appear to allow the three-dimensional analogue of de Sitter space in the $\kappa_2\to\infty$ limit, the strong coupling expansion breaks down due to large finite-size corrections around the phase transition near $\beta\approx -1$ \cite{Thorleifsson:1998qi}, which is the $\beta$ value that our tuning would require.  It would be interesting to perform the analogous strong coupling study for dimension four, since this could shed further light on where the fixed-point might appear in our phase diagram, if it exists.

Further evidence for the existence of a fixed-point comes from the work of Ref.~\cite{Benedetti:2011nn}, where the authors consider a color-tensor model, which is argued to be equivalent to dynamical triangulations in the limit $\kappa_2\to\infty$.  In this model there are two phases with a continuous (third-order) phase transition between them, suggesting that the first-order line seen in the lattice simulations ends at a continuous, higher-order critical point.  If the tensor model is truly equivalent to EDT in the $\kappa_2\to\infty$ limit, this would further strengthen the case for the existence of a UV-fixed point for the lattice theory.

Although it is difficult to show clear evidence for a higher-order fixed point in our model through standard numerical methods like finite-volume scaling, it is encouraging that the latent heat of the transition decreases as we follow the critical line to finer lattice spacings.  It is even more encouraging that what appear to be discretization effects, namely the baby-universe-like structures that lead to the long tail in the volume profile of the geometries, shrink in comparison to the mother universe as the continuum limit is approached.  

In this section we provide further evidence that a continuum limit exists in our EDT theory by computing a relative lattice spacing from a particular lattice observable and showing that it is consistent with the approach to the continuum limit that is suggested by the decrease in the latent heat and the reduction of lattice artifacts.  We also examine the spectral dimension in the long and short distance limits.  We find that the spectral dimension at long distances is compatible with four when extrapolating to the infinite-volume, continuum limit, as it must be to describe our four-dimensional world.  Our result for the short-distance spectral dimension of $D_S\approx 3/2$ has interesting implications for the ultraviolet degrees of freedom in quantum gravity and provides a possible resolution to the tension between the renormalizability of gravity as a quantum field theory and the area law for black hole entropy scaling.

\subsection{Setting the relative lattice spacing}

Ideally we would like to set the lattice spacing using a dimensionful quantity that has small discretization errors and is also largely insensitive to finite volume effects.  In quantum gravity, setting the lattice spacing amounts to determining the renormalized Planck mass.  This can be difficult because it requires making contact with semiclassical physics, while maintaining a hierarchy of scales such that the lattice spacing is much smaller than the Planck length, which in turn should be much smaller than the lattice extent.  Although we see evidence for our geometries behaving like the classical Euclidean de Sitter space solution, this is not enough to determine the lattice spacing, since Newton's constant does not appear in the classical de Sitter solution.  In Refs.~\cite{Ambjorn:2007jv, Ambjorn:2008wc}, a method for determining the lattice spacing was introduced that exploits the fact that although the classical solution for de Sitter space does not contain Newton's constant, the semiclassical fluctuations about de Sitter space do.  In the current work the lattices are too small and too coarse for a precision determination of Newton's constant, and the baby universe contamination introduces an extra systematic error, making it difficult to use this approach for a precision determination of the lattice scale.  A rough estimate of Newton's constant by this method suggests that our finest lattice spacing is close to the Planck length.  It is easier to compute the relative lattice spacing, since this can be done with an unphysical quantity that is not directly related to physics in the classical limit.  In this paper, we restrict ourselves to a calculation of the relative lattice spacing, postponing a determination of the absolute lattice spacing to future work.

\begin{figure}
\begin{center}
\includegraphics[scale=.55]{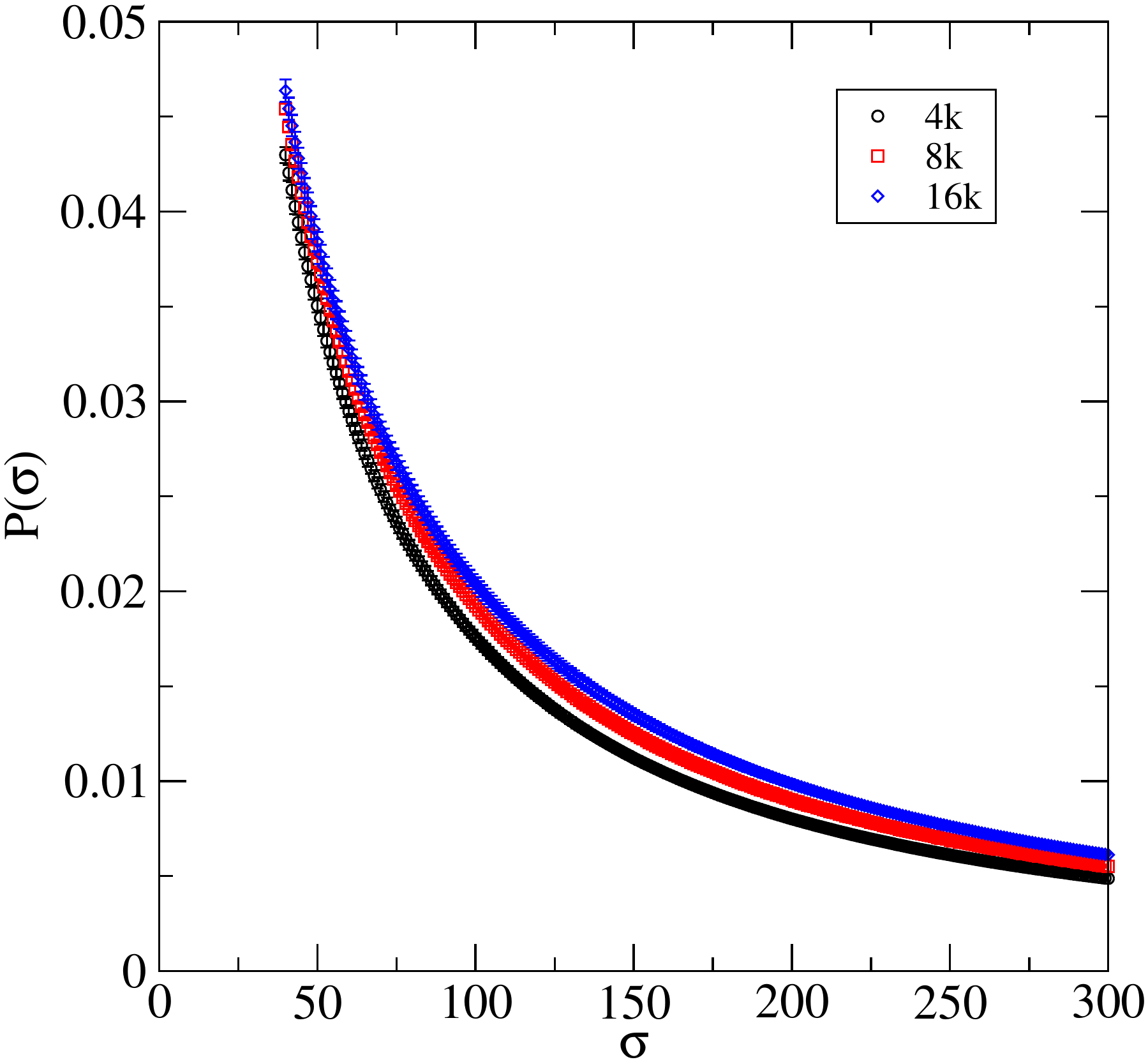}
\vspace{-3mm}
\caption{Return probability $P_S$ versus diffusion time $\sigma$ at three different volumes and fixed lattice spacing at $\beta=0$.   \label{fig:Ps_vol}}
\end{center}
\end{figure}

\begin{figure}
\begin{center}
\includegraphics[scale=.55]{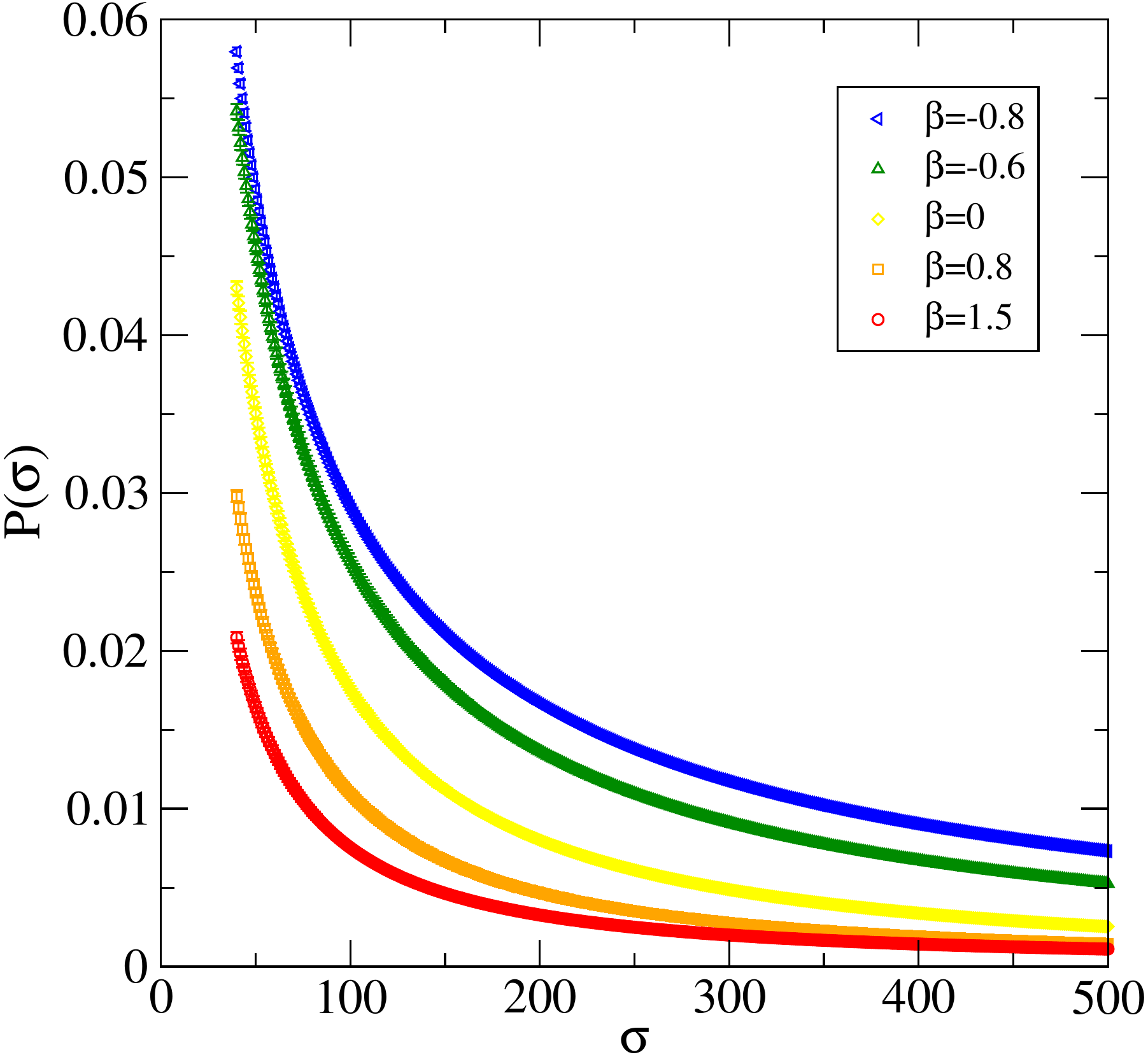}
\vspace{-3mm}
\caption{Return probability $P_S$ versus diffusion time $\sigma$ at five different lattice spacings.   \label{fig:Ps}}
\end{center}
\end{figure}

\begin{figure}
\begin{center}
\includegraphics[scale=.6]{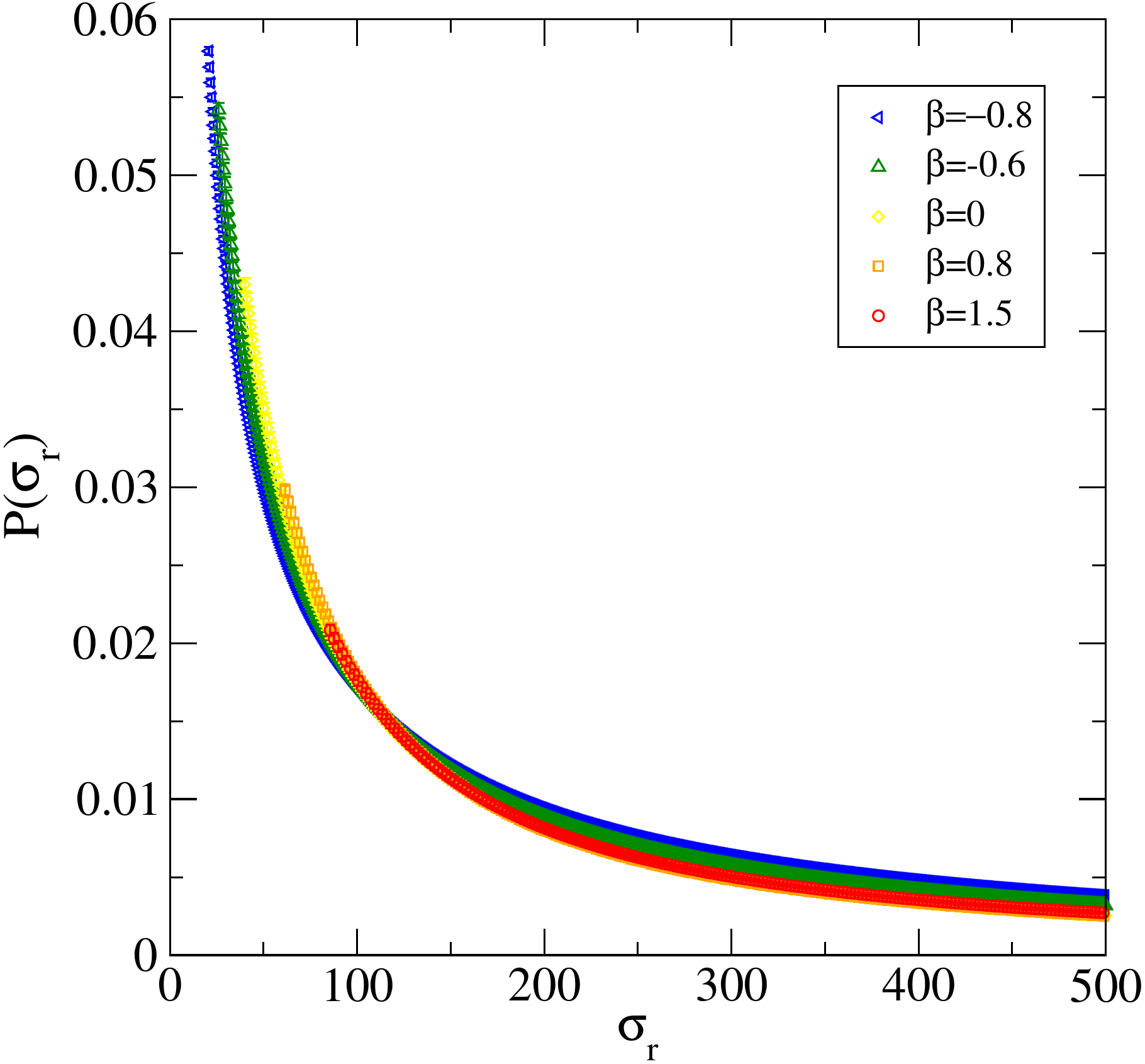}
\vspace{-3mm}
\caption{Return probability $P_S$ versus rescaled diffusion time $\sigma_{\rm r}$ at five different lattice spacings.  The $x$ axis has been rescaled so that the data points lie approximately on a universal curve.    \label{fig:Ps_rescale}}
\end{center}
\end{figure}

As an intermediate step in determining the spectral dimension, we calculate the return probability using a diffusion process.  The spectral dimension is then related to the logarithmic derivative of the return probability, as given in Eq.~(\ref{eq:spec}).  However, the return probability is actually interesting in its own right.  If we assume that the variation of the return probability with diffusion time is a universal physical quantity in our model, then we can use it to set the relative lattice spacing.  The return probability has the advantage that it is not very sensitive to volume, as we are working with small physical volumes.  Figure~\ref{fig:Ps_vol} shows the return probability $P_S$ at three different volumes at the same lattice spacing given by $\beta=0$.  The variation with volume is fairly small.  When we move along the first-order phase transition to significantly different values of $\beta$, we find a much larger difference between the various curves, as seen in Fig.~\ref{fig:Ps}.  Again, the return probability is plotted versus the diffusion time.  Since the diffusion time is a dimensionful quantity, with a dimension of distance squared, it should scale with lattice spacing squared as the continuum limit is taken, as suggested in Ref.~\cite{Coumbe:2014noa} based on the $d$-dimensional diffusion equation.  We can rescale the horizontal axis so that the data points lie on a universal curve, up to discretization and finite volume effects.  This rescaling is shown in Fig.~\ref{fig:Ps_rescale}, where the appropriate rescaling of the dimensionful $\sigma$ to convert different lattice spacings into the same physical units is $\sigma_r=\sigma a_{\rm rel}^2$.  As one can see, the rescaled points do lie on a universal curve, within differences that might be expected given finite-size effects.  We demand that the agreement be best around $\sigma_r\approx 100$; smaller distance scales could be contaminated by discretization effects and larger distances by finite-size effects.

The factor by which the horizontal axis must be rescaled between Figs.~\ref{fig:Ps} and \ref{fig:Ps_rescale} is used to infer the relative lattice spacing.  It is implicit in this determination of the relative lattice spacing that the variation of parameters in the bare action follows a line of constant physics, which would be the case if our picture of fine-tuning due to symmetry breaking and our counting of the number of relevant couplings is correct.  We find that our finest and coarsest ensembles have lattice spacings that differ by about a factor of two, with the finest ensemble in Fig.~\ref{fig:Ps} corresponding to the blue (uppermost) data set.  The lattice spacing increases monotonically with increasing $\beta$, so that the coarsest data set corresponds to the red (lowermost) curve.  The relative lattice spacings are given in Table~\ref{tab:ensembles}.  We estimate a systematic error due to finite-volume effects in this determination by comparing the return probability at different $\beta$ values with the 4k and the 16k ensembles at $\beta=0$.  We take the difference between the rescalings needed to match the two different volumes at $\beta=0$ as an estimate of the systematic error in the relative lattice spacing.

\subsection{The spectral dimension}

We are also interested in the spectral dimension itself, which can be determined from the return probability, as in Eq.~(\ref{eq:spec}).  The geometries that dominate the path integral of EDT are like the geometries that dominate the path integral in nonrelativistic quantum mechanics.  They are continuous, piecewise linear geometries that are nowhere differentiable, with a fractal structure.  The spectral dimension defines the effective dimension via a diffusion process, and it reduces to the usual definition of dimension on smooth manifolds.  In general, the Hausdorff dimension differs from the spectral dimension unless the geometry is sufficiently smooth.  Thus, it is important to study the spectral dimension of our EDT model in addition to the global Hausdorff dimension, since a value of four for the spectral dimension is also necessary to recover the correct classical limit.

The spectral dimension has received much attention over the past decade from researchers studying different approaches to quantum gravity \cite{Ambjorn:2005db, Ambjorn:2005qt, Lauscher:2005qz, Calcagni:2013dna}.  Many of these approaches have found that the spectral dimension varies as a function of the distance scale probed, and that it generically decreases from its value of 4 at large distance scales to smaller values at short distances.  The first of these calculations was that of CDT, where it was found that the spectral dimension approached a value close to 2 at short-distance scales and a value of 4 at long distance scales \cite{Ambjorn:2005db}.  A more recent investigation has found that the short-distance number is closer to 3/2 for generic points in the de Sitter phase of the model \cite{Coumbe:2014noa}.

In this section we present evidence that once the fine-tuning necessary to recover semiclassical geometry is done in our EDT model, the asymptotic values of the spectral dimension are compatible with those of CDT.  However, we find that for the relatively small physical volumes currently accessible to us, as well as the fairly coarse lattice spacings to which we are restricted, in addition to the extrapolation to large distances, we also need to extrapolate our results for the spectral dimension to infinite volume and to the continuum.  We follow the approach to our analysis of the spectral dimension of Ambjorn \textit{et al.} \cite{Ambjorn:2005qt} as closely as possible, in order to assess whether our results are compatible with theirs, and whether we can reproduce the classical space-time dimension of 4 at long distances.  In order to compute the return probability we perform a random walk on the dual lattice, such that after each step $\sigma$ in the diffusion process there is a hop to one of five (not necessarily unique) neighbors with equal probability.  In order to minimize the contamination from baby universes, we start the random walks near the center of the lattice geometries; the center is approximately determined by choosing a four-simplex at random from within the subset of four-simplices that are in the shell of maximum size that is obtained from the shelling of the geometries to find $N_4^{\rm shell}$.  This is analogous to what is done in the CDT approach, where the random walk is started in the time slice of greatest three-volume in the extended region of the geometries, away from the cutoff size stalk that extends in the time direction beyond the semiclassical bubble. 

\begin{figure}
\begin{center}
\includegraphics[scale=.6]{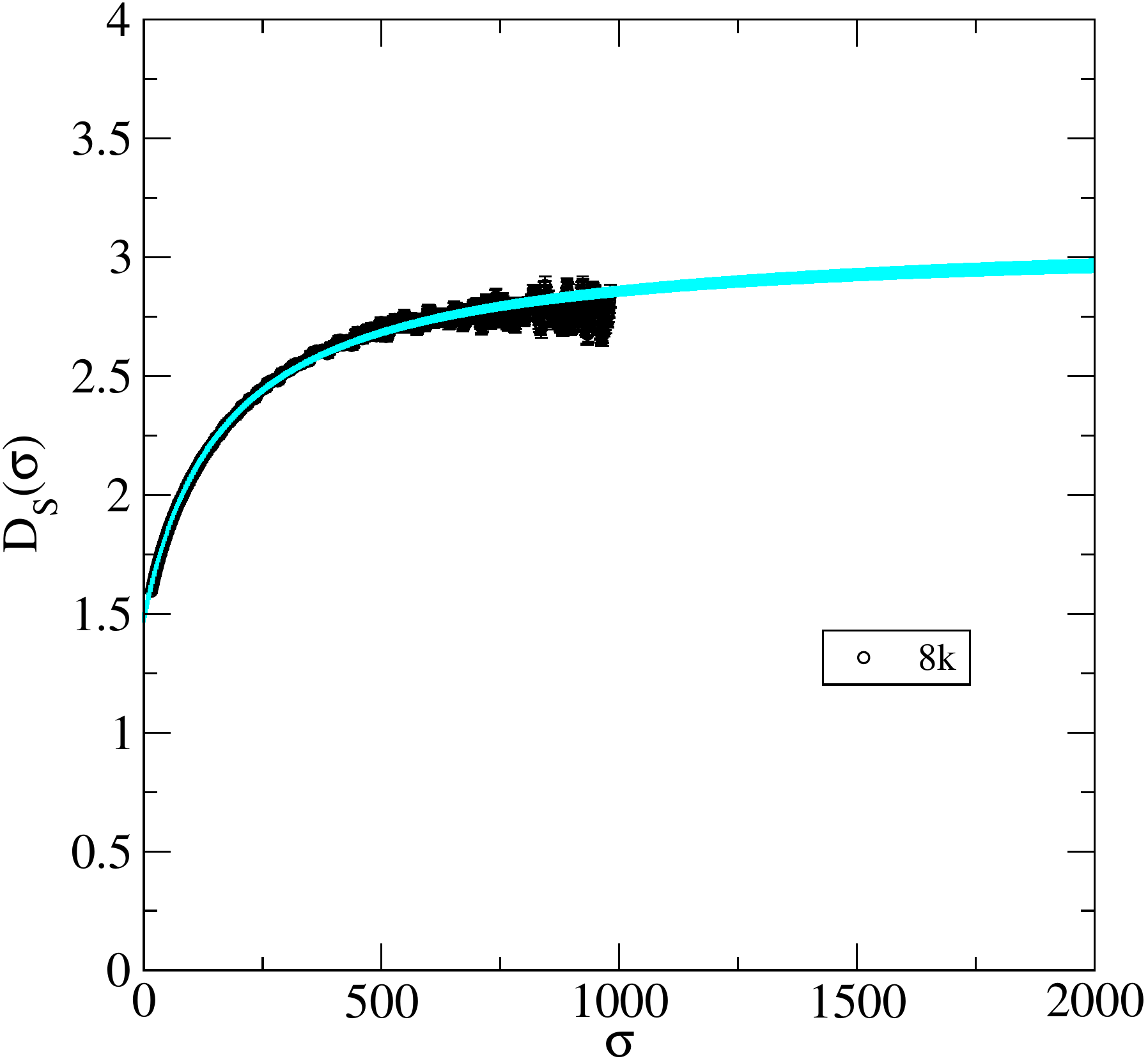}
\vspace{-3mm}
\caption{Spectral dimension $D_S$ versus diffusion time $\sigma$ for the 8k, $\beta=0$ ensemble, along with fit curve assuming the AJL ans\"atz.  The black shows the data points with jackknife error bars from a direct computation of the logarithmic derivative of the return probability.  The cyan (grey) band shows the one sigma error band of the fit curve.    \label{fig:Ds_8k}}
\end{center}
\end{figure}

\begin{table}
\begin{center}
\caption{The results of fits to the spectral dimension.  The first column, $\beta$, labels the ensemble.  The second column is the lattice volume, the third is the number of configurations sampled, the fourth is the range of $\sigma$ used in the fit, the fifth is the fit result for the extrapolation to zero distance for the spectral dimension, the sixth is the fit result for the extrapolation of $D_S$ for $\sigma\to\infty$, the seventh is the $\chi^2/{\rm d.o.f.}$ of the fit, and the last column is the $p$-value of the fit. }
\label{tab:Ds_fits}
\begin{tabular}{cccccccc}
\hline \hline
\ \ $\beta$ \ \ & \ \ $N_4$ \ \ \ & Number of configurations \ \ & \ \ $\sigma_{\rm min}$-$\sigma_{\rm max}$ \ \ & \ \ \ $D_S(0)$ \ \ \ & $D_S(\infty)$ & $\chi^2/{\rm d.o.f.}$ \ \ & $p$-value \\
\hline
$-0.8$ & 8000 & 1566 & $105-831$ & 1.445(16) & 2.75(11) & 1.14 & 0.29 \\
$-0.6$ & 4000 & 610 & $72-611$ & 1.431(29) & 2.756(89) & 1.38 & 0.11 \\
0 & 4000 & 930 & $61-556$ & 1.464(49) & 2.809(51) & 1.12 & 0.36 \\
0 & 8000 & 1250 & $61-611$ & 1.484(21) & 3.090(41) & 1.25 & 0.17 \\
0 & 16000 & 552 & $61-611$ & 1.484(37) & 3.30(12) & 0.92 & 0.76 \\
0.8 & 4000 & 523 & $61-226$ & 1.44(15) & 2.797(64) & 0.91 & 0.56 \\
1.5 & 4000 & 1815 & $61-138$ & 1.64(26) & 2.655(93) & 1.60 & 0.17 \\
\hline
\end{tabular}
\end{center}
\end{table}

Figure~\ref{fig:Ds_8k} shows data for the spectral dimension as a function of $\sigma$ on the ensemble with volume equal to 8k four simplices at $\beta=0$, along with a fit to the scale dependence of the spectral dimension using the ans\"atz introduced by Ambj{\o}rn, Jurkiewicz, and Loll (AJL) \cite{Ambjorn:2005qt} to fit the same quantity in CDT, 
\bea \label{eq:AJL}   D_S = a - \frac{b}{c+\sigma},
\eea  
with $a$, $b$, and $c$ free parameters.  In the absence of better theoretical guidance, we use Eq.~(\ref{eq:AJL}) to study what results are suggested by our data if we assume behavior similar to that of CDT.  Table~\ref{tab:Ds_fits} shows the results of our fits to the spectral dimension for all of our lattices.  The errors on $D_S(0)$ and $D_S(\infty)$ are statistical only.  We choose the fit range such that $\sigma_{\rm max}$ is as large as possible while still maintaining an acceptable $p$-value.  If $\sigma_{\rm max}$ is chosen too large, the finite-size effects that lead to a turnover in the lattice data cannot be described by the fit ans\"atz Eq.~(\ref{eq:AJL}), which is monotonic.  We take $\sigma_{\rm min}$ to be greater than $\sim60$-$100$ in order to minimize contamination from discretization effects, which is supported by studies of $D_S(\sigma)$ in the branched polymer phase \cite{Coumbe:2014nea}, for which $D_S=4/3$ is expected, regardless of scale.  Figure~\ref{fig:Ds_8k} also shows the result of our fit to the 8k, $\beta=0$ data for the spectral dimension.

The results of Table~\ref{tab:Ds_fits} show that the extrapolation to large distances using the AJL ans\"atz leads to values in the range of $D_S(\infty)=2.7-3.3$ with errors around a few percent for our data set.  Variations of the fit range that maintain acceptable $p$-values do not significantly alter this picture, and it is difficult to see how, using the AJL ans\"atz, one could find a value of $D_S(\infty)$ consistent with four on any of our individual ensembles.  However, this could be a result of finite-volume and discretization effects.  Given that we have multiple volumes and lattice spacings, we can explore the possibility that a further extrapolation to the infinite volume, continuum limit might be necessary.
Since $D_S(\infty)$ is a long distance quantity, it is at first glance difficult to see why discretization effects might be important.  However, when a regulator breaks a symmetry, even the long-distance light modes are affected.  Thus, we are motivated to consider the continuum limit as well as the infinite volume limit in our investigation of the spectral dimension at long distances.
 
\begin{figure}
\begin{center}
\includegraphics[scale=.55]{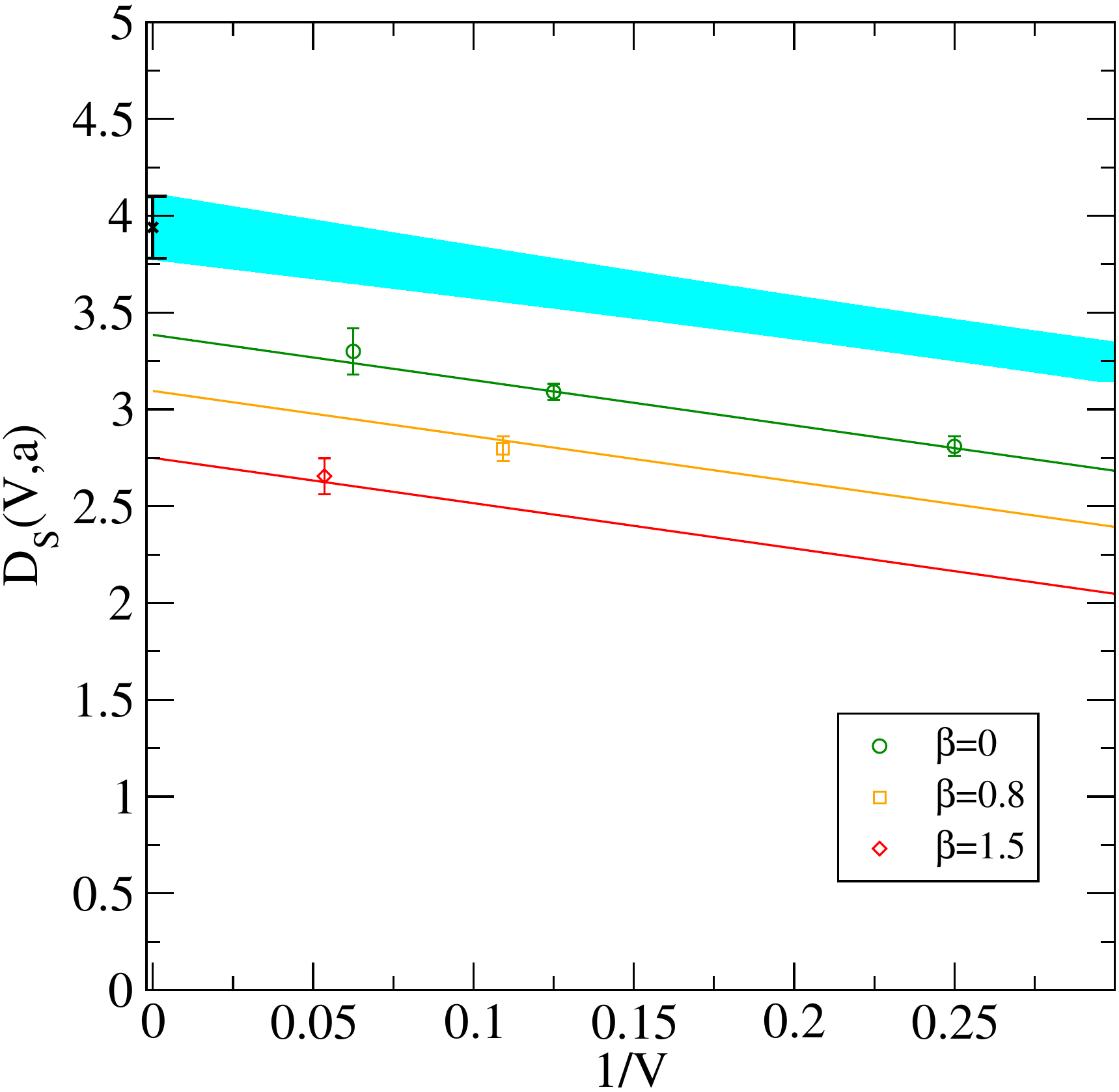}
\vspace{-3mm}
\caption{Long-distance spectral dimension $D_S$ as a function of inverse lattice volume and different lattice spacings, along with an extrapolation to the infinite volume, continuum limit.  The fit has a $\chi^2/\rm{d.o.f.}=0.79/2$, corresponding to a $p$-value of 0.67.   \label{fig:Ds_extrap}}
\end{center}
\end{figure}

\begin{figure}
\begin{center}
\includegraphics[scale=.58]{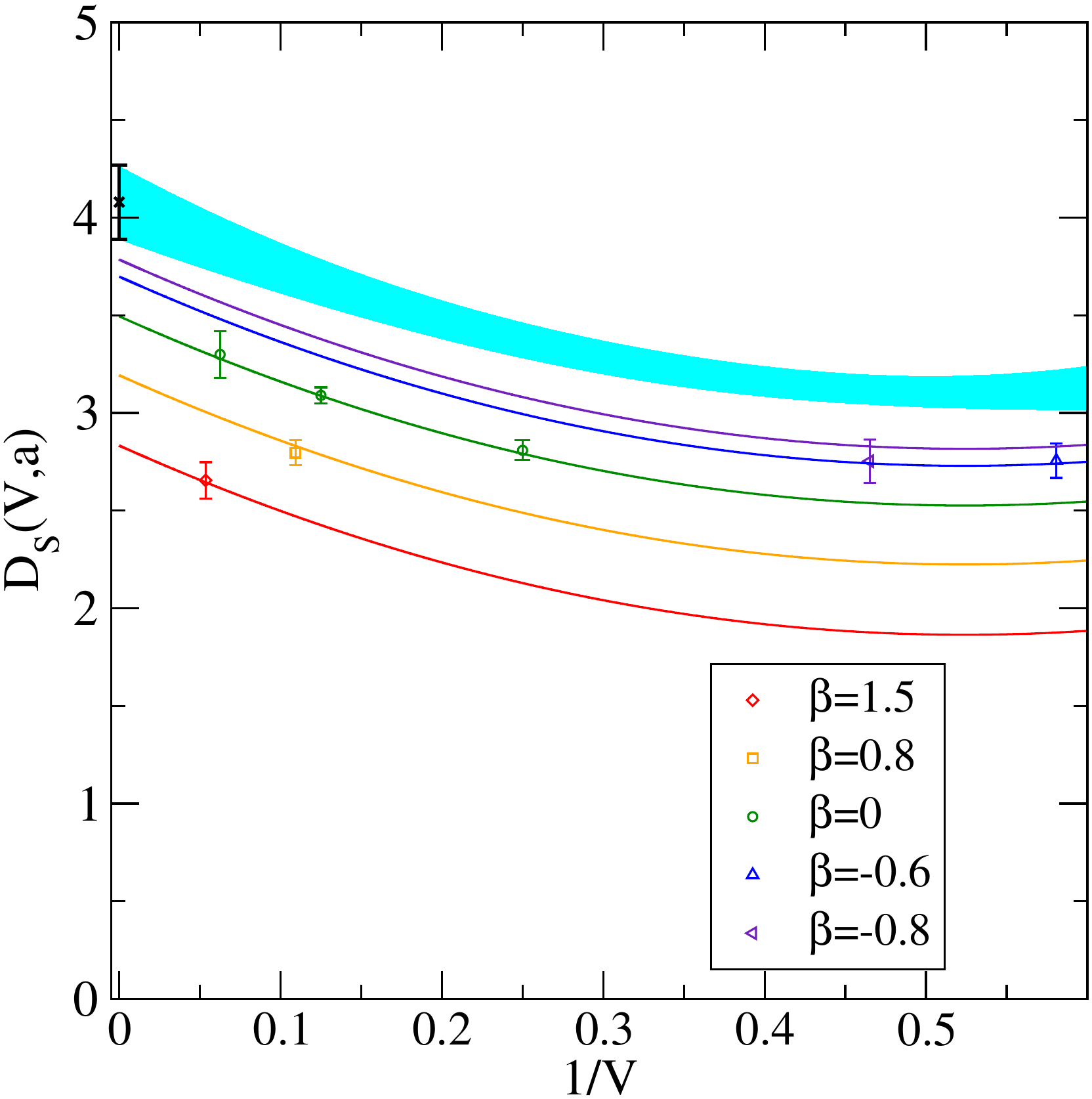}
\vspace{-3mm}
\caption{Long-distance spectral dimension $D_S$ as a function of inverse lattice volume and different lattice spacings, along with an extrapolation to the infinite volume, continuum limit.  The fit has a $\chi^2/\rm{d.o.f.}=0.96/3$, corresponding to a $p$-value of 0.81.   \label{fig:Ds_all_extrap}}
\end{center}
\end{figure}

We consider an additional extrapolation of $D_S(\infty)$ of the form
\bea\label{eq:dsfit}     D_S(\infty) = c_0 + c_1 \frac{1}{V} +c_2 a^2,
\eea
where $c_i$ is a fit parameter, $V$ is the physical volume of the ensemble, and $a$ is the lattice spacing.  We assume that the scaling with lattice spacing is $a^2$ because diffeomorphism invariance forbids terms in the action with odd dimension.  We are assuming both that we are following a renormalization group trajectory with fixed physics, and that the breaking of general coordinate invariance by the lattice regulator can be controlled by tuning the coefficient associated with the local measure term, without introducing other operators.  We find that this ans\"atz is supported by the numerical data, as the data points are linear in $1/V$ and $a^2$.  Figure~\ref{fig:Ds_extrap} shows the extrapolation to the infinite volume, continuum limit.  The cyan curve is the extrapolation to the continuum as a function of inverse volume, and the black cross shows the infinite volume limit of this curve, including the statistical error in the extrapolation.  The extrapolated value for $D_S(\infty)$ is $3.94\pm 0.16$, where the error is the statistical error only.  The fit has a $\chi^2/{\rm d.o.f.}=0.52$, corresponding to a $p$-value of 0.59.  We could also include our finer ensembles in this fit, though it does require us to include data points with significantly smaller physical volumes.  We find in this case that the data prefer a fit ans\"atz that includes a $1/V^2$ term.  Figure~\ref{fig:Ds_all_extrap} shows our fit to the full data set, including a $1/V^2$ term in Eq.~(\ref{eq:dsfit}).  This fit has a $\chi^2/{\rm d.o.f.}=0.32$, corresponding to a $p$-value of 0.81.  The value of $D_S(\infty)$ in this case is $4.08\pm0.19$.  Without the $1/V^2$ term, Eq.~(\ref{eq:dsfit}) provides a marginal fit to the data, with a $\chi^2/{\rm d.o.f.}=2.19$ corresponding to a $p$-value of 0.068 and a $D_S(\infty)=3.70\pm0.13$.  The scatter between these results gives some measure of the systematic error associated with the infinite volume, continuum extrapolation, but it is difficult to fully quantify the systematic error associated with this result, since much depends on the assumption of the fit ans\"atz for extrapolating $D_S(\sigma)$ to the limit in which $\sigma\to\infty$.  It is clear that a future analysis with more extensive data, especially at finer lattice spacings and larger volumes, is needed in order to control systematic errors associated with taking the continuum limit of $D_S(\infty)$.  Although additional work is necessary to fully control the systematic errors, our results show that it is plausible that the value of 4 might be recovered in the EDT model studied here.

\begin{figure}
\begin{center}
\includegraphics[scale=.55]{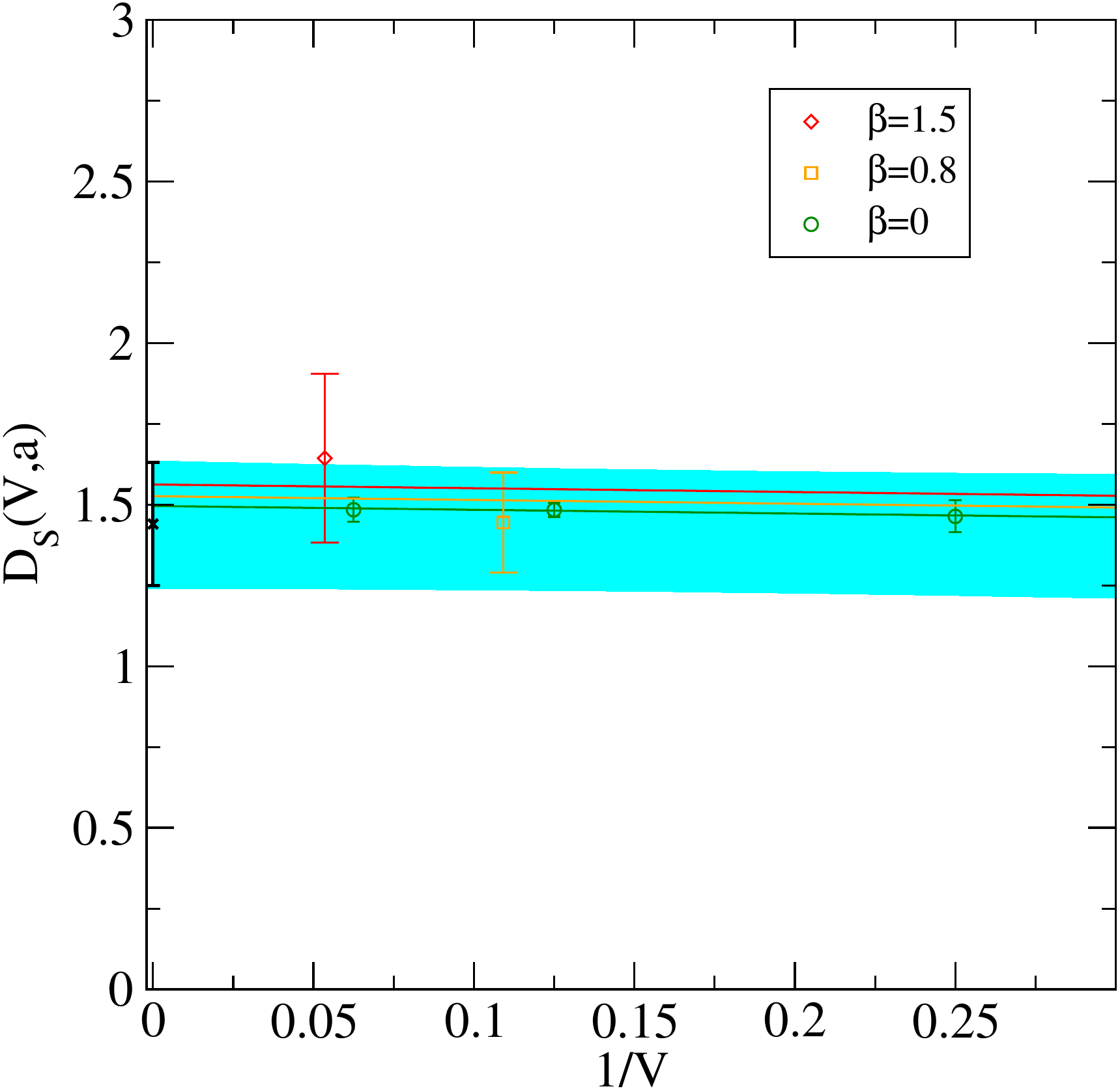}
\vspace{-3mm}
\caption{Short-distance spectral dimension $D_S$ as a function of inverse lattice volume and different lattice spacings, along with an extrapolation to the infinite volume, continuum limit with the same fit ans\"atz as the fit shown in Fig.~\ref{fig:Ds_extrap}.  The fit has a $\chi^2/{\rm d.o.f.}=0.34/2$, corresponding to a $p$-value of 0.84.  \label{fig:Ds_0_extrap}}
\end{center}
\end{figure}

\begin{figure}
\begin{center}
\includegraphics[scale=.58]{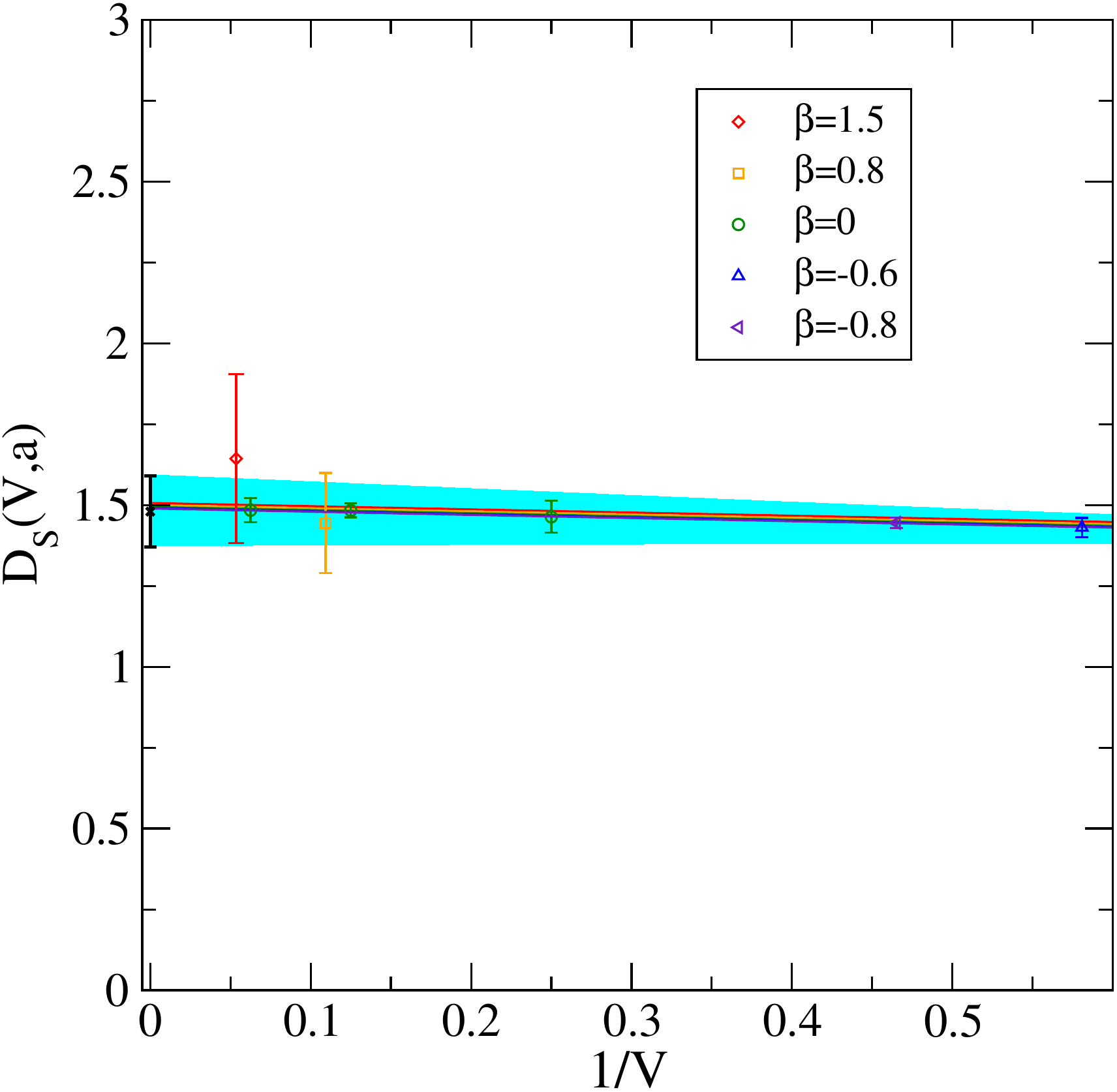}
\vspace{-3mm}
\caption{Short-distance spectral dimension $D_S$ as a function of inverse lattice volume and different lattice spacings, along with an extrapolation to the infinite volume, continuum limit with the same fit ans\"atz as the fit shown in Fig.~\ref{fig:Ds_extrap}.  The fit has a $\chi^2/{\rm d.o.f.}=0.44/4$, corresponding to a $p$-value of 0.98.  \label{fig:Ds_0_all_extrap}}
\end{center}
\end{figure}

We also examine the result of an extrapolation of $D_S(0)$ values to the infinite volume and continuum limits using the same fit ans\"atz, Eq.~(\ref{eq:dsfit}), that was used to extrapolate $D_S(\infty)$.  The result of this fit is shown in Fig.~\ref{fig:Ds_0_extrap}.  Again, the fit is excellent, with a $\chi^2/{\rm d.o.f.}=0.17$ and a $p$-value of 0.84.  The value of $D_S(0)$ obtained from this simple extrapolation is $1.44\pm 0.19$.  Figure~\ref{fig:Ds_0_all_extrap} shows a similar fit using the same fit function, Eq.~(\ref{eq:dsfit}), but this time including data from all ensembles.  The extrapolated value is $D_S(0)=1.48\pm0.11$, and the fit has a $\chi^2/{\rm d.o.f.}=0.11$, corresponding to a $p$-value of 0.98.

We find that values of $D_S$ at short distances close to 3/2 are favored by our simulations, while a value of 2 is disfavored, at least within the simple fit ans\"atze that we have assumed.  The result $D_S(0)=3/2$ is the value that could resolve the tension between asymptotic safety and black hole entropy scaling, which we can see as follows.  When a theory is formulated on a fractal space-time, it is the spectral dimension that should be used in the thermodynamic equations of state \cite{Akkermans:2010dz}, and for a black hole \cite{Carlip:2011uc}.  Thus, the value of the spectral dimension at short distances $D_S(0)$ is the quantity that should be identified with $d$ in Eqs.~(\ref{eq:CFT}) and (\ref{eq:GR}). As can be easily verified, 3/2 is the unique value of the dimension in which the scaling of Eqs.~(\ref{eq:CFT}) and (\ref{eq:GR}) agree.  If this result holds up to further scrutiny, the Banks argument may become an argument in favor of asymptotic safety rather than an argument against it.  We compare our value for the spectral dimension with the results obtained using other methods in Sec.~\ref{sect:others}.


\section{Running of couplings and dimension of UV critical surface}
\label{sec:running}

Given the nontrivial tests that have been passed by the EDT model investigated here, it is interesting to examine the flow of couplings as a function of lattice spacing.  We are especially interested in the dimension of the ultraviolet critical surface, since this tells us how many experimental inputs are required in order to use the theory to make predictions.  

We argue in this section that of the three parameters that we adjust in our bare action, one is redundant, and two are relevant, in the renormalization group sense, but one of these relevant parameters would not be necessary if the regulator maintained the continuum coordinate invariance.  A subtraction of unphysical cutoff size contributions will be necessary, just as for the bare mass for Wilson fermions in QCD.  To understand the implications for the number of relevant couplings in the theory we turn to a discussion of redundant operators in quantum gravity.

\subsection{Redundant operators and field strength renormalization}

There is a standard test for identifying the redundant parameters in a theory \cite{Weinberg:1980gg}.  The coupling $\gamma$ appearing in a Lagrangian ${\cal L}$ is redundant if and only if $\frac{\partial {\cal L}}{\partial \gamma}$ vanishes or is a total derivative when one uses the Euler-Lagrange equations of motion.  Specializing to the case of gravity, the Euler-Lagrange equations are
\bea  \frac{\partial {\cal L}}{\partial g_{\mu\nu}} -\partial_\alpha \frac{\partial {\cal L}}{\partial(\partial_\alpha g_{\mu\nu})} = 0,
\eea
and given the Einstein-Hilbert Lagrangian they lead to the Einstein equations.

For convenience in the following argument, we can rewrite the Lagrangian corresponding to the Einstein-Hilbert action [Eq.~(\ref{eq:ERcont})] as
\bea  {\cal L} = \frac{\omega}{16\pi}\sqrt{g}(R-2\omega\Lambda'),
\eea
where $\omega$ and $\Lambda'$ replace $G$ and $\Lambda$.
Then
\bea  \frac{\partial {\cal L}}{\partial \omega} = \frac{1}{16\pi}\sqrt{g}(R-4\omega\Lambda').
\eea
The classical equations of motion following from the Euler-Lagrange equations are just the vacuum Einstein equations with a cosmological constant, which in our new notation imply $R-4\omega\Lambda'=0$, leading to $\partial {\cal L}/\partial\omega =0$ when applying the equations of motion.  Thus $\omega = 1/G$ is a redundant parameter.  
The only parameter that is not redundant is $\Lambda'$, as $\partial {\cal L}/\partial \Lambda'$ does not vanish and is not a total derivative.  In the more standard notation of Eq.~(\ref{eq:ERcont}), the nonredundant parameter is $G \Lambda$, the cosmological constant in Planck units.  Note that a nonredundant coupling may be relevant or irrelevant, depending on its renormalization group flow in the vicinity of a fixed point.  A similar argument can be used to show that the field-strength renormalization constant in a quantum field theory is a redundant parameter; it can be rescaled by a simple field redefinition. Indeed, the quantity $\omega$ appears as the field-strength renormalization constant for the gravitational field.  The analogue of this derivation also holds when matter fields are included.

It was argued originally by the authors of Ref.~\cite{Gastmans:1977ad} that $G$ could still be a nonredundant coupling in the presence of boundary terms, where the coupling appearing in the coefficient of the boundary terms differs from that of the bulk term.  This argument does not apply if the metric vanishes sufficiently fast at distances approaching infinity.  Our formulation does not introduce a boundary, so the condition on the metric is implicit, and the argument of Ref.~\cite{Gastmans:1977ad} does not apply.  Even so, we find evidence for the existence of a continuum limit without boundary terms.

We now discuss the implications for the lattice theory.  We can study the dependence of $G$ and $\Lambda$ as a function of lattice spacing by reading off the values of $G$ and $\Lambda$ (or rather their dimensionless counterparts) used to produce a given ensemble and comparing the relative lattice spacings of the various ensembles as outlined in Sec.~\ref{sec:continuum}.  There is a complication, however.  The field redefinition that is needed to show that $G$ and $\Lambda$ cannot be separately relevant couplings requires general coordinate invariance to hold.  Because the lattice regulator appears to break this symmetry, we do not necessarily have the freedom to perform this rescaling in the lattice approach without fine-tuning; this could obscure the interpretation of the running of couplings in the four-dimensional lattice theory.  Assuming that the coordinate invariance is broken by the regulator, a fine-tuning would then be necessary to recover the correct continuum physics.  This may be accomplished by introducing the measure term and its accompanying parameter $\beta$ to our simulations and fine-tuning $\beta$ for each new value of $G\Lambda$. 
We do not have the freedom to adjust $\Lambda$ independently of $G$ and $\beta$.
This is because it is necessary to tune the bare parameter $\Lambda$ to its pseudocritical value as a function of $G$ and $\beta$ in order to take the infinite lattice volume limit.  In view of the above discussion, we can view this tuning of $\Lambda$ as a particular choice of field-strength renormalization.  The remaining combinations of parameters $G\Lambda$ and $\beta$ are then relevant couplings, which must be tuned to take the continuum limit.

We note here that the tuning of the bare cosmological constant is not by itself enough to recover a small renormalized cosmological constant.  Both the branched polymer phase and the collapsed phase have a cutoff size renormalized cosmological constant, since they can be interpreted as universes with sizes of the order of the cutoff.  In the EDT formulation, {\it both} the measure parameter $\beta$ and the bare cosmological constant must be tuned to their critical values at a first-order phase transition in order to recover a universe with a large physical volume, as opposed to merely a large lattice volume with physical size of the order of the cutoff.

It is convenient to define the dimensionless versions of the Newton's coupling and cosmological constant,
\bea  \hat{G} = G/a^2, \ \ \ \  \hat{\Lambda} = a^2\Lambda,
\eea
since it is the dimensionless coupling that must approach a constant at the fixed point in the asymptotic safety scenario.  Here $a$ is the lattice spacing;  i.e. the length of the link in a four-simplex.  We show below that a subtraction of the bare cosmological constant is necessary to recover the expected flow of $\hat{G}\hat{\Lambda}= G\Lambda$ in the four-dimensional theory, and that there is evidence that this quantity (after subtraction) approaches a constant in what appears to be the vicinity of the fixed point in the simulations.  Notably, the dimensionless bare $\hat{G}$ and (subtracted) $\hat{\Lambda}$ do not.  Since only relevant couplings need to approach a fixed point for a theory to be asymptotically safe, but redundant couplings do not \cite{Weinberg:1980gg}, this does not undermine the case for asymptotic safety for this theory.  It does, however, provide further evidence that $\hat{G}$ and $\hat{\Lambda}$ are not separately relevant couplings.

\subsection{The subtraction condition and the running of $G\Lambda$}

\begin{figure}
\begin{center}
\includegraphics[scale=.55]{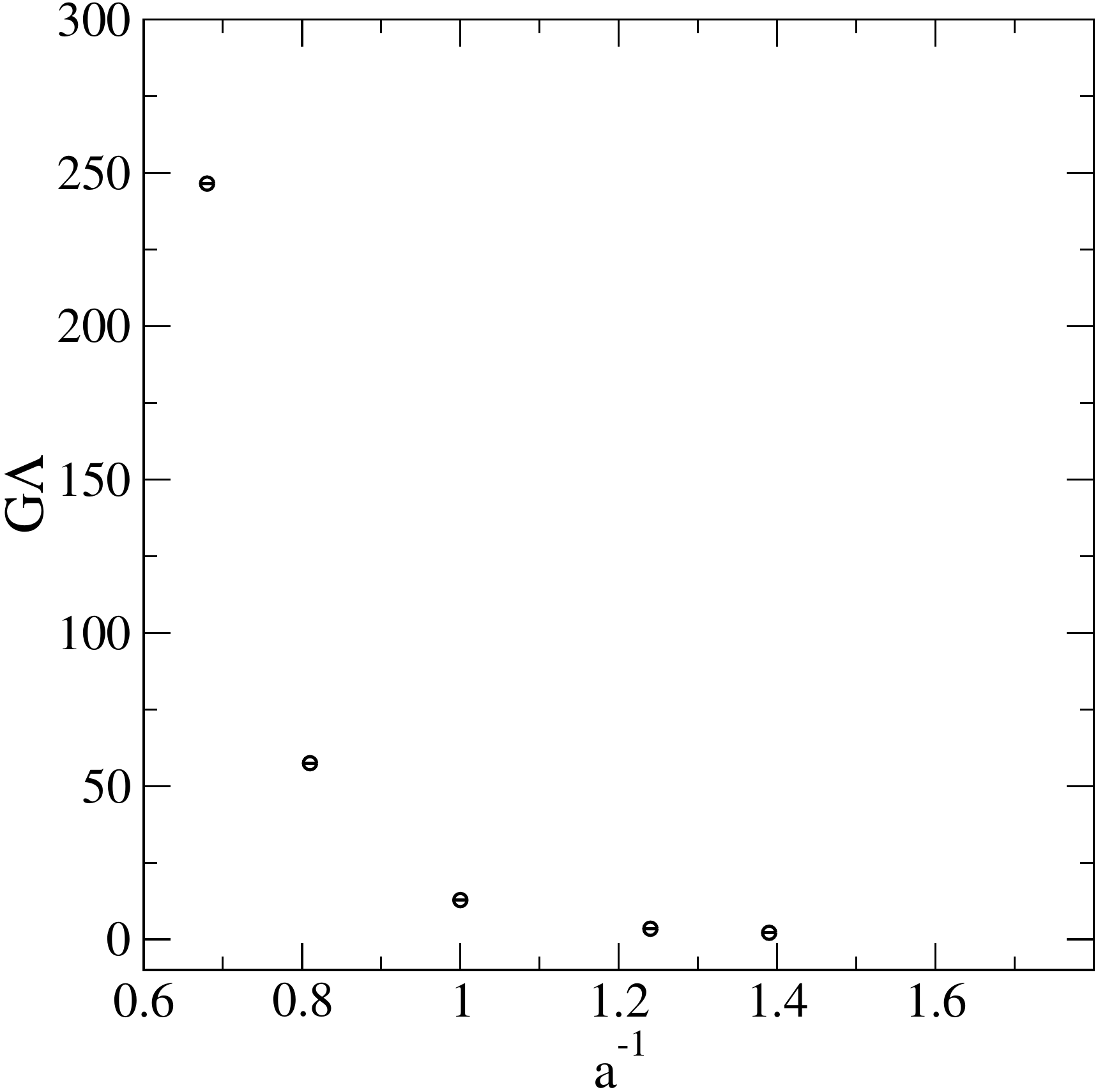}
\vspace{-3mm}
\caption{The bare $G\Lambda$ as a function of inverse lattice spacing.  \label{fig:GL_unsub}}
\end{center}
\end{figure}

Figure~\ref{fig:GL_unsub} shows our bare $G\Lambda$ without subtraction plotted as a function of inverse lattice spacing.  The behavior of the coupling in this plot is unphysical, since the cosmological constant in Planck units is increasing from its fixed-point value and approaching a large number in its flow to the infrared.  This is the opposite of our expectations from a direct study of the lattice geometries.  Our results suggest that at long distances the geometries are dominated by the de Sitter instanton solution, where in the (semi)classical theory the cosmological constant determines the size of the universe.  Since we find approximately semiclassical de Sitter universes in our simulations, the renormalized cosmological constant should be fixed by the lattice volume, which serves as an infrared cutoff.  Thus, we expect that $G\Lambda$ should approach a small value as it flows to the infrared, rather than a very large value $( \gg 1)$ in that limit.  However, since the measure term renormalizes the cosmological constant, and since the running of $\beta$ in the measure term is unphysical, the dependence of $\Lambda$ on lattice spacing also has an unphysical contribution to it.  Even so, we show that it is possible to introduce a subtraction procedure for $\Lambda$ and $\beta$ where physical running for the couplings is restored.

We implement the subtraction procedure by assuming that $\beta$ is not a relevant coupling in the continuum theory, but rather a fixed constant.  This is consistent with the role usually played by a nonuniform measure term, i.e. a Jacobian of the integration variables, with no new associated coupling constants.  We assume that the tuning of $\beta$ is done to restore the coordinate invariance at finite lattice spacing, and that as a by-product, it adjusts the bare cosmological constant to compensate for unphysical cutoff effects.  Thus, we posit that the difference between the subtracted bare cosmological constant and the bare input value used in the simulations can be obtained by determining the value of the bare cosmological constant when $\beta$ is held fixed as a function of lattice spacing at some appropriate value.    

\begin{figure}
\begin{center}
\includegraphics[scale=.65]{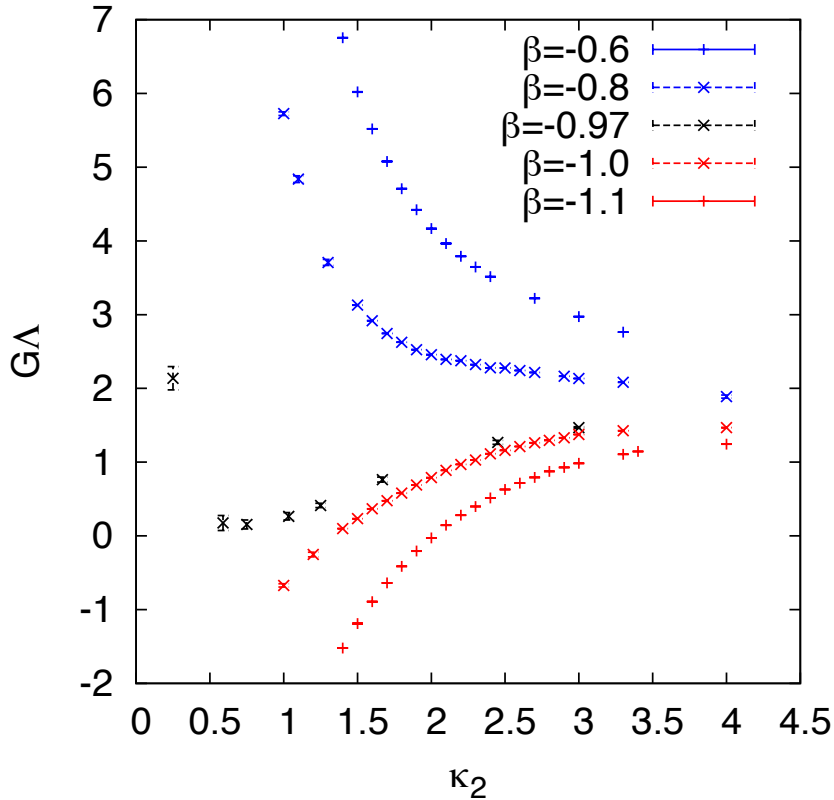}
\vspace{-26mm}
\caption{Bare $G\Lambda$ as a function of $\kappa_2$ for several different values of $\beta$.  \label{fig:GL_diff_beta}}
\end{center}
\end{figure}

It remains to specify the fixed value of $\beta$ that one must use in order to define the subtracted bare cosmological constant.  The behavior of $G\Lambda$ as a function of $\kappa_2$ for various $\beta$ values is illustrated in Fig.~\ref{fig:GL_diff_beta}.  The parameter $\kappa_2$ serves as a proxy for lattice spacing here, with larger values of $\kappa_2$ corresponding to smaller lattice spacings.  It is impractical to determine the actual relative lattice spacing at all of these points.  When we look at the subtracted value of $G\Lambda$ as a function of $\kappa_2$ for various fixed values of $\beta$, we find that for $\beta$ around $-0.8$ or greater, the subtracted $G\Lambda$ grows in the infrared, the opposite of the expected behavior.  For $\beta$ less than around $-1$, the subtracted $G\Lambda$ decreases from a positive constant to large negative values in the infrared.  In between, the behavior is more interesting.  For values between around $-0.9$ and $-1$ the value decreases from a positive ${\cal O}(1)$ constant and reaches a local minimum before rapidly increasing.  This suggests the subtraction condition that we describe below.  

\begin{figure}
\begin{center}
\includegraphics[scale=.65]{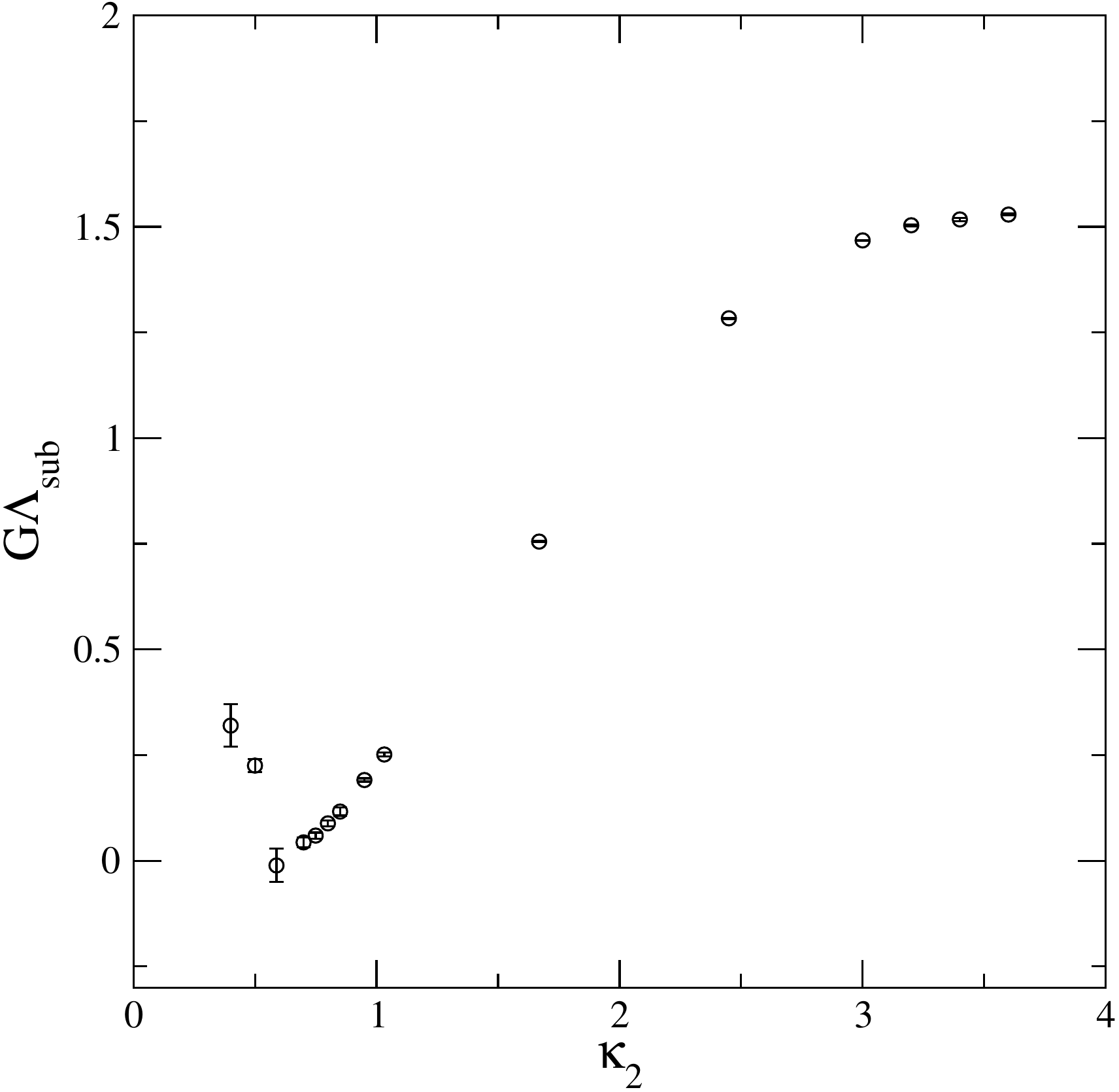}
\vspace{-3mm}
\caption{$G\Lambda_{\rm sub}$ as a function of $\kappa_2$.  Increasing $\kappa_2$ corresponds to finer lattice spacings.  The value of $\beta$ is tuned so that the local minimum (around $\kappa_2=0.5$) coincides with the zero of $G\Lambda_{\rm sub}$. \label{fig:GLambda_k2}}
\end{center}
\end{figure}

\begin{figure}
\begin{center}
\includegraphics[scale=.55]{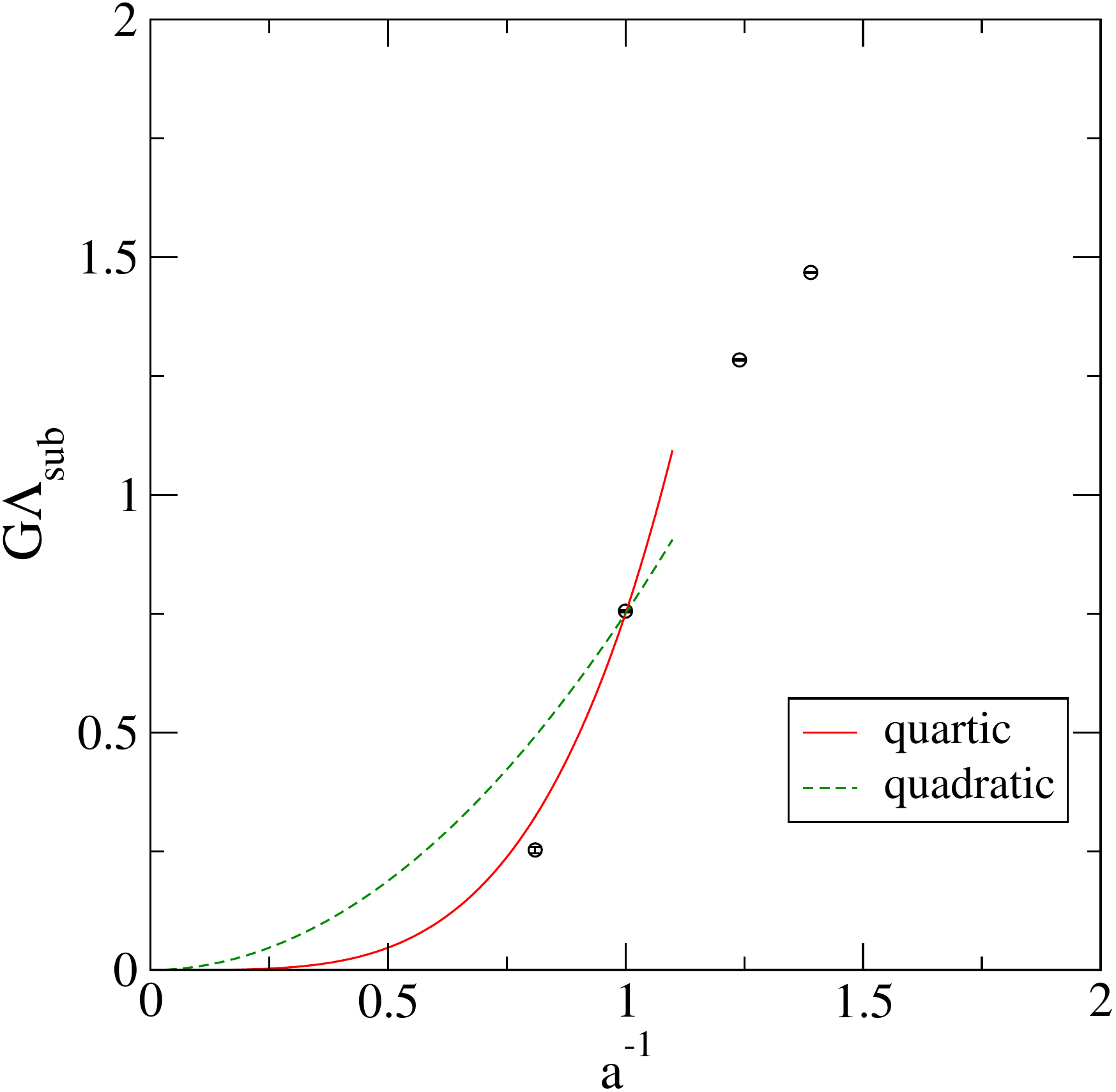}
\vspace{-3mm}
\caption{Bare $G\Lambda_{\rm sub}$ as a function of inverse lattice spacing.  The solid line is a quartic constrained to pass through the second to leftmost data point.  The dashed line is a quadratic constrained to pass through the same point.  \label{fig:GL_sub}}
\end{center}
\end{figure}

In order to motivate our choice of $\beta$ for the subtraction condition, we first recall the semiclassical behavior of the lattice geometries.  The reasonably good agreement between our results for the lattice geometry and Euclidean de Sitter space tells us that the finite-volume cutoff determines the renormalized cosmological constant in the infrared.  The larger the volume, the smaller the cosmological constant, such that $\Lambda \propto 1/\sqrt{V_4}$; this is simply the classical relation between the cosmological constant and the four-volume of the universe for de Sitter space.  In order to reach this classical value deep in the infrared where classical physics is expected to be a good description, the running coupling must decrease according to a power law in the renormalization scale (in our case the inverse lattice spacing) with a power that is quadratic or greater.  Otherwise, the cosmological constant would not be small enough at scales of the order of the Hubble scale for the classical scaling relation between four-volume and the cosmological constant to hold.  Thus, both the value of $G\Lambda$ and its derivative with respect to the renormalization scale should go to zero deep in the infrared.  Our subtraction condition is that the local minimum in our subtracted $G\Lambda$ must coincide with the zero of the subtracted $G\Lambda$.  That this happens for any value of the subtracted cosmological constant is nontrivial, and the fact that it happens far in the infrared is a good consistency check of the condition.  Figure~\ref{fig:GLambda_k2} shows $G\Lambda_{\rm sub}$ as a function of $\kappa_2$, our proxy for inverse lattice spacing, and Fig.~\ref{fig:GL_sub} shows $G\Lambda_{\rm sub}$ as a function of inverse lattice spacing.  Our subtraction condition necessarily suffers from discretization effects, since the condition must be satisfied in the infrared, where lattice artifacts could be large.  Our subtraction procedure should, however, be systematically improvable.  This systematic improvement could be done using Monte Carlo renormalization-group techniques, which would allow us to approach more closely the renormalized trajectory, where discretization effects are absent even at finite lattice spacing.  In order to perform simulations on the renormalized trajectory we would require a perfect action.  However, a perfect action is not practical to simulate, since it would require an infinite number of terms in the action.  Still, an imperfect action that is improved over the current action should lead to a subtraction condition that yields a running for $G\Lambda_{\rm sub}$ that is closer to that of the renormalized trajectory.  Further improvements to the action should lead to an approach to the renormalized trajectory and the correct running of $G\Lambda_{\rm sub}$, even deep into the infrared.

\begin{figure}
\begin{center}
\includegraphics[scale=1.15]{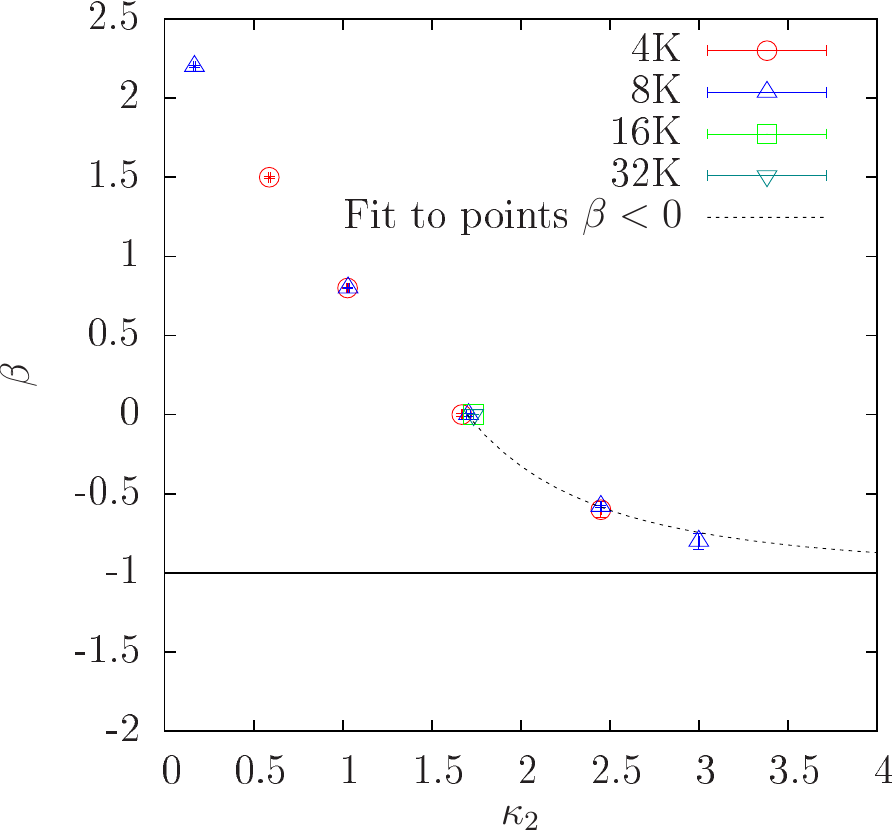}
\vspace{-3mm}
\caption{The phase diagram in the $\kappa_2$, $\beta$ plane.  A simple fit to the data along the critical line with $\beta<0$ is shown as the dashed line.  This fit is constrained to asymptote to $-1$.   \label{fig:phase}}
\end{center}
\end{figure}

There are cross-checks that the subtraction is behaving as we would like it to.  First, we expect that as the continuum limit is approached, coordinate invariance should be restored, so the value of $\beta$ in the continuum should match the tuned value from our subtraction procedure, up to discretization effects.  Figure~\ref{fig:phase} shows the detailed phase diagram, with the extrapolation to large $\kappa_2$ defining the continuum limit.  The diagram is consistent with a value of $\beta\approx -1$ in the continuum limit, though it is difficult from this data alone to pin down the precise value for the critical $\beta$ in that limit.  This is close to the value of $\beta\approx-0.97$ that is obtained from the tuning for the subtraction condition.  Second, we can compare the running of $G\Lambda$ to the behavior of the scalar curvature, which acts as an effective cosmological constant at scales of the order of the cutoff, to which we turn in the next subsection.

\subsection{The scalar curvature}

As a cross-check of these ideas, we look at the expectation value of the lattice curvature, which we identify with an effective cosmological constant,
\bea\label{eq:cc_eff}   a^2\langle R \rangle \equiv \frac{a^2\langle\int d^4x \sqrt{g}R\rangle}{\langle\int d^4x \sqrt{g}\rangle} \propto a^2\Lambda_{\rm eff}.
\eea
We have explicitly included a power of the lattice spacing $a$, since these are dimensionless lattice quantities.  The Regge curvature on the lattice is determined from the deficit angle about a 
$d-2$ simplex, and corresponds to the curvature evaluated at a scale of the order of the lattice cutoff.  The identification between the Regge curvature and the effective cosmological constant is made using the tree-level relation $\langle R\rangle=4\Lambda$, which we expect to hold approximately if the scale at which the coupling is evaluated is chosen to minimize radiative corrections.  We could then convert the effective cosmological constant from lattice units to physical units if we knew the value of the Planck mass (or equivalently, Newton's constant) in lattice units.  We would then multiply the right side of Eq.~(\ref{eq:cc_eff}) by $G_N/a^2$, where the subscript $N$ denotes the fact that this is the physical, renormalized Newton's coupling, not the bare one.  There is a caveat here in that we do not know $G_N$ in lattice units, but we do know the relative lattice spacing from our studies of the return probability, so that we can construct $\frac{a^2}{a_{\rm fid}^2}\langle R \rangle$ in physical units, where $a_{\rm fid}$ is our fiducial lattice spacing at $\beta=0$.  This should match $G_N\langle R\rangle$ up to discretization effects and an overall normalization.

\begin{figure}
\begin{center}
\includegraphics[scale=.55]{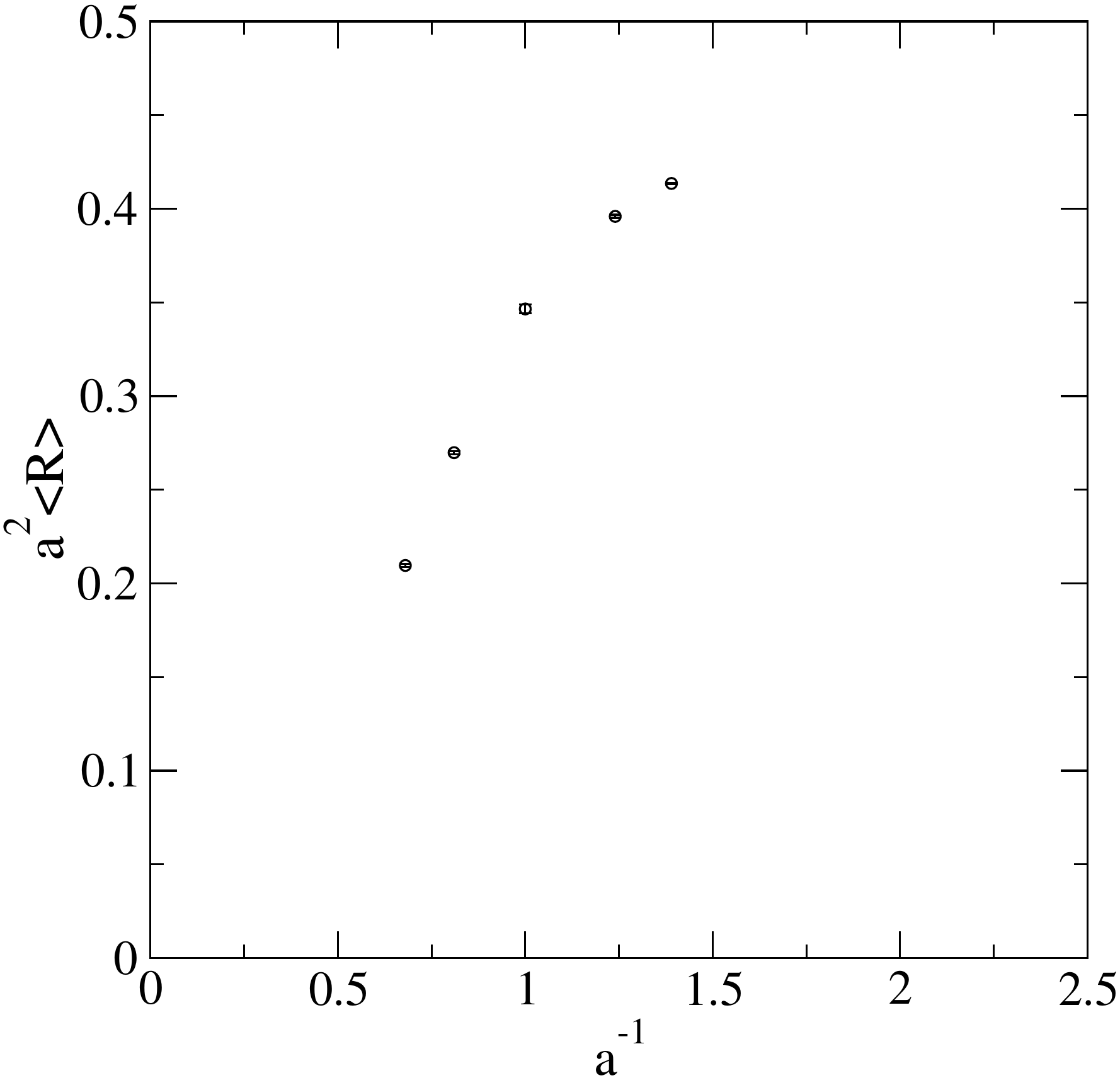}
\vspace{-3mm}
\caption{$a^2\langle R\rangle$ as a function of inverse lattice spacing.  \label{fig:a2R}}
\end{center}
\end{figure}

\begin{figure}
\begin{center}
\includegraphics[scale=.55]{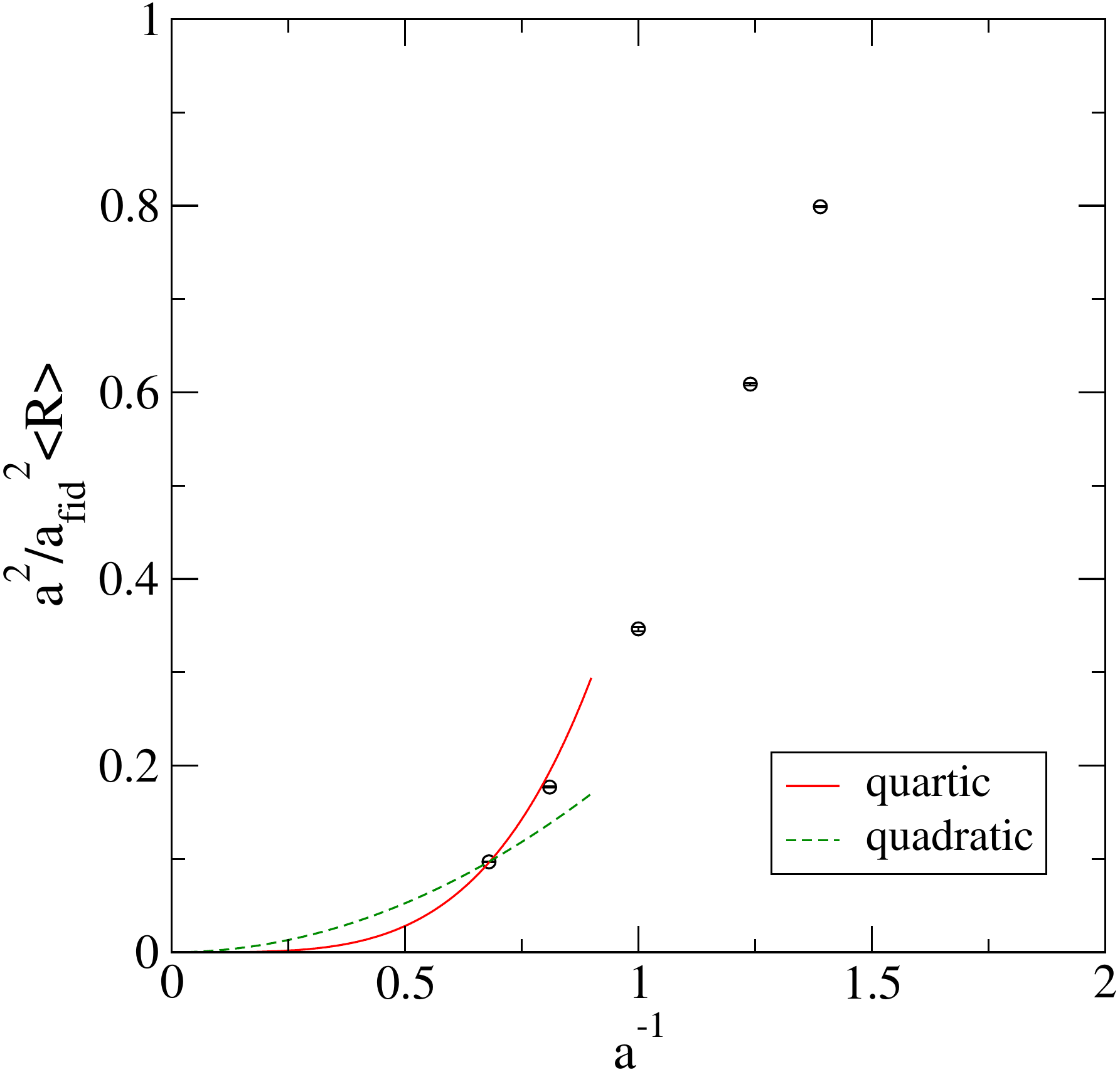}
\vspace{-3mm}
\caption{$\frac{a^2}{a_{\rm fid}^2} \langle R\rangle$ as a function of inverse lattice spacing.  The solid line is a quartic constrained to pass through the leftmost data point.  The dashed line is a quadratic constrained to pass through the same point.  \label{fig:GR}}
\end{center}
\end{figure}

Figure~\ref{fig:a2R} shows the dimensionless curvature scalar as a function of inverse lattice spacing.  There is some indication that this quantity is approaching a constant in the vicinity of the fixed point at large inverse lattice spacing; it drops to small values in the infrared.  If the continuum limit is approached by taking $\kappa_2$ large, then $N_2/N_4$ approaches its kinematic limit, which forces the value of $a^2\langle R\rangle$ to a constant around 0.43 in the fixed-point regime.  Figure~\ref{fig:GR} shows $a^2/a_{\rm fid}^2 \langle R\rangle$, the effective curvature converted into physical units, plotted as a function of inverse lattice spacing.  In this case, the effective curvature drops even more rapidly in the infrared.  It continues to rise deep in the ultraviolet, suggesting that the effective curvature defined at the cutoff diverges as the continuum limit is taken.  This is not by itself in conflict with asymptotic safety, but is analogous to the divergence of the kinetic energy measured at ever decreasing time intervals in nonrelativistic quantum mechanics.  We can compare the behavior of the effective curvature with the expectation from renormalization group calculations \cite{Reuter:2001ag}.  Deep in the infrared, the running of the cosmological constant is governed by the Gaussian (free) fixed point.  This implies that the effective curvature is described by the perturbative running, which is proportional to the cutoff scale to the fourth power.  Since the cosmological constant of the lattice simulations in the deep infrared is fixed by the finite volume cutoff, it is quite small compared to the effective curvature at the cutoff and can be neglected here.  Thus the running governed by the Gaussian fixed point is 
\bea\label{eq:GL_run}   \frac{a^2}{a_{\rm fid}^2} \langle R\rangle =  \frac{c}{a^4},
\eea
with $c$ some constant.  When we constrain this quartic to pass through the first data point of $\frac{a^2}{a_{\rm fid}^2} \langle R\rangle$, we obtain the solid curve shown in Fig.~\ref{fig:GR}.  The agreement between this curve and the data is only qualitative, and a more definitive test would require a comparison deeper in the infrared.  Still, the quartic curve shows better agreement with the second data point from the left than the corresponding curve with a quadratic dependence on $\mu$, shown as the dashed curve in Fig.~\ref{fig:GR}.

We expect that the bare $G\Lambda_{\rm sub}$ should have a behavior similar to that of the effective curvature discussed above as they run deep into the infrared away from the fixed point, though they should not be identical, since the latter is an effective coupling inferred from a lattice observable and the former is a bare coupling in a particular lattice regulator scheme.  Figure~\ref{fig:GL_sub} shows the running of $G\Lambda_{\rm sub}$ as a function of inverse lattice spacing.  Just as in the case of the scalar curvature, we see a steep drop in this parameter under flow to the infrared.  Although the fact that the coupling approaches zero in the infrared is a consequence of our subtraction condition, and in that sense is put in by hand, the precise form of the running is not.  We see from the plot that the running of $G\Lambda_{\rm sub}$ is reasonably consistent with a quartic running.  A quadratic running leads to substantially worse qualitative agreement.  Note that the quartic running fits with our expectations from effective field theory.  The fact that the scalar curvature, which may be interpreted as an effective cosmological constant, shows a similar behavior in the infrared bolsters our argument that the subtraction procedure for $G\Lambda$ is both necessary and leads to sensible physical results compatible with other properties of our geometries.  

\subsection{The dimension of the UV critical surface}

We have argued that although there are three couplings that must be adjusted in the lattice theory, one of them is redundant and two of them are relevant.  However, only one coupling is relevant in the theory when the breaking of a symmetry by the regulator is compensated by a fine-tuning and subtraction.  In the subtracted theory, the coupling $\beta$ associated with the nonuniform measure term is a fixed constant, around $-1$.  The evidence for this is that the subtraction procedure successfully resolves two questions:  One, why does the exponent in the measure term run, when one would not expect such a parameter to run with renormalization scale?  Two, why does the running of the unsubtracted $G\Lambda$ not appear to be consistent with the physics of the emergent geometries of the simulations, namely geometries with small curvature radius and large spatial extent?  In the subtracted theory these problems are resolved.  The subtraction is well motivated, since although we do not expect $\beta$ to run, it appears that we need to tune it in order to restore the coordinate invariance as the continuum limit is approached.  This tuning becomes necessary because the regulator breaks the symmetry that would be needed to absorb a rescaling of the bare Newton's constant into a redefinition of the fields.  We suggest that the subtraction ``undoes" the adjustment of $\beta$ needed for the fine-tuning, fixing it to a value close to its continuum value. 

\begin{figure}
\begin{center}
\includegraphics[scale=.65]{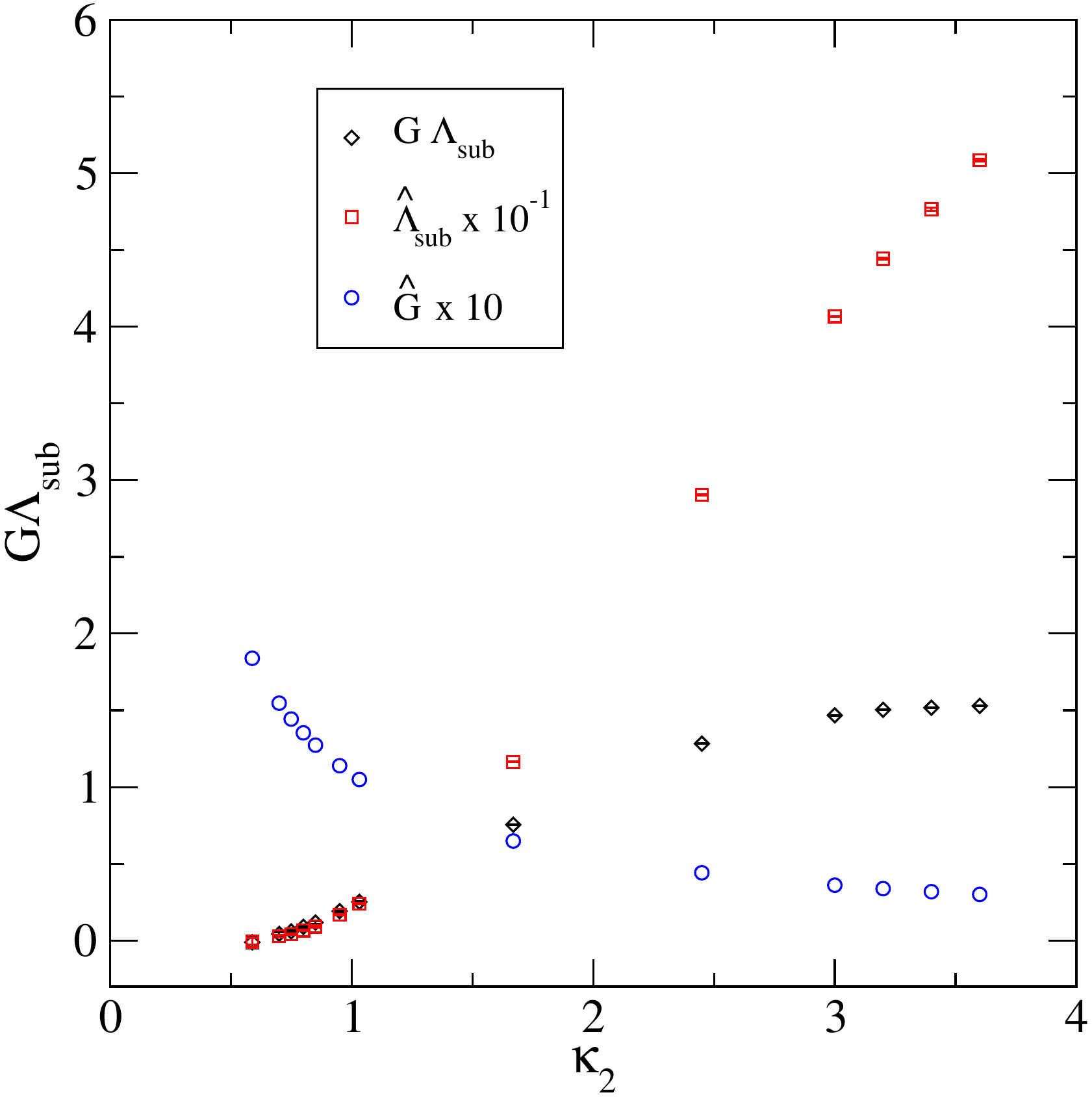}
\vspace{-3mm}
\caption{$\hat{G}\times 10$, $\hat{\Lambda}_{\rm sub}\times 10^{-1}$, and $G\Lambda_{\rm sub}$ as a function of $\kappa_2$.  Increasing $\kappa_2$ corresponds to finer lattice spacings.  \label{fig:GLambda}}
\end{center}
\end{figure}

Crucial to the argument that there is only one relevant coupling in the theory is that the bare cosmological constant and the bare Newton's constant are not separately relevant couplings.  Figure~\ref{fig:GLambda} shows the running of $G\Lambda_{\rm sub}$ as a function of the bare lattice parameter $\kappa_2$.  In this figure, $\kappa_2$ is a stand-in for the lattice scale.  Large $\kappa_2$ corresponds to fine lattice spacings, but it is difficult to make the lattices large enough at extremely fine lattice spacings that we can use them to reliably determine the relative lattice spacing.  Nonetheless, as we go to large values of $\kappa_2$, the volume dependence of the tuned bare cosmological constant decreases, so that we are able to extract the bare couplings along this path with some confidence.  The bare $G\Lambda_{\rm sub}$ shows the behavior that we expect for a relevant dimensionless coupling in the asymptotic safety scenario.  Figure~\ref{fig:GLambda} shows that at very fine lattice spacings, this coupling approaches a constant of around 1.5 in the deep ultraviolet, and it flows to small values in the infrared.  Figure~\ref{fig:GLambda} also shows the running of the couplings $\hat{G}$ and $\hat{\Lambda}_{\rm sub}$ as a function of $\kappa_2$.  Here we see that in the regime of very fine lattice spacings where $G\Lambda_{\rm sub}$ is approaching a constant that $\hat{\Lambda}_{\rm sub}$ is still increasing roughly linearly, and $\hat{G}$ is decreasing towards zero.  We have argued that this does not invalidate the asymptotic safety scenario, as long as $\hat{\Lambda}_{\rm sub}$ is not, by itself, a relevant parameter.  We also do not expect $\hat{G}$ to run to zero, since this would imply that the Einstein-Hilbert theory is perturbatively  renormalizable and asymptotically free.  We know from explicit perturbative calculations that this is not the case \cite{Goroff:1985th}.  These potential problems with the individual couplings are resolved if our model is an asymptotically safe theory with only one relevant coupling, $G\Lambda$.  Although it is possible that $\hat{G}$ and $\hat{\Lambda}_{\rm sub}$ eventually also approach constants for sufficiently large $\kappa_2$, it would be strange for some constants to approach their fixed point values long after other couplings and observables suggest that the fixed-point regime has been reached.  According to the argument in Sec.~\ref{sec:continuum}, the fixed point should be approached as $\kappa_2\to\infty$ if our picture involving symmetry breaking and fine-tuning is correct, in which case $\hat{\Lambda}_{\rm sub}$ diverges and $\hat{G}$ goes to zero as $a\to 0$, but with $\hat{G}\hat{\Lambda}_{\rm sub}$ approaching a constant as required.

\section{Connection of this work to other approaches}\label{sect:others}

Over the past several years, evidence has been found that gravity in four dimensions may be asymptotically safe.  This evidence comes mainly from continuum functional renormalization group methods \cite{Reuter:2001ag,Lauscher:2001ya,Litim:2003vp,Codello:2007bd,Codello:2008vh,Benedetti:2009rx} and lattice calculations \cite{Ambjorn:2005qt,Ambjorn:1998xu,Ambjorn:2008wc}.  It has been found that in a variety of truncations of the renormalization group equations that the dimension of the ultraviolet critical surface is finite.  Different types of truncations using more than three independent couplings indicate that the critical surface is three dimensional, even when many couplings are included in the truncation \cite{Codello:2007bd,Codello:2008vh,Benedetti:2009rx,Falls:2014tra}.  However, it is difficult to systematically assess the reliability of the results obtained by solving the renormalization group equations once they have been truncated, and a method that allows calculations with controlled systematic errors is desirable.  This is, in principle, possible with lattice methods.

There has been progress on the lattice using the CDT formulation.  As discussed in Sec.~\ref{sect:intro}, a number of interesting results have been obtained from numerical simulations using this approach, including the emergence of a four-dimensional semiclassical phase resembling Euclidean de Sitter space \cite{Ambjorn:2004qm, Ambjorn:2005qt}, and a spectral dimension that runs as a function of the distance scale probed \cite{Ambjorn:2005db}.  A second-order critical line has been found in the phase diagram of the theory \cite{Ambjorn:2011cg}, though subsequent work has shown that this line does not border the physical de Sitter phase, except maybe at a point, because of the discovery of a new phase in the CDT phase diagram called the ``bifurcation" phase \cite{Coumbe:2015oaa}.  Even so, there may exist a second-order line between the physical phase and the new bifurcation phase \cite{Coumbe:2015oaa, Ambjorn:2017tnl} where a continuum limit might be defined.  Even with these significant achievements of the CDT formulation, one might worry that the preferred foliation of the CDT approach is at odds with general covariance and may not reproduce general relativity in the infrared.  However, the foliation may amount to a partial gauge fixing.  Recent work in three dimensions suggests that it is possible to relax the restriction of a fixed foliation in CDT using an approach called locally causal dynamical triangulations \cite{Jordan:2013awa}, but a different treatment of spacelike and timelike links is still necessary.  An extension of the local CDT approach to four dimensions appears challenging.

It might be that CDT can recover the space-time symmetric quantum theory of general relativity, possibly with some fine-tuning.  The results of this work provide evidence that the EDT formulation can reproduce the correct classical limit.  It is then interesting to ask whether the two approaches are in the same universality class because they have complementary features.  The EDT approach is explicitly space-time symmetric, so the correctness of this formulation would lend support to the notion that the fixed foliation of CDT amounts to a partial gauge fixing, rather than a truncation of physical degrees of freedom.  On the other hand, it has been shown that the CDT approach is unitary at finite lattice spacing, since the existence of a reflection positive transfer matrix in the CDT approach ensures unitarity \cite{Ambjorn:2000dv}.  If it turns out that EDT is in the same universality class as CDT, then the question of unitarity would be decided in EDT as well.  This would be desirable, since it is more difficult to address the question of unitarity directly in the EDT approach.

We now turn to a discussion of the results of this work for the spectral dimension and for the number of relevant couplings, and we compare our results with the results from other methods.  

\subsection{Results for the spectral dimension}

Researchers using other approaches have considered the spectral dimension at short-distance scales.  Reference~\cite{Lauscher:2005qz} argues in the context of the renormalization group that the spectral dimension at short distances in any asymptotically safe theory of gravity must be equal to two exactly and that this result is independent of any truncation.  Our result is close to 3/2, with a value of 2 disfavored.  What are we to make of the emerging discrepancy between the renormalization group result and our result?  On one hand, there are still systematics in our determination of this quantity that need to be quantified, especially the errors associated with the extrapolation to $\sigma=0$, so that our error is surely underestimated.  For example, it is difficult to distinguish between a value for $D_S(0)=3/2$ and other scenarios if the behavior of $D_S(\sigma)$ is not monotonic in $\sigma$.  In Ref.~\cite{Reuter:2011ah}, Reuter and Saueressig show that in the simplest truncation of the renormalization group equations, the spectral dimension starts at 4 at very long distances, approaches a plateau of 4/3 at shorter distance scales, and then eventually jumps up to 2 at the sub-Planckian distance scales in the vicinity of the ultraviolet fixed point.  Other behavior is seen with additional terms included in the truncation, including the possibility of a plateau at 3/2 before approaching a value of 2 near the fixed point \cite{Rechenberger:2012pm}.  This nonmonotonic behavior is not evident in our data, but the value that we obtain for $D_S(0)$ depends on the assumptions that we make for the extrapolation to short distances, and we will most likely need much finer lattices in addition to a good estimate of the absolute lattice spacing to settle its value in the EDT approach.

On the other hand, the value of the spectral dimension obtained from the renormalization group arguments is tied to the scaling of the couplings in the fixed point regime, where, in particular, the cosmological constant and Newton's constant are treated as separately relevant couplings.  There is evidence that this is not the case, as we discuss in Sec.~\ref{sec:running}.  If the result of $D_S(0)=3/2$ holds up to further scrutiny, it could resolve the tension between the holographic entropy scaling expected in a theory containing black holes and a renormalizable theory with conformal asymptotic scaling in the ultraviolet.  A recent determination of the spectral dimension within CDT also found that the value at short distances was compatible with 3/2 and that 2 was strongly disfavored \cite{Coumbe:2014noa}.  This provides further evidence that not only can holography and asymptotic safety be reconciled, but that CDT and EDT may be in the same universality class,  given the nontrivial agreement between CDT and EDT on this nonperturbative quantity.  The agreement also bolsters the case that EDT is unitary, since it has already been established that CDT is unitary \cite{Ambjorn:2001cv}, as discussed above.

\subsection{The number of relevant parameters}

We have argued in this paper that the only relevant coupling in our theory is $G\Lambda$, and that this implies that the theory is maximally predictive.  Key to this argument is the fact that the breaking of general coordinate invariance obscures this possibility, requiring the introduction of a new parameter that must be fine-tuned to take the continuum limit.  A subtraction must be performed in order to recover the continuum running.  We turn now to other formulations of asymptotic safety, and compare their results for the number of relevant parameters with those of this work.

It has been argued in the context of the renormalization group equations that the fact that $G$  is dimensionful implies that $G$ and $\Lambda$ must be treated as separately relevant parameters and that the demonstration in Sec.~\ref{sec:running} that only the combination $G\Lambda$ is relevant does not apply \cite{Percacci:2004sb}.  This is shown explicitly in truncations of the renormalization group equations, and a demonstration that it is also true in the full theory without truncation is given \cite{Lauscher:2001rz, Lauscher:2005qz} \footnote{See however, Ref.~\cite{Rosten:2011mf}, where it is argued in the context of the renormalization group that for the truncations that have been used so far, this result has been implicitly built in as an assumption and that more sophisticated truncations might lead to a different conclusion.}.  In Ref.~\cite{Hamber:2013rb}, Hamber and Toriumi claim that this argument is incorrect, and that the breaking of diffeomorphism invariance by the regulator in the renormalization group approach could lead to spurious renormalization group running for $G$ and $\Lambda$ separately.  These authors recall the case of quantum gravity in $2+\epsilon$ dimensions in order to illustrate their point.  We briefly review this result here.

The theory in $2+\epsilon$ dimensions \cite{Weinberg:1980gg}, with $\epsilon$ small, can be solved within a joint expansion in $\epsilon$ and $G$ using continuum perturbation theory.  It has been known for some time that gravity in $2+\epsilon$ dimensions possesses a nontrivial UV fixed point, realizing the asymptotic safety scenario \cite{Kawai:1989yh, Kawai:1993mb}.  As such, it is a useful toy model, but it cannot by itself decide the question of whether asymptotic safety holds in four dimensions because $\epsilon =2$ is not a small perturbation about $d=2$.  However, the existence of an analytically tractable theory of gravity in lower dimensions that exhibits asymptotic safety provides a valuable test bed for understanding aspects of the theory in four dimensions.  In this toy model, the renormalization group running of $G$ and $\Lambda$ separately is found to be gauge dependent \cite{Kawai:1989yh, Hamber:2013rb}.  This is unlike the running of, e.g. the strong coupling $\alpha_s$ or the quark masses in perturbative QCD, where the $\beta$ function and quark-mass anomalous dimension are gauge independent.  The running of a coupling in a gauge theory, in order for that coupling to not be redundant, should not depend on the gauge choice.  Field-strength renormalization constants are redundant couplings, and those appearing in gauge theories are gauge dependent.  In the case of $2+\epsilon$ gravity one can use the freedom to perform a field redefinition to rescale the metric, and it can be shown that the running of the combination $G \Lambda^{\frac{d-2}{2}}$ is gauge invariant as expected, even though the running of $G$ and $\Lambda$ separately is not.  The coupling $G \Lambda^{\frac{d-2}{2}}$ is not redundant, and given its renormalization group running, is a relevant coupling of the ultraviolet fixed point.  The calculations in the $2+\epsilon$ theory use dimensional regularization, which preserves the continuum diffeomorphism invariance, an essential ingredient in the above argument.   This supports the notion that $G\Lambda^{\frac{d-2}{2}}$ (or $G\Lambda$ in the special case of four dimensions) is a relevant coupling, while $G$ and $\Lambda$ separately are not.\footnote{Note, however, that the recent work of Falls \cite{Falls:2017cze} argues that this gauge dependence is an artifact of perturbation theory and that a re-summation eliminates it.}

\section{Conclusions and Outlook}\label{sec:concl}

In this work we have presented evidence that a quantum field theory of general relativity (without matter) is renormalizable when formulated nonperturbatively, and that the cosmological constant in Planck units is the only relevant parameter in the theory.  A summary of the results that we have presented in support of these conclusions follows below.

We have presented evidence that lattice quantum gravity, as defined by Euclidean dynamical triangulations, requires a fine-tuning in order to restore a target symmetry that is broken by the lattice regulator.  This tuning is necessary to take a continuum limit and to obtain the correct semiclassical limit.  We identify this target symmetry as the continuum general coordinate invariance, which is needed to ensure that only physical states are included in the path-integral sum.  A form for the local measure that is motivated by continuum approaches is of the type $\prod_x \sqrt{g}^\beta$.  Studies in the continuum suggest that the parameter $\beta$ is fixed by BRST symmetry, which implements the continuum general coordinate invariance in the quantum theory, ensuring the absence of gauge anomalies \cite{Fujikawa:1983im}.  However, we argue that this symmetry is broken at finite lattice spacing by EDT.  In that case, it is possible that $\beta$ must be fine-tuned as a function of lattice spacing.  Only in the continuum limit would it then be possible to recover general coordinate invariance.  

Experience with other lattice formulations where a symmetry is broken suggests that we must fine-tune the couplings to simulate close to a first-order transition line in order to restore a symmetry.  The phase transition in our model softens as we follow the transition line to larger $\kappa_2$, suggesting a continuous phase transition at the end of the first-order line (possibly at $\kappa_2\to \infty$).  Our calculations of the relative lattice spacing using a diffusion process show that the lattice spacing decreases as $\kappa_2$ increases.  There appears to be no obstruction to taking the continuum limit as we run parallel to the first-order line towards larger $\kappa_2$.  This evidence for a continuum limit lends nontrivial support to Weinberg's asymptotic safety scenario for quantum gravity.  

Another crucial test that the model must pass if it is to realize the asymptotic safety scenario is whether it can reproduce the correct semiclassical physics.  As we argue in this work, one should focus on the behavior of geometries close to the first-order critical line.  Our simulations close to this line produce geometries with global Hausdorff dimension compatible with four, as required to recover the correct classical limit.  These geometries also resemble Euclidean de Sitter space, which is the maximally symmetric classical solution of Einstein's equations for a universe with a positive cosmological constant.  The agreement with de Sitter gets better as the continuum limit is approached, and the baby-universe-like structures that branch off of the mother universe and dominate at coarser couplings are a much smaller fraction of the mother universe at finer lattice spacings.  This strongly suggests that the baby universes are a cutoff effect, and that if we are to recover the correct classical limit we are required to take the continuum limit as well as the large volume limit.  Discretization effects can effect long distance quantities when the lattice regulator breaks a symmetry, and recovering the correct long distance behavior can require first taking the continuum limit; such effects are familiar from lattice gauge theory.

We have studied the return probability and the spectral dimension, which are defined via a diffusion process on our random geometries.  There is evidence that in the large-volume, continuum limit the spectral dimension becomes four at long distances, as it must in order to reproduce the classical theory.  At short distances, the spectral dimension is close to 3/2.  It has been argued that the value 3/2 is special, because it could resolve the tension between a renormalizable theory of gravity and the holographic entropy scaling of black holes \cite{Laiho:2011ya}.  Our new result of 3/2 at short distances matches that of a recent calculation using CDT \cite{Coumbe:2014noa}, providing evidence that the two formulations are in the same universality class.  The agreement lends support to the idea that the foliation of CDT amounts to a gauge fixing, rather than an exclusion of physical degrees of freedom.  If EDT is in the same universality class as CDT, then we could draw the important conclusion that EDT is unitary, since it has already been shown that CDT is unitary \cite{Ambjorn:2000dv}.  

Finally, we have examined the number of relevant parameters in the EDT formulation.  We have reviewed arguments that the cosmological constant and Newton's constant are not separately relevant couplings, and we presented a new one of our own.  There is evidence that, if the continuum limit exists, the bare dimensionless cosmological constant does not approach a constant, though the (dimensionless) combination $G\Lambda$ does approach a constant.  If only the latter is a relevant coupling then there is no contradiction with asymptotic safety, since in that scenario only relevant dimensionless couplings must approach a constant fixed-point value.  The fact that the bare (subtracted) dimensionless cosmological constant appears to diverge in this limit in our scheme is not a problem if it is a redundant coupling.  We have also argued that the tuning required for the parameter $\beta$ associated with the measure term is due to the breaking of general coordinate invariance by the regulator, and that $\beta$ would not be a relevant coupling if the symmetry were preserved.  This is due to the following:  if a symmetry that is broken by the regulator is exact in the quantum theory, then a reduction of the number of relevant parameters can be expected if one adopts a symmetry preserving regulator.  In the absence of such a symmetry preserving regulator, a fine-tuning and subtraction are required, as we argue in this work.

We point out that the numerical evidence for both a semiclassical limit and a continuum limit presented here could stand apart from the theoretical picture that we have proposed to understand it, and that these results by themselves provide support for the asymptotic safety scenario for quantum gravity.  Even so, we have attempted to explain these results with, in our view, the most economical picture possible, from the need to fine-tune the parameters in the simulations, to the fact that the regulator does appear to break the key symmetry (general coordinate invariance) of general relativity, to the interpretation of the measure term and the need for a subtraction to make sense of the running of the bare couplings.  Indeed, much of the theoretical picture preceded and motivated the numerical simulations performed here.  Still, other approaches to asymptotically safe gravity do not find this result; e.g. most truncations of the renormalization group equations appear to require at least three relevant couplings \cite{Codello:2008vh, Benedetti:2009rx, Falls:2014tra}.  It is possible that other operators, in particular higher curvature operators, might be accompanied by relevant couplings in the EDT formulation.  However, if one can show the existence of a continuum limit that does not require the introduction of such terms, this would argue against their relevance, since it is unlikely that their fixed-point values would be zero in the absence of a symmetry ensuring that is the case.  Although we regard this as unlikely, it remains a possibility and cannot be ruled out, particularly if our measure term is interpreted as an operator involving higher powers of the scalar curvature.  We consider it an important outstanding issue to resolve the differences implied by our calculations for the number of relevant couplings and those of other approaches.

To reiterate, we have provided evidence that properly interpreted, the EDT formulation provides a UV complete theory of quantum gravity with only one relevant parameter.  A one-dimensional ultraviolet critical surface is the most desirable outcome because it implies that the theory is maximally predictive.  Consider, for example, the cosmological constant.  In the usual effective field theory picture, the bare cosmological constant is additively renormalized by physics up to and including the Planck scale.  The bare cosmological constant therefore has to be tuned very precisely in order to cancel this additive renormalization so that the long-distance renormalized cosmological constant is very close to zero.  This is the cosmological constant fine-tuning problem.  In our approach, however, we argue that the tuning that needs to be done is due solely to the breaking of the continuum diffeomorphism invariance by the regulator.  In the subtracted theory there is only one relevant coupling, and with only one relevant coupling, there are no adjustable parameters.  Thus, if this picture is correct, all observables should be predictions of the theory, including the renormalized cosmological constant at long distances, once the value of the Planck scale is established to convert lattice units into dimensionful units.  Although this may be the case, we have not explicitly exhibited a mechanism for the screening of the cosmological constant.  Perhaps an idea along the lines of Refs.~\cite{Polyakov:2000fk, Jackiw:2005yc} invoking conformal fluctuations of the metric could provide the underlying mechanism.

It should be noted that for renormalization group flow in the presence of an explicit infrared (IR) cutoff, the IR cutoff itself is a new parameter, not known {\it a priori}.  In our gravity calculations, it fixes the overall size of the simulated universes, and thus the value of the cosmological constant in the IR.  Although the theory appears to have a one-dimensional UV critical surface, with the renormalized trajectory starting on a UV fixed point and flowing into the infrared, an explicit IR cutoff on the renormalized trajectory would require a new input parameter to specify the IR cutoff scale.  In practice, one might be able to associate this IR cutoff with the Hubble scale, which determines the size of the observable universe.  Since the Hubble scale has been different at different epochs of the universe's history, the possibility of making predictions still exists.  Whatever the predictions of pure gravity, they are likely to be modified by the matter sector, which comes with additional relevant parameters like gauge couplings, Higgs-Yukawa couplings, etc.  It may be possible, following this framework, to determine the value of the cosmological constant from first principles, once the theory is adapted to include the matter sector.  This assumes that the theory remains asymptotically safe, with results qualitatively similar to that of the pure gravity theory, after the addition of the matter sector.

There are, of course, many questions that remain.  Our determination of the Hausdorff dimension and the spectral dimension are provisional, and would benefit from larger volumes at finer lattice spacings.  Improved algorithms and actions would be useful to meet this goal.  Finer lattice spacings and larger volumes are also needed to test whether the spectral dimension at short distances is $3/2$, or some other value, perhaps with nonmonotonic dependence on the length of the diffusion path $\sigma$.  A determination of the absolute lattice spacing would be useful, as it is needed for a reliable estimation of continuum extrapolation errors for this and other quantities.  A crucial test would be to add matter content to the lattice studies, to test within the EDT approach whether the asymptotic safety scenario holds in the presence of matter.  It is important to continue to refine the various different approaches to asymptotically safe gravity, to further test them against the picture presented here.  The hope is that this will help us make progress towards our ultimate goal, to make unambiguous predictions for quantum gravitational effects in observational cosmology.

\acknowledgments

We thank J. Ambjorn, C. Bernard, S. Carlip, S. Catterall, B. Dittrich, K. Falls, J. Hubisz, A. Kronfeld, R. Loll, D. Miller, and C. White for useful discussions, and C. Bernard and A. Kronfeld for comments on the manuscript.
Computations for this work were carried out in part on facilities of
the USQCD Collaboration, which are funded by the Office of Science of
the U.S. Department of Energy, on the Darwin Supercomputer as part of STFC's DiRAC facility jointly funded by STFC, BIS, and the Universities of Cambridge and Glasgow, and on the Syracuse University condor cluster, which is funded by the National Science Foundation under Grant No.~ACI-1341006.
This work was supported in part by the National Science Foundation under Grant No.~PHY14-17805 (S.B.,D.D.,J.L.,J.N.), by the National Science Centre Poland under Grant No. DEC-2012/06/A/ST2/00389 (D.C.), by the ERC-Advance Grant No. 291092, ``Exploring the Quantum Universe" (D.C.), and by the U.S. Department of Energy, Office of Science, Office of High Energy Physics, under Award No. $\textrm{DE-SC0009998}$ (S.B., J.L.). 

\appendix*
\section{Parallel rejection}

Parallel rejection is a simple algorithm that turns the execution of a series of Markov-chain Monte Carlo (MCMC) updates into parallel updates, which is helpful when the acceptance rate is low. The algorithm is based on the idea that most of the MCMC moves will not be accepted, so the attempted moves can be pre-calculated in parallel. Let $M$ be the size of the series to simulate, which can be evenly distributed to $n$ parallel streams. The $i$th stream is responsible for calculating the attempted moves of $i$, $n+i$, $2n+i$,...$(r-1)n+i$, where $r=M/n$.  Whenever a move is accepted, all streams stop. Any precalculated attempted moves after the accepted move (in the original series sequence) will be voided. The simulation then repeats from the accepted update.

It was verified for our implementation of lattice gravity that the parallel rejection code gives the exact same result as the scalar code, configuration by configuration.  Since the Metropolis acceptance rate is very low in our application ($10^{-3}$ or lower), this is a problem that can benefit significantly from parallel rejection.  Scaling on machines with up to eight cores per socket is good, leading to a factor of 6.3 speed-up in program execution.  We see similar scaling in a single Intel Xeon Phi card with 61 cores.  The efficient implementation of this algorithm requires that the parallelization be done on a shared memory architecture, using for example OPENMP.

A pseudocode using OPENMP is given below as an example:
\bigskip
\bigskip

	\begin{lstlisting}
N <- attempted moves per sweep, divided into blocks
M <- size of the blocks, distributed by threads
nCol <- number of threads
nRow = M/ncol <- tasks per thread per block
nSweeps <- number of sweeps to simulate

globalStopLoc <- shared variable to signal quitting a loop
globalSweepStop <- signal of quitting a loop due to a sweep
globalUpdateStop <- signal of quitting a loop due to an update
globalMoveParams <- parameters to uniquely determine a move

parallel start
load lattice
id <- thread id
s = 0
lastStop = 0
barrier

while s < nSweeps:
	sweepStop = false
	updateStop = false
	critical
		globalStopLoc = 0
		globalSweepStop = false
		globalUpdateStop = false
		updatedLoc = M

	for i=0 to nRow-1:
		loc = id + nCol*i
		if lastStop+loc == N-1:
			sweepStop = true
	
		result, moveParams = attempted_move(loc)
		if result == 'accepted':
			updateStop = true
	
		if sweepStop or updateStop:
			atomic
				globalStopLoc += (loc+1) 
			break

		if globalStopLoc != 0 and loc > globalStopLoc - 1:
			break
	barrier

	if not sweepStop and not updateStop:
		loc = M
	if loc < updatedLoc:
		critical
			if loc < updatedLoc:
				globalSweepStop = sweepStop
				globalUpdateStop = updateStop
				updatedLoc = loc
				globalMoveParams = moveParams
	barrier

	if updatedLoc == M:
		lastStop += M
	else:
		if globalUpdateStop:
			updateLattice(globalMoveParams )
			lastStop += updatedLoc + 1
	
		if globalSweepStop:
			master
				lattice -> outToFile
			s += 1
			lastStop = 0
	barrier
\end{lstlisting}

	\clearpage

\bibliography{SuperBib}

\begin{thebibliography}{84}
\expandafter\ifx\csname natexlab\endcsname\relax\def\natexlab#1{#1}\fi
\expandafter\ifx\csname bibnamefont\endcsname\relax
  \def\bibnamefont#1{#1}\fi
\expandafter\ifx\csname bibfnamefont\endcsname\relax
  \def\bibfnamefont#1{#1}\fi
\expandafter\ifx\csname citenamefont\endcsname\relax
  \def\citenamefont#1{#1}\fi
\expandafter\ifx\csname url\endcsname\relax
  \def\url#1{\texttt{#1}}\fi
\expandafter\ifx\csname urlprefix\endcsname\relax\def\urlprefix{URL }\fi
\providecommand{\bibinfo}[2]{#2}
\providecommand{\eprint}[2][]{\url{#2}}

\bibitem[{\citenamefont{Goroff and Sagnotti}(1986)}]{Goroff:1985th}
\bibinfo{author}{\bibfnamefont{M.~H.} \bibnamefont{Goroff}} \bibnamefont{and}
  \bibinfo{author}{\bibfnamefont{A.}~\bibnamefont{Sagnotti}},
  \bibinfo{journal}{Nucl.Phys.} \textbf{\bibinfo{volume}{B266}},
  \bibinfo{pages}{709} (\bibinfo{year}{1986}).

\bibitem[{\citenamefont{'t~Hooft and Veltman}(1974)}]{'tHooft:1974bx}
\bibinfo{author}{\bibfnamefont{G.}~\bibnamefont{'t~Hooft}} \bibnamefont{and}
  \bibinfo{author}{\bibfnamefont{M.}~\bibnamefont{Veltman}},
  \bibinfo{journal}{Annales Poincare Phys.Theor.}
  \textbf{\bibinfo{volume}{A20}}, \bibinfo{pages}{69} (\bibinfo{year}{1974}).

\bibitem[{\citenamefont{Donoghue}(1997)}]{Donoghue:1997hx}
\bibinfo{author}{\bibfnamefont{J.}~\bibnamefont{Donoghue}}, pp.
  \bibinfo{pages}{26--39} (\bibinfo{year}{1997}), \eprint{gr-qc/9712070}.

\bibitem[{\citenamefont{Weinberg}(2000)}]{Weinberg:2000yb}
\bibinfo{author}{\bibfnamefont{S.}~\bibnamefont{Weinberg}}, in
  \emph{\bibinfo{booktitle}{{Sources and detection of dark matter and dark
  energy in the universe. Proceedings, 4th International Symposium, DM 2000,
  Marina del Rey, USA, February 23-25, 2000}}} (\bibinfo{year}{2000}), pp.
  \bibinfo{pages}{18--26}, \eprint{astro-ph/0005265},
  \urlprefix\url{http://www.slac.stanford.edu/spires/find/books/www?cl=QB461:I57:2000}.

\bibitem[{\citenamefont{Weinberg}(1980)}]{Weinberg:1980gg}
\bibinfo{author}{\bibfnamefont{S.}~\bibnamefont{Weinberg}}, in
  \emph{\bibinfo{booktitle}{General Relativity: An Einstein Centenary Survey}}
  (\bibinfo{year}{1980}), pp. \bibinfo{pages}{790--831}.

\bibitem[{\citenamefont{Olive et~al.}(2014)}]{Agashe:2014kda}
\bibinfo{author}{\bibfnamefont{K.~A.} \bibnamefont{Olive}} \bibnamefont{et~al.}
  (\bibinfo{collaboration}{Particle Data Group}), \bibinfo{journal}{Chin.
  Phys.} \textbf{\bibinfo{volume}{C38}}, \bibinfo{pages}{090001}
  (\bibinfo{year}{2014}).

\bibitem[{\citenamefont{Aoki et~al.}(2014)}]{Aoki:2013ldr}
\bibinfo{author}{\bibfnamefont{S.}~\bibnamefont{Aoki}} \bibnamefont{et~al.},
  \bibinfo{journal}{Eur. Phys. J.} \textbf{\bibinfo{volume}{C74}},
  \bibinfo{pages}{2890} (\bibinfo{year}{2014}), \eprint{1310.8555}.

\bibitem[{\citenamefont{Ambjorn and Jurkiewicz}(1992)}]{Ambjorn:1991pq}
\bibinfo{author}{\bibfnamefont{J.}~\bibnamefont{Ambjorn}} \bibnamefont{and}
  \bibinfo{author}{\bibfnamefont{J.}~\bibnamefont{Jurkiewicz}},
  \bibinfo{journal}{Phys.Lett.} \textbf{\bibinfo{volume}{B278}},
  \bibinfo{pages}{42} (\bibinfo{year}{1992}), \bibinfo{note}{revision of
  NBI-HE-91-47}.

\bibitem[{\citenamefont{Agishtein and Migdal}(1992)}]{Agishtein:1991cv}
\bibinfo{author}{\bibfnamefont{M.~E.} \bibnamefont{Agishtein}}
  \bibnamefont{and} \bibinfo{author}{\bibfnamefont{A.~A.}
  \bibnamefont{Migdal}}, \bibinfo{journal}{Mod. Phys. Lett.}
  \textbf{\bibinfo{volume}{A7}}, \bibinfo{pages}{1039} (\bibinfo{year}{1992}).

\bibitem[{\citenamefont{Ambjorn}(2002)}]{Ambjorn:2002uk}
\bibinfo{author}{\bibfnamefont{J.}~\bibnamefont{Ambjorn}},
  \bibinfo{journal}{Grav. Cosmol.} \textbf{\bibinfo{volume}{8}},
  \bibinfo{pages}{144} (\bibinfo{year}{2002}).

\bibitem[{\citenamefont{de~Bakker and Smit}(1995)}]{deBakker:1994zf}
\bibinfo{author}{\bibfnamefont{B.~V.} \bibnamefont{de~Bakker}}
  \bibnamefont{and} \bibinfo{author}{\bibfnamefont{J.}~\bibnamefont{Smit}},
  \bibinfo{journal}{Nucl. Phys.} \textbf{\bibinfo{volume}{B439}},
  \bibinfo{pages}{239} (\bibinfo{year}{1995}), \eprint{hep-lat/9407014}.

\bibitem[{\citenamefont{Ambjorn and Jurkiewicz}(1995)}]{Ambjorn:1995dj}
\bibinfo{author}{\bibfnamefont{J.}~\bibnamefont{Ambjorn}} \bibnamefont{and}
  \bibinfo{author}{\bibfnamefont{J.}~\bibnamefont{Jurkiewicz}},
  \bibinfo{journal}{Nucl. Phys.} \textbf{\bibinfo{volume}{B451}},
  \bibinfo{pages}{643} (\bibinfo{year}{1995}), \eprint{hep-th/9503006}.

\bibitem[{\citenamefont{Catterall et~al.}(1994)\citenamefont{Catterall, Kogut,
  and Renken}}]{Catterall:1994pg}
\bibinfo{author}{\bibfnamefont{S.}~\bibnamefont{Catterall}},
  \bibinfo{author}{\bibfnamefont{J.~B.} \bibnamefont{Kogut}}, \bibnamefont{and}
  \bibinfo{author}{\bibfnamefont{R.}~\bibnamefont{Renken}},
  \bibinfo{journal}{Phys. Lett.} \textbf{\bibinfo{volume}{B328}},
  \bibinfo{pages}{277} (\bibinfo{year}{1994}), \eprint{hep-lat/9401026}.

\bibitem[{\citenamefont{Egawa et~al.}(1997)\citenamefont{Egawa, Hotta,
  Izubuchi, Tsuda, and Yukawa}}]{Egawa:1996fu}
\bibinfo{author}{\bibfnamefont{H.~S.} \bibnamefont{Egawa}},
  \bibinfo{author}{\bibfnamefont{T.}~\bibnamefont{Hotta}},
  \bibinfo{author}{\bibfnamefont{T.}~\bibnamefont{Izubuchi}},
  \bibinfo{author}{\bibfnamefont{N.}~\bibnamefont{Tsuda}}, \bibnamefont{and}
  \bibinfo{author}{\bibfnamefont{T.}~\bibnamefont{Yukawa}},
  \bibinfo{journal}{Prog. Theor. Phys.} \textbf{\bibinfo{volume}{97}},
  \bibinfo{pages}{539} (\bibinfo{year}{1997}), \eprint{hep-lat/9611028}.

\bibitem[{\citenamefont{Bialas et~al.}(1996)\citenamefont{Bialas, Burda,
  Krzywicki, and Petersson}}]{Bialas:1996wu}
\bibinfo{author}{\bibfnamefont{P.}~\bibnamefont{Bialas}},
  \bibinfo{author}{\bibfnamefont{Z.}~\bibnamefont{Burda}},
  \bibinfo{author}{\bibfnamefont{A.}~\bibnamefont{Krzywicki}},
  \bibnamefont{and}
  \bibinfo{author}{\bibfnamefont{B.}~\bibnamefont{Petersson}},
  \bibinfo{journal}{Nucl.Phys.} \textbf{\bibinfo{volume}{B472}},
  \bibinfo{pages}{293} (\bibinfo{year}{1996}), \eprint{hep-lat/9601024}.

\bibitem[{\citenamefont{de~Bakker}(1996)}]{deBakker:1996zx}
\bibinfo{author}{\bibfnamefont{B.~V.} \bibnamefont{de~Bakker}},
  \bibinfo{journal}{Phys.Lett.} \textbf{\bibinfo{volume}{B389}},
  \bibinfo{pages}{238} (\bibinfo{year}{1996}), \eprint{hep-lat/9603024}.

\bibitem[{\citenamefont{Ambjorn et~al.}(2013)\citenamefont{Ambjorn, Glaser,
  Goerlich, and Jurkiewicz}}]{Ambjorn:2013eha}
\bibinfo{author}{\bibfnamefont{J.}~\bibnamefont{Ambjorn}},
  \bibinfo{author}{\bibfnamefont{L.}~\bibnamefont{Glaser}},
  \bibinfo{author}{\bibfnamefont{A.}~\bibnamefont{Goerlich}}, \bibnamefont{and}
  \bibinfo{author}{\bibfnamefont{J.}~\bibnamefont{Jurkiewicz}},
  \bibinfo{journal}{JHEP} \textbf{\bibinfo{volume}{1310}}, \bibinfo{pages}{100}
  (\bibinfo{year}{2013}), \eprint{1307.2270}.

\bibitem[{\citenamefont{Coumbe and Laiho}(2015)}]{Coumbe:2014nea}
\bibinfo{author}{\bibfnamefont{D.}~\bibnamefont{Coumbe}} \bibnamefont{and}
  \bibinfo{author}{\bibfnamefont{J.}~\bibnamefont{Laiho}},
  \bibinfo{journal}{JHEP} \textbf{\bibinfo{volume}{04}}, \bibinfo{pages}{028}
  (\bibinfo{year}{2015}), \eprint{1401.3299}.

\bibitem[{\citenamefont{Rindlisbacher and
  de~Forcrand}(2015)}]{Rindlisbacher:2015ewa}
\bibinfo{author}{\bibfnamefont{T.}~\bibnamefont{Rindlisbacher}}
  \bibnamefont{and}
  \bibinfo{author}{\bibfnamefont{P.}~\bibnamefont{de~Forcrand}},
  \bibinfo{journal}{JHEP} \textbf{\bibinfo{volume}{05}}, \bibinfo{pages}{138}
  (\bibinfo{year}{2015}), \eprint{1503.03706}.

\bibitem[{\citenamefont{Bruegmann and Marinari}(1993)}]{Bruegmann:1992jk}
\bibinfo{author}{\bibfnamefont{B.}~\bibnamefont{Bruegmann}} \bibnamefont{and}
  \bibinfo{author}{\bibfnamefont{E.}~\bibnamefont{Marinari}},
  \bibinfo{journal}{Phys.Rev.Lett.} \textbf{\bibinfo{volume}{70}},
  \bibinfo{pages}{1908} (\bibinfo{year}{1993}), \eprint{hep-lat/9210002}.

\bibitem[{\citenamefont{Bilke et~al.}(1998)\citenamefont{Bilke, Burda,
  Krzywicki, Petersson, Tabaczek et~al.}}]{Bilke:1998vj}
\bibinfo{author}{\bibfnamefont{S.}~\bibnamefont{Bilke}},
  \bibinfo{author}{\bibfnamefont{Z.}~\bibnamefont{Burda}},
  \bibinfo{author}{\bibfnamefont{A.}~\bibnamefont{Krzywicki}},
  \bibinfo{author}{\bibfnamefont{B.}~\bibnamefont{Petersson}},
  \bibinfo{author}{\bibfnamefont{J.}~\bibnamefont{Tabaczek}},
  \bibnamefont{et~al.}, \bibinfo{journal}{Phys.Lett.}
  \textbf{\bibinfo{volume}{B432}}, \bibinfo{pages}{279} (\bibinfo{year}{1998}),
  \eprint{hep-lat/9804011}.

\bibitem[{\citenamefont{Ambjorn and Loll}(1998)}]{Ambjorn:1998xu}
\bibinfo{author}{\bibfnamefont{J.}~\bibnamefont{Ambjorn}} \bibnamefont{and}
  \bibinfo{author}{\bibfnamefont{R.}~\bibnamefont{Loll}},
  \bibinfo{journal}{Nucl.Phys.} \textbf{\bibinfo{volume}{B536}},
  \bibinfo{pages}{407} (\bibinfo{year}{1998}), \eprint{hep-th/9805108}.

\bibitem[{\citenamefont{Ambjorn et~al.}(2004)\citenamefont{Ambjorn, Jurkiewicz,
  and Loll}}]{Ambjorn:2004qm}
\bibinfo{author}{\bibfnamefont{J.}~\bibnamefont{Ambjorn}},
  \bibinfo{author}{\bibfnamefont{J.}~\bibnamefont{Jurkiewicz}},
  \bibnamefont{and} \bibinfo{author}{\bibfnamefont{R.}~\bibnamefont{Loll}},
  \bibinfo{journal}{Phys. Rev. Lett.} \textbf{\bibinfo{volume}{93}},
  \bibinfo{pages}{131301} (\bibinfo{year}{2004}), \eprint{hep-th/0404156}.

\bibitem[{\citenamefont{Ambjorn
  et~al.}(2005{\natexlab{a}})\citenamefont{Ambjorn, Jurkiewicz, and
  Loll}}]{Ambjorn:2005qt}
\bibinfo{author}{\bibfnamefont{J.}~\bibnamefont{Ambjorn}},
  \bibinfo{author}{\bibfnamefont{J.}~\bibnamefont{Jurkiewicz}},
  \bibnamefont{and} \bibinfo{author}{\bibfnamefont{R.}~\bibnamefont{Loll}},
  \bibinfo{journal}{Phys. Rev.} \textbf{\bibinfo{volume}{D72}},
  \bibinfo{pages}{064014} (\bibinfo{year}{2005}{\natexlab{a}}),
  \eprint{hep-th/0505154}.

\bibitem[{\citenamefont{Ambjorn
  et~al.}(2005{\natexlab{b}})\citenamefont{Ambjorn, Jurkiewicz, and
  Loll}}]{Ambjorn:2005db}
\bibinfo{author}{\bibfnamefont{J.}~\bibnamefont{Ambjorn}},
  \bibinfo{author}{\bibfnamefont{J.}~\bibnamefont{Jurkiewicz}},
  \bibnamefont{and} \bibinfo{author}{\bibfnamefont{R.}~\bibnamefont{Loll}},
  \bibinfo{journal}{Phys. Rev. Lett.} \textbf{\bibinfo{volume}{95}},
  \bibinfo{pages}{171301} (\bibinfo{year}{2005}{\natexlab{b}}),
  \eprint{hep-th/0505113}.

\bibitem[{\citenamefont{Jordan and Loll}(2013)}]{Jordan:2013awa}
\bibinfo{author}{\bibfnamefont{S.}~\bibnamefont{Jordan}} \bibnamefont{and}
  \bibinfo{author}{\bibfnamefont{R.}~\bibnamefont{Loll}},
  \bibinfo{journal}{Phys.Lett.} \textbf{\bibinfo{volume}{B724}},
  \bibinfo{pages}{155} (\bibinfo{year}{2013}), \eprint{1305.4582}.

\bibitem[{\citenamefont{Laiho and Coumbe}(2011)}]{Laiho:2011ya}
\bibinfo{author}{\bibfnamefont{J.}~\bibnamefont{Laiho}} \bibnamefont{and}
  \bibinfo{author}{\bibfnamefont{D.}~\bibnamefont{Coumbe}},
  \bibinfo{journal}{Phys. Rev. Lett.} \textbf{\bibinfo{volume}{107}},
  \bibinfo{pages}{161301} (\bibinfo{year}{2011}), \eprint{1104.5505}.

\bibitem[{\citenamefont{Wilson}(1975)}]{Wilson:1975id}
\bibinfo{author}{\bibfnamefont{K.~G.} \bibnamefont{Wilson}}, in
  \emph{\bibinfo{booktitle}{{Boston Conf. 1975:99, Erice
  Subnucl.Phys.1975:0069}}} (\bibinfo{year}{1975}), p.~\bibinfo{pages}{99},
  \bibinfo{note}{[,0069(1975)]}.

\bibitem[{\citenamefont{Banks}(2010)}]{Banks:2010tj}
\bibinfo{author}{\bibfnamefont{T.}~\bibnamefont{Banks}} (\bibinfo{year}{2010}),
  \eprint{arXiv:1007.4001}.

\bibitem[{\citenamefont{Romer and Zahringer}(1986)}]{Romer:1985rc}
\bibinfo{author}{\bibfnamefont{H.}~\bibnamefont{Romer}} \bibnamefont{and}
  \bibinfo{author}{\bibfnamefont{M.}~\bibnamefont{Zahringer}},
  \bibinfo{journal}{Class. Quant. Grav.} \textbf{\bibinfo{volume}{3}},
  \bibinfo{pages}{897} (\bibinfo{year}{1986}).

\bibitem[{\citenamefont{Hasenfratz and Niedermayer}(1994)}]{Hasenfratz:1993sp}
\bibinfo{author}{\bibfnamefont{P.}~\bibnamefont{Hasenfratz}} \bibnamefont{and}
  \bibinfo{author}{\bibfnamefont{F.}~\bibnamefont{Niedermayer}},
  \bibinfo{journal}{Nucl. Phys.} \textbf{\bibinfo{volume}{B414}},
  \bibinfo{pages}{785} (\bibinfo{year}{1994}), \eprint{hep-lat/9308004}.

\bibitem[{\citenamefont{Fujikawa}(1983)}]{Fujikawa:1983im}
\bibinfo{author}{\bibfnamefont{K.}~\bibnamefont{Fujikawa}},
  \bibinfo{journal}{Nucl. Phys.} \textbf{\bibinfo{volume}{B226}},
  \bibinfo{pages}{437} (\bibinfo{year}{1983}).

\bibitem[{\citenamefont{Shomer}(2007)}]{Shomer:2007vq}
\bibinfo{author}{\bibfnamefont{A.}~\bibnamefont{Shomer}}
  (\bibinfo{year}{2007}), \eprint{arXiv:0709.3555}.

\bibitem[{\citenamefont{Percacci and Vacca}(2010)}]{Percacci:2010af}
\bibinfo{author}{\bibfnamefont{R.}~\bibnamefont{Percacci}} \bibnamefont{and}
  \bibinfo{author}{\bibfnamefont{G.~P.} \bibnamefont{Vacca}},
  \bibinfo{journal}{Class. Quant. Grav.} \textbf{\bibinfo{volume}{27}},
  \bibinfo{pages}{245026} (\bibinfo{year}{2010}), \eprint{1008.3621}.

\bibitem[{\citenamefont{Falls and Litim}(2014)}]{Falls:2012nd}
\bibinfo{author}{\bibfnamefont{K.}~\bibnamefont{Falls}} \bibnamefont{and}
  \bibinfo{author}{\bibfnamefont{D.~F.} \bibnamefont{Litim}},
  \bibinfo{journal}{Phys. Rev.} \textbf{\bibinfo{volume}{D89}},
  \bibinfo{pages}{084002} (\bibinfo{year}{2014}), \eprint{1212.1821}.

\bibitem[{\citenamefont{Weinberg}(1989)}]{Weinberg:1988cp}
\bibinfo{author}{\bibfnamefont{S.}~\bibnamefont{Weinberg}},
  \bibinfo{journal}{Rev. Mod. Phys.} \textbf{\bibinfo{volume}{61}},
  \bibinfo{pages}{1} (\bibinfo{year}{1989}).

\bibitem[{\citenamefont{Regge}(1961)}]{Regge:1961px}
\bibinfo{author}{\bibfnamefont{T.}~\bibnamefont{Regge}},
  \bibinfo{journal}{Nuovo Cim.} \textbf{\bibinfo{volume}{19}},
  \bibinfo{pages}{558} (\bibinfo{year}{1961}).

\bibitem[{\citenamefont{Bilke and Thorleifsson}(1999)}]{Bilke:1998bn}
\bibinfo{author}{\bibfnamefont{S.}~\bibnamefont{Bilke}} \bibnamefont{and}
  \bibinfo{author}{\bibfnamefont{G.}~\bibnamefont{Thorleifsson}},
  \bibinfo{journal}{Phys.Rev.} \textbf{\bibinfo{volume}{D59}},
  \bibinfo{pages}{124008} (\bibinfo{year}{1999}), \eprint{hep-lat/9810049}.

\bibitem[{\citenamefont{Fradkin and Vilkovisky}(1973)}]{Fradkin:1974df}
\bibinfo{author}{\bibfnamefont{E.~S.} \bibnamefont{Fradkin}} \bibnamefont{and}
  \bibinfo{author}{\bibfnamefont{G.~A.} \bibnamefont{Vilkovisky}},
  \bibinfo{journal}{Phys. Rev.} \textbf{\bibinfo{volume}{D8}},
  \bibinfo{pages}{4241} (\bibinfo{year}{1973}).

\bibitem[{\citenamefont{Unz}(1986)}]{Unz:1985wq}
\bibinfo{author}{\bibfnamefont{R.~K.} \bibnamefont{Unz}},
  \bibinfo{journal}{Nuovo Cim.} \textbf{\bibinfo{volume}{A92}},
  \bibinfo{pages}{397} (\bibinfo{year}{1986}).

\bibitem[{\citenamefont{Codello et~al.}(2008)\citenamefont{Codello, Percacci,
  and Rahmede}}]{Codello:2007bd}
\bibinfo{author}{\bibfnamefont{A.}~\bibnamefont{Codello}},
  \bibinfo{author}{\bibfnamefont{R.}~\bibnamefont{Percacci}}, \bibnamefont{and}
  \bibinfo{author}{\bibfnamefont{C.}~\bibnamefont{Rahmede}},
  \bibinfo{journal}{Int.J.Mod.Phys.} \textbf{\bibinfo{volume}{A23}},
  \bibinfo{pages}{143} (\bibinfo{year}{2008}), \eprint{arXiv:0705.1769}.

\bibitem[{\citenamefont{Codello et~al.}(2009)\citenamefont{Codello, Percacci,
  and Rahmede}}]{Codello:2008vh}
\bibinfo{author}{\bibfnamefont{A.}~\bibnamefont{Codello}},
  \bibinfo{author}{\bibfnamefont{R.}~\bibnamefont{Percacci}}, \bibnamefont{and}
  \bibinfo{author}{\bibfnamefont{C.}~\bibnamefont{Rahmede}},
  \bibinfo{journal}{Annals Phys.} \textbf{\bibinfo{volume}{324}},
  \bibinfo{pages}{414} (\bibinfo{year}{2009}), \eprint{arXiv:0805.2909}.

\bibitem[{\citenamefont{Benedetti et~al.}(2009)\citenamefont{Benedetti,
  Machado, and Saueressig}}]{Benedetti:2009rx}
\bibinfo{author}{\bibfnamefont{D.}~\bibnamefont{Benedetti}},
  \bibinfo{author}{\bibfnamefont{P.~F.} \bibnamefont{Machado}},
  \bibnamefont{and}
  \bibinfo{author}{\bibfnamefont{F.}~\bibnamefont{Saueressig}},
  \bibinfo{journal}{Mod.Phys.Lett.} \textbf{\bibinfo{volume}{A24}},
  \bibinfo{pages}{2233} (\bibinfo{year}{2009}), \eprint{arXiv:0901.2984}.

\bibitem[{\citenamefont{Falls et~al.}(2014)\citenamefont{Falls, Litim,
  Nikolakopoulos, and Rahmede}}]{Falls:2014tra}
\bibinfo{author}{\bibfnamefont{K.}~\bibnamefont{Falls}},
  \bibinfo{author}{\bibfnamefont{D.~F.} \bibnamefont{Litim}},
  \bibinfo{author}{\bibfnamefont{K.}~\bibnamefont{Nikolakopoulos}},
  \bibnamefont{and} \bibinfo{author}{\bibfnamefont{C.}~\bibnamefont{Rahmede}}
  (\bibinfo{year}{2014}), \eprint{1410.4815}.

\bibitem[{\citenamefont{Ambjorn et~al.}(1999)\citenamefont{Ambjorn,
  Anagnostopoulos, and Jurkiewicz}}]{Ambjorn:1999ix}
\bibinfo{author}{\bibfnamefont{J.}~\bibnamefont{Ambjorn}},
  \bibinfo{author}{\bibfnamefont{K.~N.} \bibnamefont{Anagnostopoulos}},
  \bibnamefont{and}
  \bibinfo{author}{\bibfnamefont{J.}~\bibnamefont{Jurkiewicz}},
  \bibinfo{journal}{JHEP} \textbf{\bibinfo{volume}{08}}, \bibinfo{pages}{016}
  (\bibinfo{year}{1999}), \eprint{hep-lat/9907027}.

\bibitem[{\citenamefont{'t~Hooft}(1979)}]{'tHooft:1978id}
\bibinfo{author}{\bibfnamefont{G.}~\bibnamefont{'t~Hooft}},
  \bibinfo{journal}{NATO Sci. Ser. B} \textbf{\bibinfo{volume}{44}},
  \bibinfo{pages}{323} (\bibinfo{year}{1979}).

\bibitem[{\citenamefont{Ambjorn et~al.}(1997)\citenamefont{Ambjorn, Durhuus,
  and Jonsson}}]{Ambjorn:1997di}
\bibinfo{author}{\bibfnamefont{J.}~\bibnamefont{Ambjorn}},
  \bibinfo{author}{\bibfnamefont{B.}~\bibnamefont{Durhuus}}, \bibnamefont{and}
  \bibinfo{author}{\bibfnamefont{T.}~\bibnamefont{Jonsson}}
  (\bibinfo{year}{1997}).

\bibitem[{\citenamefont{Gross and Varsted}(1992)}]{Gross:1991je}
\bibinfo{author}{\bibfnamefont{M.}~\bibnamefont{Gross}} \bibnamefont{and}
  \bibinfo{author}{\bibfnamefont{S.}~\bibnamefont{Varsted}},
  \bibinfo{journal}{Nucl. Phys.} \textbf{\bibinfo{volume}{B378}},
  \bibinfo{pages}{367} (\bibinfo{year}{1992}).

\bibitem[{\citenamefont{Smit}(2013)}]{Smit:2013wua}
\bibinfo{author}{\bibfnamefont{J.}~\bibnamefont{Smit}}, \bibinfo{journal}{JHEP}
  \textbf{\bibinfo{volume}{08}}, \bibinfo{pages}{016} (\bibinfo{year}{2013}),
  \bibinfo{note}{[Erratum: JHEP09,048(2015)]}, \eprint{1304.6339}.

\bibitem[{\citenamefont{Aoki}(1984)}]{Aoki:1983qi}
\bibinfo{author}{\bibfnamefont{S.}~\bibnamefont{Aoki}}, \bibinfo{journal}{Phys.
  Rev.} \textbf{\bibinfo{volume}{D30}}, \bibinfo{pages}{2653}
  (\bibinfo{year}{1984}).

\bibitem[{\citenamefont{Sharpe and Singleton}(1998)}]{Sharpe:1998xm}
\bibinfo{author}{\bibfnamefont{S.~R.} \bibnamefont{Sharpe}} \bibnamefont{and}
  \bibinfo{author}{\bibfnamefont{R.~L.} \bibnamefont{Singleton}},
  \bibinfo{journal}{Phys. Rev.} \textbf{\bibinfo{volume}{D58}},
  \bibinfo{pages}{074501} (\bibinfo{year}{1998}), \eprint{hep-lat/9804028}.

\bibitem[{\citenamefont{Bernard}(2005)}]{Bernard:2004ab}
\bibinfo{author}{\bibfnamefont{C.}~\bibnamefont{Bernard}},
  \bibinfo{journal}{Phys. Rev.} \textbf{\bibinfo{volume}{D71}},
  \bibinfo{pages}{094020} (\bibinfo{year}{2005}), \eprint{hep-lat/0412030}.

\bibitem[{\citenamefont{Ambjorn
  et~al.}(2008{\natexlab{a}})\citenamefont{Ambjorn, Gorlich, Jurkiewicz, and
  Loll}}]{Ambjorn:2008wc}
\bibinfo{author}{\bibfnamefont{J.}~\bibnamefont{Ambjorn}},
  \bibinfo{author}{\bibfnamefont{A.}~\bibnamefont{Gorlich}},
  \bibinfo{author}{\bibfnamefont{J.}~\bibnamefont{Jurkiewicz}},
  \bibnamefont{and} \bibinfo{author}{\bibfnamefont{R.}~\bibnamefont{Loll}},
  \bibinfo{journal}{Phys. Rev.} \textbf{\bibinfo{volume}{D78}},
  \bibinfo{pages}{063544} (\bibinfo{year}{2008}{\natexlab{a}}),
  \eprint{arXiv:0807.4481}.

\bibitem[{\citenamefont{Peixoto}(2014)}]{peixoto_graph-tool_2014}
\bibinfo{author}{\bibfnamefont{T.~P.} \bibnamefont{Peixoto}},
  \bibinfo{journal}{figshare}  (\bibinfo{year}{2014}),
  \urlprefix\url{http://figshare.com/articles/graph_tool/1164194}.

\bibitem[{\citenamefont{Ambjorn}(Private communication)}]{Ambjorn:pc}
\bibinfo{author}{\bibfnamefont{J.}~\bibnamefont{Ambjorn}}
  (\bibinfo{year}{Private communication}).

\bibitem[{\citenamefont{Renken et~al.}(1998)\citenamefont{Renken, Catterall,
  and Kogut}}]{Renken:1997na}
\bibinfo{author}{\bibfnamefont{R.~L.} \bibnamefont{Renken}},
  \bibinfo{author}{\bibfnamefont{S.~M.} \bibnamefont{Catterall}},
  \bibnamefont{and} \bibinfo{author}{\bibfnamefont{J.~B.} \bibnamefont{Kogut}},
  \bibinfo{journal}{Nucl. Phys.} \textbf{\bibinfo{volume}{B523}},
  \bibinfo{pages}{553} (\bibinfo{year}{1998}), \eprint{hep-lat/9712011}.

\bibitem[{\citenamefont{Thorleifsson et~al.}(1999)\citenamefont{Thorleifsson,
  Bialas, and Petersson}}]{Thorleifsson:1998qi}
\bibinfo{author}{\bibfnamefont{G.}~\bibnamefont{Thorleifsson}},
  \bibinfo{author}{\bibfnamefont{P.}~\bibnamefont{Bialas}}, \bibnamefont{and}
  \bibinfo{author}{\bibfnamefont{B.}~\bibnamefont{Petersson}},
  \bibinfo{journal}{Nucl. Phys.} \textbf{\bibinfo{volume}{B550}},
  \bibinfo{pages}{465} (\bibinfo{year}{1999}), \eprint{hep-lat/9812022}.

\bibitem[{\citenamefont{Benedetti and Gurau}(2012)}]{Benedetti:2011nn}
\bibinfo{author}{\bibfnamefont{D.}~\bibnamefont{Benedetti}} \bibnamefont{and}
  \bibinfo{author}{\bibfnamefont{R.}~\bibnamefont{Gurau}},
  \bibinfo{journal}{Nucl. Phys.} \textbf{\bibinfo{volume}{B855}},
  \bibinfo{pages}{420} (\bibinfo{year}{2012}), \eprint{1108.5389}.

\bibitem[{\citenamefont{Ambjorn
  et~al.}(2008{\natexlab{b}})\citenamefont{Ambjorn, Gorlich, Jurkiewicz, and
  Loll}}]{Ambjorn:2007jv}
\bibinfo{author}{\bibfnamefont{J.}~\bibnamefont{Ambjorn}},
  \bibinfo{author}{\bibfnamefont{A.}~\bibnamefont{Gorlich}},
  \bibinfo{author}{\bibfnamefont{J.}~\bibnamefont{Jurkiewicz}},
  \bibnamefont{and} \bibinfo{author}{\bibfnamefont{R.}~\bibnamefont{Loll}},
  \bibinfo{journal}{Phys.Rev.Lett.} \textbf{\bibinfo{volume}{100}},
  \bibinfo{pages}{091304} (\bibinfo{year}{2008}{\natexlab{b}}),
  \eprint{arXiv:0712.2485}.

\bibitem[{\citenamefont{Coumbe and Jurkiewicz}(2015)}]{Coumbe:2014noa}
\bibinfo{author}{\bibfnamefont{D.~N.} \bibnamefont{Coumbe}} \bibnamefont{and}
  \bibinfo{author}{\bibfnamefont{J.}~\bibnamefont{Jurkiewicz}},
  \bibinfo{journal}{JHEP} \textbf{\bibinfo{volume}{03}}, \bibinfo{pages}{151}
  (\bibinfo{year}{2015}), \eprint{1411.7712}.

\bibitem[{\citenamefont{Lauscher and Reuter}(2005)}]{Lauscher:2005qz}
\bibinfo{author}{\bibfnamefont{O.}~\bibnamefont{Lauscher}} \bibnamefont{and}
  \bibinfo{author}{\bibfnamefont{M.}~\bibnamefont{Reuter}},
  \bibinfo{journal}{JHEP} \textbf{\bibinfo{volume}{0510}}, \bibinfo{pages}{050}
  (\bibinfo{year}{2005}), \eprint{hep-th/0508202}.

\bibitem[{\citenamefont{Calcagni et~al.}(2014)\citenamefont{Calcagni, Oriti,
  and ThŸrigen}}]{Calcagni:2013dna}
\bibinfo{author}{\bibfnamefont{G.}~\bibnamefont{Calcagni}},
  \bibinfo{author}{\bibfnamefont{D.}~\bibnamefont{Oriti}}, \bibnamefont{and}
  \bibinfo{author}{\bibfnamefont{J.}~\bibnamefont{ThŸrigen}},
  \bibinfo{journal}{Class. Quant. Grav.} \textbf{\bibinfo{volume}{31}},
  \bibinfo{pages}{135014} (\bibinfo{year}{2014}), \eprint{1311.3340}.

\bibitem[{\citenamefont{Akkermans et~al.}(2010)\citenamefont{Akkermans, Dunne,
  and Teplyaev}}]{Akkermans:2010dz}
\bibinfo{author}{\bibfnamefont{E.}~\bibnamefont{Akkermans}},
  \bibinfo{author}{\bibfnamefont{G.~V.} \bibnamefont{Dunne}}, \bibnamefont{and}
  \bibinfo{author}{\bibfnamefont{A.}~\bibnamefont{Teplyaev}},
  \bibinfo{journal}{Phys.Rev.Lett.} \textbf{\bibinfo{volume}{105}},
  \bibinfo{pages}{230407} (\bibinfo{year}{2010}), \eprint{arXiv:1010.1148}.

\bibitem[{\citenamefont{Carlip and Grumiller}(2011)}]{Carlip:2011uc}
\bibinfo{author}{\bibfnamefont{S.}~\bibnamefont{Carlip}} \bibnamefont{and}
  \bibinfo{author}{\bibfnamefont{D.}~\bibnamefont{Grumiller}},
  \bibinfo{journal}{Phys. Rev.} \textbf{\bibinfo{volume}{D84}},
  \bibinfo{pages}{084029} (\bibinfo{year}{2011}), \eprint{1108.4686}.

\bibitem[{\citenamefont{Gastmans et~al.}(1978)\citenamefont{Gastmans, Kallosh,
  and Truffin}}]{Gastmans:1977ad}
\bibinfo{author}{\bibfnamefont{R.}~\bibnamefont{Gastmans}},
  \bibinfo{author}{\bibfnamefont{R.}~\bibnamefont{Kallosh}}, \bibnamefont{and}
  \bibinfo{author}{\bibfnamefont{C.}~\bibnamefont{Truffin}},
  \bibinfo{journal}{Nucl. Phys.} \textbf{\bibinfo{volume}{B133}},
  \bibinfo{pages}{417} (\bibinfo{year}{1978}).

\bibitem[{\citenamefont{Reuter and Saueressig}(2002)}]{Reuter:2001ag}
\bibinfo{author}{\bibfnamefont{M.}~\bibnamefont{Reuter}} \bibnamefont{and}
  \bibinfo{author}{\bibfnamefont{F.}~\bibnamefont{Saueressig}},
  \bibinfo{journal}{Phys. Rev.} \textbf{\bibinfo{volume}{D65}},
  \bibinfo{pages}{065016} (\bibinfo{year}{2002}), \eprint{hep-th/0110054}.

\bibitem[{\citenamefont{Lauscher and
  Reuter}(2002{\natexlab{a}})}]{Lauscher:2001ya}
\bibinfo{author}{\bibfnamefont{O.}~\bibnamefont{Lauscher}} \bibnamefont{and}
  \bibinfo{author}{\bibfnamefont{M.}~\bibnamefont{Reuter}},
  \bibinfo{journal}{Phys.Rev.} \textbf{\bibinfo{volume}{D65}},
  \bibinfo{pages}{025013} (\bibinfo{year}{2002}{\natexlab{a}}),
  \eprint{hep-th/0108040}.

\bibitem[{\citenamefont{Litim}(2004)}]{Litim:2003vp}
\bibinfo{author}{\bibfnamefont{D.~F.} \bibnamefont{Litim}},
  \bibinfo{journal}{Phys.Rev.Lett.} \textbf{\bibinfo{volume}{92}},
  \bibinfo{pages}{201301} (\bibinfo{year}{2004}), \eprint{hep-th/0312114}.

\bibitem[{\citenamefont{Ambjorn et~al.}(2011)\citenamefont{Ambjorn, Jordan,
  Jurkiewicz, and Loll}}]{Ambjorn:2011cg}
\bibinfo{author}{\bibfnamefont{J.}~\bibnamefont{Ambjorn}},
  \bibinfo{author}{\bibfnamefont{S.}~\bibnamefont{Jordan}},
  \bibinfo{author}{\bibfnamefont{J.}~\bibnamefont{Jurkiewicz}},
  \bibnamefont{and} \bibinfo{author}{\bibfnamefont{R.}~\bibnamefont{Loll}},
  \bibinfo{journal}{Phys. Rev. Lett.} \textbf{\bibinfo{volume}{107}},
  \bibinfo{pages}{211303} (\bibinfo{year}{2011}), \eprint{1108.3932}.

\bibitem[{\citenamefont{Coumbe et~al.}(2016)\citenamefont{Coumbe,
  Gizbert-Studnicki, and Jurkiewicz}}]{Coumbe:2015oaa}
\bibinfo{author}{\bibfnamefont{D.~N.} \bibnamefont{Coumbe}},
  \bibinfo{author}{\bibfnamefont{J.}~\bibnamefont{Gizbert-Studnicki}},
  \bibnamefont{and}
  \bibinfo{author}{\bibfnamefont{J.}~\bibnamefont{Jurkiewicz}},
  \bibinfo{journal}{JHEP} \textbf{\bibinfo{volume}{02}}, \bibinfo{pages}{144}
  (\bibinfo{year}{2016}), \eprint{1510.08672}.

\bibitem[{\citenamefont{Ambjorn et~al.}(2017)\citenamefont{Ambjorn, Coumbe,
  Gizbert-Studnicki, Gorlich, and Jurkiewicz}}]{Ambjorn:2017tnl}
\bibinfo{author}{\bibfnamefont{J.}~\bibnamefont{Ambjorn}},
  \bibinfo{author}{\bibfnamefont{D.}~\bibnamefont{Coumbe}},
  \bibinfo{author}{\bibfnamefont{J.}~\bibnamefont{Gizbert-Studnicki}},
  \bibinfo{author}{\bibfnamefont{A.}~\bibnamefont{Gorlich}}, \bibnamefont{and}
  \bibinfo{author}{\bibfnamefont{J.}~\bibnamefont{Jurkiewicz}},
  \bibinfo{journal}{Phys. Rev.} \textbf{\bibinfo{volume}{D95}},
  \bibinfo{pages}{124029} (\bibinfo{year}{2017}), \eprint{1704.04373}.

\bibitem[{\citenamefont{Ambjorn et~al.}(2000)\citenamefont{Ambjorn, Jurkiewicz,
  and Loll}}]{Ambjorn:2000dv}
\bibinfo{author}{\bibfnamefont{J.}~\bibnamefont{Ambjorn}},
  \bibinfo{author}{\bibfnamefont{J.}~\bibnamefont{Jurkiewicz}},
  \bibnamefont{and} \bibinfo{author}{\bibfnamefont{R.}~\bibnamefont{Loll}},
  \bibinfo{journal}{Phys. Rev. Lett.} \textbf{\bibinfo{volume}{85}},
  \bibinfo{pages}{924} (\bibinfo{year}{2000}), \eprint{hep-th/0002050}.

\bibitem[{\citenamefont{Reuter and Saueressig}(2011)}]{Reuter:2011ah}
\bibinfo{author}{\bibfnamefont{M.}~\bibnamefont{Reuter}} \bibnamefont{and}
  \bibinfo{author}{\bibfnamefont{F.}~\bibnamefont{Saueressig}},
  \bibinfo{journal}{JHEP} \textbf{\bibinfo{volume}{12}}, \bibinfo{pages}{012}
  (\bibinfo{year}{2011}), \eprint{1110.5224}.

\bibitem[{\citenamefont{Rechenberger and
  Saueressig}(2012)}]{Rechenberger:2012pm}
\bibinfo{author}{\bibfnamefont{S.}~\bibnamefont{Rechenberger}}
  \bibnamefont{and}
  \bibinfo{author}{\bibfnamefont{F.}~\bibnamefont{Saueressig}},
  \bibinfo{journal}{Phys. Rev.} \textbf{\bibinfo{volume}{D86}},
  \bibinfo{pages}{024018} (\bibinfo{year}{2012}), \eprint{1206.0657}.

\bibitem[{\citenamefont{Ambjorn et~al.}(2001)\citenamefont{Ambjorn, Jurkiewicz,
  and Loll}}]{Ambjorn:2001cv}
\bibinfo{author}{\bibfnamefont{J.}~\bibnamefont{Ambjorn}},
  \bibinfo{author}{\bibfnamefont{J.}~\bibnamefont{Jurkiewicz}},
  \bibnamefont{and} \bibinfo{author}{\bibfnamefont{R.}~\bibnamefont{Loll}},
  \bibinfo{journal}{Nucl. Phys.} \textbf{\bibinfo{volume}{B610}},
  \bibinfo{pages}{347} (\bibinfo{year}{2001}), \eprint{hep-th/0105267}.

\bibitem[{\citenamefont{Percacci and Perini}(2004)}]{Percacci:2004sb}
\bibinfo{author}{\bibfnamefont{R.}~\bibnamefont{Percacci}} \bibnamefont{and}
  \bibinfo{author}{\bibfnamefont{D.}~\bibnamefont{Perini}},
  \bibinfo{journal}{Class. Quant. Grav.} \textbf{\bibinfo{volume}{21}},
  \bibinfo{pages}{5035} (\bibinfo{year}{2004}), \eprint{hep-th/0401071}.

\bibitem[{\citenamefont{Lauscher and
  Reuter}(2002{\natexlab{b}})}]{Lauscher:2001rz}
\bibinfo{author}{\bibfnamefont{O.}~\bibnamefont{Lauscher}} \bibnamefont{and}
  \bibinfo{author}{\bibfnamefont{M.}~\bibnamefont{Reuter}},
  \bibinfo{journal}{Class.Quant.Grav.} \textbf{\bibinfo{volume}{19}},
  \bibinfo{pages}{483} (\bibinfo{year}{2002}{\natexlab{b}}),
  \eprint{hep-th/0110021}.

\bibitem[{\citenamefont{Rosten}(2011)}]{Rosten:2011mf}
\bibinfo{author}{\bibfnamefont{O.~J.} \bibnamefont{Rosten}}
  (\bibinfo{year}{2011}), \eprint{1106.2544}.

\bibitem[{\citenamefont{Hamber and Toriumi}(2013)}]{Hamber:2013rb}
\bibinfo{author}{\bibfnamefont{H.~W.} \bibnamefont{Hamber}} \bibnamefont{and}
  \bibinfo{author}{\bibfnamefont{R.}~\bibnamefont{Toriumi}},
  \bibinfo{journal}{Int. J. Mod. Phys.} \textbf{\bibinfo{volume}{D22}},
  \bibinfo{pages}{1330023} (\bibinfo{year}{2013}), \eprint{1301.6259}.

\bibitem[{\citenamefont{Kawai and Ninomiya}(1990)}]{Kawai:1989yh}
\bibinfo{author}{\bibfnamefont{H.}~\bibnamefont{Kawai}} \bibnamefont{and}
  \bibinfo{author}{\bibfnamefont{M.}~\bibnamefont{Ninomiya}},
  \bibinfo{journal}{Nucl. Phys.} \textbf{\bibinfo{volume}{B336}},
  \bibinfo{pages}{115} (\bibinfo{year}{1990}).

\bibitem[{\citenamefont{Kawai et~al.}(1993)\citenamefont{Kawai, Kitazawa, and
  Ninomiya}}]{Kawai:1993mb}
\bibinfo{author}{\bibfnamefont{H.}~\bibnamefont{Kawai}},
  \bibinfo{author}{\bibfnamefont{Y.}~\bibnamefont{Kitazawa}}, \bibnamefont{and}
  \bibinfo{author}{\bibfnamefont{M.}~\bibnamefont{Ninomiya}},
  \bibinfo{journal}{Nucl. Phys.} \textbf{\bibinfo{volume}{B404}},
  \bibinfo{pages}{684} (\bibinfo{year}{1993}), \eprint{hep-th/9303123}.

\bibitem[{\citenamefont{Falls}(2017)}]{Falls:2017cze}
\bibinfo{author}{\bibfnamefont{K.}~\bibnamefont{Falls}} (\bibinfo{year}{2017}),
  \eprint{1702.03577}.

\bibitem[{\citenamefont{Polyakov}(2001)}]{Polyakov:2000fk}
\bibinfo{author}{\bibfnamefont{A.~M.} \bibnamefont{Polyakov}},
  \bibinfo{journal}{Phys. Atom. Nucl.} \textbf{\bibinfo{volume}{64}},
  \bibinfo{pages}{540} (\bibinfo{year}{2001}), \bibinfo{note}{[Int. J. Mod.
  Phys.A16,4511(2001)]}, \eprint{hep-th/0006132}.

\bibitem[{\citenamefont{Jackiw et~al.}(2005)\citenamefont{Jackiw, Nunez, and
  Pi}}]{Jackiw:2005yc}
\bibinfo{author}{\bibfnamefont{R.}~\bibnamefont{Jackiw}},
  \bibinfo{author}{\bibfnamefont{C.}~\bibnamefont{Nunez}}, \bibnamefont{and}
  \bibinfo{author}{\bibfnamefont{S.~Y.} \bibnamefont{Pi}},
  \bibinfo{journal}{Phys. Lett.} \textbf{\bibinfo{volume}{A347}},
  \bibinfo{pages}{47} (\bibinfo{year}{2005}), \eprint{hep-th/0502215}.

\end{thebibliography}

\end{document}